\documentclass[aps,prd,twocolumn,noeprint,showpacs,nofootinbib,superscriptaddress]{revtex4}

\usepackage{graphicx}
\usepackage{color}
\usepackage{natbib}
\usepackage{epsfig}
\usepackage{float}
\usepackage{xcolor}
\usepackage{amsmath,amssymb,amsfonts}
\usepackage[normalem]{ulem}

\newif\iffull
\fulltrue

\newif\ifversA
\versAfalse

\newif\ifversB
\versBtrue

\newcommand{\apjl}{Astrophys. J. Lett.}%
\newcommand{\apjs}{Astrophys. J. Supp.}%
\newcommand{\aap}{Astron. Astrophys.}%
\newcommand{\mnras}{Mon. Not. Roy. Astron. Soc.}%
\newcommand{\lrr}{Living Reviews in Relativity}%
\newcommand{\pasa}{Publications of the Astronomical Society of Australia}
\newcommand{\prx}{Phys. Rev. X}
\newcommand{\araa}{Annual Review of Astronomy and Astrophysics}
\newcommand{\physrep}{Physics Reports}
\begin{document}

\title{Systematics of prompt black-hole formation in neutron star mergers}

\author{Andreas Bauswein}
\affiliation{GSI Helmholtzzentrum f\"ur Schwerionenforschung, Planckstra{\ss}e 1, 64291 Darmstadt, Germany}
\affiliation{Helmholtz Research Academy Hesse for FAIR (HFHF), GSI Helmholtz Center for Heavy Ion Research, Campus Darmstadt, Planckstra{\ss}e 1, 64291 Darmstadt, Germany}

\author{Sebastian Blacker}
\affiliation{GSI Helmholtzzentrum f\"ur Schwerionenforschung, Planckstra{\ss}e 1, 64291 Darmstadt, Germany}
\affiliation{Institut f\"ur Kernphysik, Technische Universit\"at Darmstadt, 64289 Darmstadt, Germany}

\author{Georgios Lioutas}
\affiliation{GSI Helmholtzzentrum f\"ur Schwerionenforschung, Planckstra{\ss}e 1, 64291 Darmstadt, Germany}

\author{Theodoros Soultanis}
\affiliation{Heidelberg Institute for Theoretical Studies, Schloss-Wolfsbrunnenweg 35, 69118 Heidelberg, Germany}
\affiliation{Max-Planck-Institut f\"ur Astronomie, K\"onigstuhl 17, 69117 Heidelberg, Germany}

\author{Vimal Vijayan}
\affiliation{GSI Helmholtzzentrum f\"ur Schwerionenforschung, Planckstra{\ss}e 1, 64291 Darmstadt, Germany}

\author{Nikolaos Stergioulas}
\affiliation{Department of Physics, Aristotle University of Thessaloniki, 54124 Thessaloniki, Greece}

\date{\today}

% Abbkuerzungen:
%NS
%BH
%GW
%EoS

\begin{abstract} 
This study addresses the collapse behavior of neutron star (NS) mergers expressed through the binary threshold mass $M_\mathrm{thres}$ for prompt black hole (BH) formation, which we determine by relativistic hydrodynamical simulations for a set of 40 equation of state  (EoS) models of NS matter. $M_\mathrm{thres}$ can be well described by various fit formulae involving stellar parameters of nonrotating NSs, which are employed to characterize the EoS models. Using these relations we compute which constraints on NS radii and the tidal deformability are set by current and future merger detections that reveal information about the merger product. We systematically investigate the impact of the binary mass ratio $q=M_1/M_2$ and assemble various fits, which make different assumptions about a-priori knowlegde. This includes fit formulae for $M_\mathrm{thres}$ for a fixed mass ratio or a range of $q$ if this parameter is not known very well. Also, we construct relations describing the threshold to prompt collapse for different classes of candidate EoSs, which for instance do or do not include models with a phase transition to quark matter. In particular, we find fit formulae for $M_\mathrm{thres}$ including an explicit $q$ dependence, which are valid in a broad range of $0.7\leq q \leq 1$ and which are nearly as tight as relations for fixed mass ratios. For most EoS models except for some extreme cases the threshold mass of asymmetric mergers is equal or smaller than the one of equal-mass binaries. Generally, the impact of the binary mass asymmetry on $M_\mathrm{thres}$ becomes stronger with more extreme mass ratios, while $M_\mathrm{thres}$ is approximately constant for small deviations from $q=1$, i.e. for $0.85 \leq q \leq 1$. The magnitude of the reduction of $M_\mathrm{thres}$ with the binary mass asymmetry follows a systematic EoS dependence. We also describe in more detail that a phase transition to deconfined quark matter can leave a characteristic imprint on the collapse behavior of NS mergers. The occurrence of quark matter can reduce the stability of the remnant and thus the threshold mass relative to a purely hadronic reference model. Comparing specifically the threshold mass and the combined tidal deformability $\tilde{\Lambda}_\mathrm{thres}$ of a system with $M_\mathrm{thres}$ can yield peculiar combinations of those two quantities, where $M_\mathrm{thres}$ is particularly small in relation to $\tilde{\Lambda}_\mathrm{thres}$. Since no purely hadronic EoS can yield such a combination of $M_\mathrm{thres}$ and $\tilde{\Lambda}_\mathrm{thres}$, a combined measurement or a constraint on both quantities can indicate the onset of quark deconfinement. Finally, we point out new univariate relations between $M_\mathrm{thres}$ and stellar properties of high-mass NSs, which can be employed for direct EoS constraints or consistency checks in combination with other measurements.

\end{abstract}


% Gravitaion waves 04.30.−w
% 04.30.Tv Gravitational-wave astrophysics
% 12.38.−t Quantum chromodynamics
% 12.38.Aw General properties of QCD (dynamics, confinement, etc.)
% 12.38.Mh Quark-gluon plasma
% 26.50.+x Nuclear physics aspects of novae, supernovae, and other explosive environments
% 26.60.+c Nuclear matter aspects of neutron stars
% 95.30.Lz Hydrodynamics
% 95.30.Sf Relativity and gravitation
% 95.85.Sz Gravitational radiation, magnetic fields, and other observations
% 97.60.Jd Neutron stars

\maketitle   

\section{Introduction}

% Abbkuerzungen:
%NS
%BH
%GW
%EoS

The merging of two neutron stars (NSs) leads to either the formation of a black hole (BH) or a NS remnant \cite{Shibata2005,Shibata2006,Baiotti2008}. A direct gravitational collapse takes place if the total mass of the system is higher than some threshold mass beyond which the forming remnant cannot be stabilized. For lower total binary masses rapid differential rotation and thermal pressure support the central object against prompt collapse even if the total mass exceeds the maximum mass of nonrotating NSs. Such systems may undergo a delayed collapse to a BH as a result of angular momentum losses and redistribution and cooling. The life time of the meta-stable remnant decreases with higher total masses up to the threshold where a BH forms on a dynamical time scale of about a millisecond~\cite{Faber2012,Baiotti2017,Paschalidis2017,Friedman2018,Bauswein2019b,Baiotti2019,Duez2019,Lucca2019,Shibata2019,Bernuzzi2020a,Radice2020,Friedman2020}. The outcome of a merger is predominantly determined by the total mass of the binary $M_\mathrm{tot}$ relative to the threshold binary mass for prompt BH formation $M_\mathrm{thres}$, which is thus the crucial parameter to characterize the merger product~\cite{Hotokezaka2011,Bauswein2013}.

The immediate merger outcome is highly important because it determines the properties of postmerger gravitational waves (GWs) and the features of the electromagnetic counterpart, for instance, the quasi-thermal emission powered by the radioactive decays of nuclei synthesized by the rapid neutron-capture process (so-called kilonovae) and radiation produced by relativistic outflows, e.g.~\cite{Metzger2017,Ruiz2017,Yang2018,Margalit2019,Coughlin2019a,Nakar2019,Paschalidis2019,Lue2019,Abbott2019,Gill2019,Metzger2019,Cowan2019,Foley2020,Coughlin2020,Agathos2020,Chen2020,Krueger2020,Abbott2020,Antier2020,Nathanail2020}. Therefore, it is important to understand the collapse behavior of NS mergers, which is quantitatively expressed through the threshold binary mass.

The dynamics of a NS merger in general depend on the incompletely known equation of state (EoS) of high-density matter~\cite{Oezel2016,Lattimer2016,Oertel2017}. Therefore, the EoS determines the merger product for a given binary configuration as one of the most basic characteristics. The actual threshold binary mass $M_\mathrm{thres}$ for prompt BH formation is thus currently unknown and has to be specified for every EoS model. The EoS dependence of the threshold mass can be expressed by stellar parameters of nonrotating NSs, which are uniquely linked to the EoS. In Ref.~\cite{Bauswein2013} we pointed out that the threshold mass can be well described by the maximum mass and radii of nonrotating NSs. Specifically, we found tight relations $M_\mathrm{thres}=M_\mathrm{thres}(M_\mathrm{max},R_{1.6})$ and $M_\mathrm{thres}=M_\mathrm{thres}(M_\mathrm{max},R_\mathrm{max})$ with $R_{1.6}$ and $R_\mathrm{max}$ being the radius of 1.6~$M_\odot$ NS and the radius of the nonrotating maximum-mass configuration, respectively. Because of its EoS dependence, a measurement of $M_\mathrm{thres}$, in turn, constrains the EoS of NS matter. This prospect is in particular interesting because the threshold mass probes the very high-density regime of the EoS. For instance, the inversion of $M_\mathrm{thres}(M_\mathrm{max},R_{1.6})$ with some information on $R_{1.6}$ can yield the maximum mass $M_\mathrm{max}$ of nonrotating NSs. Also, constraints on $M_\mathrm{thres}$ can be employed for current and future multi-messenger interpretations of merger events to infer NS properties like radii and tidal deformabilities as in~\cite{Bauswein2017,Radice2018,Radice2019,Koeppel2019,Bauswein2019a,Bauswein2019b,Bauswein2019c,Kiuchi2019,Capano2020,Bauswein2020a}.

The merger outcome leaves a very strong imprint on different observables, like the postmerger GW signal or kilonova features. The binary masses of a merger event are measurable with good precision at sufficiently close distances~\cite{Rodriguez2014,Farr2016} implying that $M_\mathrm{thres}$ can be determined by a number of events with different binary masses and information on the respective merger product. The system with the lowest total mass and observational indications for a prompt collapse gives an upper limit on $M_\mathrm{thres}$, whereas a detection with characteristics excluding a prompt collapse limits $M_\mathrm{thres}$ from below. In practice, only the chirp mass $\mathcal{M}=(M_1 M_2 )^{3/5} /(M_1+M_2)^{1/5}$ is measured with high precision, which provides an estimate of the total binary mass in combination with a constraint on the binary mass ratio $q=M_1/M_2$. The mass ratio may only be given with limited accuracy for events at larger distances~\cite{Rodriguez2014,Farr2016}. It is thus critical to address the influence of the mass ratio on the collapse behavior as in some cases only the total mass or the chirp mass will be known with good precision.

In Ref.~\cite{Bauswein2020a} we extended the analysis of the collapse behavior: 1) We considered a very large set of candidate EoSs to essentially cover the full range of viable hadronic models with a fine sampling. 2) We systematically determined the threshold mass for asymmetric binary mergers, which we further detail here. 3) We expressed the collapse behavior by various new functions, i.e. fit formulae linking $M_\mathrm{thres}$, $M_\mathrm{max}$ and other stellar parameters. We devised bivariate relations with $M_\mathrm{max}$ as an independent variable to be determined or constrained from measurements, but as bi-linear functions the relations can be easily inverted. These new relations are more accurate and consider different dependent variables, which might be measured more precisely. This includes in particular the (combined) tidal deformability, which is the stellar parameter describing EoS effects during the inspiral and is thus the quantity which can be directly obtained from a GW measurement, e.g.~\cite{Hinderer2008,Hinderer2010,Damour2010,Read2009,Chatziioannou2020}. The combined tidal deformability of a binary system is defined by $\tilde{\Lambda}=\frac{16}{13(M_1+M_2)^5}((M_1+12 M_2)M_1^4\Lambda_1+(M_2+12 M_1)M_2^4\Lambda_2)$, where the tidal deformabilities of the individual binary components are given by $\Lambda_{1,2}=\frac{2}{3}k_2(M_{1,2})\left(\frac{R(M_{1,2})}{M_{1,2}}\right)^5$ with the stellar radius $R$ and the tidal Love number $k_2$. These new expressions of the collapse behavior are directly relevant for more accurate EoS constraints~\cite{Bauswein2017,Radice2018,Radice2019,Koeppel2019,Bauswein2019a,Bauswein2019b,Bauswein2019c,Capano2020} from multi-messenger observations and for the implementation in analysis pipelines~\cite{Yang2018,Margalit2019,Tsang2019,Coughlin2019a,Agathos2020,Foley2020,Capano2020,Paschalidis2019,Chen2020,Coughlin2020}. 4) Finally, in~\cite{Bauswein2020a} we identified a new signature, which is indicative of the occurrence of the hadron-quark phase transition, which may or may not take place in NSs. By considering a large sample of hybrid EoSs with a first-order phase transition to deconfined quark matter~\cite{Bastian2018,Cierniak2018,Fischer2018,Bauswein2019,Bastian2020}, we find that some of those models can lead to relatively low threshold masses in comparison to the combined tidal deformability $\tilde{\Lambda}_\mathrm{thres}$ at the threshold, i.e. the tidal deformability of the system with total binary mass $M_\mathrm{tot}=M_\mathrm{thres}$. Specifically, the comparison between $M_\mathrm{thres}$ and $\tilde{\Lambda}_\mathrm{thres}$ reveals combinations of both quantities where $M_\mathrm{thres}$ is particularly small for the given $\tilde{\Lambda}_\mathrm{thres}$. Such combinations of $M_\mathrm{thres}$ and $\tilde{\Lambda}_\mathrm{thres}$ do not occur for any purely hadronic EoS within our large representative sample of models. Therefore, a combined measurement of both quantities can yield evidence for the appearance of a phase of deconfined quark matter in NSs. We refer to Ref.~\cite{Bauswein2020a} for a more detailed motivation and discussion of these different aspects and their applications. 

Here we discuss more details of the collapse behavior of NS mergers as follow-up to our previous study. First, we describe a more complete and more comprehensive set of fit formulae for the binary threshold mass for prompt BH formation. As one example, these fit formulae can be used for constraints on NS properties like the radius that can be inferred from merger events which provide information about the merger outcome. Second, we explicitly quantify which EoS constraints are implied by a measurement of a merger with a given total mass. Hence, the resulting formula can be immediately applied to any new detection. Third, we quantify the impact of the binary mass ratio on the stability of the merger remnant. In particular, we develop fit formulae for the threshold mass valid for a range of binary mass ratios, which are as tight as relations for a fixed mass ratio. We also provide a tentative explanation of the mass ratio impact on the merger stability based on a semi-analytic model. Fourth, we discuss in more detail the recently proposed signature of the hadron-quark phase transition, which is based on the collapse behavior of NS mergers~\cite{Bauswein2020a}. We extend these considerations to asymmetric binaries and we describe general EoS dependencies of the collapse behavior, which may be useful for additional constraints on the EoS. Fifth, we point out new univariate relations between the threshold mass $M_\mathrm{thres}$ and stellar properties of high-mass NSs, which can be employed for EoS constraints and consistency checks in GW data analysis. Interestingly, these relations are insensitive to the presence of a phase transition.
 
One of our main findings is that $M_\mathrm{thres}$ depends on the binary mass ratio in a systematic way, while previous studies have only tentatively assessed the impact of $q$~\cite{Bauswein2013,Bauswein2017a,Kiuchi2019,Bernuzzi2020}. For most EoS models the threshold mass of asymmetric mergers is smaller than or equal to the one of the equal-mass binary. Importantly, the difference in the threshold mass between equal-mass binaries and asymmetric mergers is influenced by the EoS. We find general formulae for the threshold binary mass which include the mass ratio as a parameter and still yield an accurate description of $M_\mathrm{thres}$ with a precision comparable to that of formulae for fixed $q$.

Understanding the impact of $q$ on the collapse behavior is important for at least two reasons. First, as mentioned the binary mass ratio may not be inferred with high precision from a measurement with lower signal-to-noise ratio. Hence, in practice relations for $M_\mathrm{thres}$ which are valid for a range of mass ratios are often required. Recall that for instance the mass ratio of GW170817 was found to be in the range $0.7\leq q \leq 1$~\cite{Abbott2017,Abbott2019}. Second, for determining the threshold mass for prompt collapse, different detections with information on the merger outcome and measured total binary mass (or chirp mass) have to be combined. Even if those measurements yield relatively precise $q$ estimates, the different events may have different mass ratios. Therefore, it is indispensable to quantify the detailed influence of $q$ on $M_\mathrm{thres}$ especially if the threshold mass is employed for EoS constraints and for the interpretation of merger observations (including the implementation in analysis pipelines). See for instance~\cite{Bauswein2017,Yang2018,Margalit2019,Tsang2019,Coughlin2019a,Koeppel2019,Breschi2019,Agathos2020,Foley2020,Capano2020,Paschalidis2019,Chen2020,Coughlin2020} for a direct application of relations describing $M_\mathrm{thres}$. 

This paper is organized as follows. Sect.~\ref{sec:eos} provides background information on the simulations and in particular on the EoS sample studied in this paper and in Ref.~\cite{Bauswein2020a}. We introduce different classes of EoSs, which are motivated by which additional information on the EoS may be available. We describe different relations quantifying the collapse behavior of NS mergers and describe their application for EoS constraints in Sect.~\ref{sec:mthr}. This section also includes a discussion of current and future constraints on NS properties inferred from the collapse behavior. In Sect.~\ref{sec:q} we focus on the impact of the mass ratio on $M_\mathrm{thres}$ and develop generalized fit formulae, which explicitly include $q$. Section~\ref{sec:toy} presents an intuitive toy model, which reproduces our findings. Section~\ref{sec:pt} discusses the aforementioned signature of the hadron-quark phase transition, which is based on a comparison between $M_\mathrm{thres}$ and $\tilde{\Lambda}_\mathrm{thres}$. We provide more details on the impact of the mass ratio and describe the influence of general EoS properties. We present some additional, useful univariate relations linking the collapse behavior of NS mergers, i.e. $M_\mathrm{thres}$, and properties of high-mass NSs in Sect.~\ref{sec:add}. We summarize in Sect.~\ref{sec:sum}.

In this paper masses refer to the gravitational mass. Binary masses including $M_\mathrm{thres}$ are considered at infinite orbital separation. We define the mass ratio as $q=M_1/M_2$ with $M_1 \leq M_2$.

\section{Equation of state sample and simulation data} \label{sec:eos}

In this paper we extend our study of the collapse behavior of NS mergers in~\cite{Bauswein2020a} by providing a more detailed discussion and pointing out additional relations. We describe results for the same set of NS merger simulations considering the same 40 EoS models~\cite{Banik2014,Fortin2018,Marques2017,Hempel2010,Typel2010,Typel2005,Alvarez-Castillo2016,Akmal1998,Goriely2010,Wiringa1988,Lattimer1991,Shen2011,Lalazissis1997a,Hempel2012,Douchin2001,Steiner2013,Muther1987,Alford2005,Engvik1996,Schneider2019,Read2009a,Lackey2006,Glendenning1985,Shen2011,Lattimer1991,Lalazissis1997a,Sugahara1994a,Toki1995,Alvarez-Castillo2016,Kaltenborn2017,Bastian2018,Cierniak2018,Fischer2018,Bauswein2019,Bastian2020}. These high-density models include a subset of EoSs with a phase transition to deconfined quark matter. We refer the reader to Ref.~\cite{Bauswein2020a} and its Supplemental Material, where more details about the simulations and in particular the EoS models can be found. 

For the sake of completeness we summarize some main features of the simulation data discussed here. We group the 40 EoS models in three subsets.

(a) We identify a ``base sample'' containing 23 purely hadronic models which are compatible with current astrophysical constraints~\cite{Banik2014,Fortin2018,Marques2017,Hempel2010,Typel2010,Typel2005,Alvarez-Castillo2016,Akmal1998,Goriely2010,Wiringa1988,Lattimer1991,Shen2011,Lalazissis1997a,Hempel2012,Douchin2001,Steiner2013,Muther1987,Alford2005,Engvik1996,Schneider2019,Read2009a}. Specifically, we require those models to be consistent with the limits on the maximum mass on nonrotating NSs inferred from pulsar observations~\cite{Antoniadis2013,Arzoumanian2018a} and with constraints on the tidal deformability from GW170817~\cite{Abbott2017}.

(b) We extend this set of models with an ``excluded hadronic sample'' which contains 8 EoSs~\cite{Lackey2006,Glendenning1985,Shen2011,Lattimer1991,Lalazissis1997a,Sugahara1994a,Hempel2012,Toki1995}. These models describe purely hadronic matter and are incompatible with the aforementioned astrophysical constraints. Mostly, they are in tension with the tidal deformability limit and at least marginally in agreement with current knowledge about $M_\mathrm{max}$~\cite{Antoniadis2013,Arzoumanian2018a}. Generally, these EoSs are rather stiff and lead to relatively large NS radii. 

(c) Finally, we consider a ``hybrid sample'' of 9 models which include a phase transition to deconfined quark matter~\cite{Alvarez-Castillo2016,Kaltenborn2017,Bastian2018,Cierniak2018,Fischer2018,Bauswein2019,Bastian2020}. These EoSs employ the same description of hadronic matter~\cite{Typel2005,Typel2010,Alvarez-Castillo2016}, i.e. at densities below the phase transition density. The models make different assumptions about the properties of quark matter, which leads to different stellar parameters of hybrid stars. Note that also the onset density of the phase transition varies among the models. All EoSs of this set are compatible with the astrophysical constraints mentioned above; and the hadronic part at lower and moderate densities is consistent with experimental constraints from~\cite{Danielewicz2002,Tsang2018,Lattimer2013,Oertel2017,Krueger2013}.

The classification of the different EoS models can be found in Tab.~\ref{tab:data} in Appendix~\ref{app:data} that contains also the corresponding references and acronyms, which we will employ here. The table lists also certain stellar properties and the simulation results. Considering the stellar properties we argue that our set of hadronic EoSs covers the full range of possible models in the sense that the true EoS should have stellar parameters which are close to one of the models within our sample. This does not apply to the hybrid sample, since all those models employ the same low-density EoS and the possibility of a phase transition introduces additional degrees of freedom, e.g. the onset density, the latent heat and the exact properties of the quark phase. However, the hadronic part of the hybrid EoSs falls roughly in the middle of the allowed range of stellar parameters. This choice thus represents an EoS from which only moderate deviations may be expected. The quark matter EoSs of the hybrid sample vary significantly leading to a broad variety of different stellar parameters (see also Figs.~4 and~5 in~\cite{Bauswein2019a}). We also refer to Fig.~1 of the Supplemental Material of~\cite{Bauswein2020a} illustrating the variety of EoS models by their mass-radius relations.

For all EoS models we determine the threshold binary mass $M_\mathrm{thres}$ for fixed mass ratios $q=M_1/M_2=1$, $q=0.85$ and $q=0.7$. For this we simulate specific binary setups by running calculations with a relativistic smooth particle hydrodynamics (SPH) code, which adopts the conformal flatness condition to solve the Einstein field equations~\cite{Isenberg1980,Wilson1996,Oechslin2002,Oechslin2007,Bauswein2010,Bauswein2012a}. {For every calculation with a given total binary mass we check the evolution of the minimum lapse function $\alpha_\mathrm{min}$. If $\alpha_\mathrm{min}$ continuously decreases, we classify the merger as a prompt collapse event. An increasing $\alpha_\mathrm{min}$ after merging instead indicates a bounce of the merger components and we regard the outcome as no prompt collapse. Within our sample of calculations for different total binary masses $M_\mathrm{tot}$ and fixed mass ratio $q$ we then identify the lightest system with $M_\mathrm{tot}=M_\mathrm{unstab}$ leading to a prompt collapse and the most massive binary with $M_\mathrm{tot}=M_\mathrm{stab}$ which does not undergo a direct collapse. As in previous publications we define $M_\mathrm{thres}=0.5(M_\mathrm{unstab}+M_\mathrm{stab})$. We thus determine $M_\mathrm{thres}$ with an accuracy of $\pm 0.5(M_\mathrm{unstab}-M_\mathrm{stab})$, which is at least $\pm 0.025~M_\odot$ for every EoS model and binary mass ratio in this study.} The results and more details about the simulations and their setup can be found in the Supplemental Material of~\cite{Bauswein2020a} and references therein. We run in total more than 400 simulations with about 300,000 SPH particles to obtain $M_\mathrm{thres}$ for the different setups. For some selected EoS models we run additional simulations for mass ratios $q=0.9$, $q=0.8$, $=0.6$ and $q=0.5$, which will be further discussed in Sect.~\ref{sec:q}.

We emphasize that most EoS models are temperature dependent. This holds in particular for the EoS tables of the hybrid sample. Other EoSs which are only provided as barotropic relations at zero-temperature are treated with an approximate inclusion of thermal pressure. For these models we set the thermal ideal-gas index $\Gamma_\mathrm{th}=1.75$ (see~\cite{Bauswein2010} for an assessment and justification of this choice).

To develop a sense for the importance of thermal effects in determining $M_\mathrm{thres}$, we perform a number of simulations with the approximate treatment of thermal pressure for EoS models for which the full temperature dependence is available. We consider the purely hadronic SFHX EoS and the hybrid model DD2F-SF3 and adopt different choices for the thermal ideal-gas index $\Gamma_\mathrm{th}$, which controlls the strength of thermal pressure support. Usually, $\Gamma_\mathrm{th}=1.75$ is a good choice to reproduce results of fully temperature-dependent tables, and, furthermore, $\Gamma_\mathrm{th}=1.75$ approximately equals an average value of $\Gamma_\mathrm{th}$ which one would directly extract from a temperature-dependent EoS table of hadronic models (see~\cite{Bauswein2010}).

For the purely hadronic EoS SFHX we find that $\Gamma_\mathrm{th}=1.75$ leads to the same $M_\mathrm{thres}$ within the precision to which we determine the threshold mass in this study\footnote{For both EoS models we simulate systems with $q=1$ and the same total binary masses as those we considered to determine $M_\mathrm{thres}$ with the full tables, i.e. we check the merger outcome for 2.95~$M_\odot$ and 3.0~$M_\odot$ for SFHX, and 2.8~$M_\odot$ and 2.85~$M_\odot$ for DD2F-SF3.}. $\Gamma_\mathrm{th}=1.5$ slightly underestimates the threshold mass in comparison to the simulation with the full temperature dependence. For the hybrid model DD2F-SF3, the calculation with $\Gamma_\mathrm{th}=1.75$ yields a too high threshold mass, and only reducing the thermal ideal-gas index to 1.5 or 1.334 yields the same result as the temperature-dependent EoS table. This is understandable because the quark phase has a weak thermal pressure support with $\Gamma_\mathrm{th}$ closer to 4/3, which is much lower than $\Gamma_\mathrm{th}$ in the hadronic phase. Based on previous work, e.g.~\cite{Bauswein2010}, and this limited set of additional calculations we conclude that $\Gamma_\mathrm{th}=1.75$ works well for hadronic models, but choosing $\Gamma_\mathrm{th}$ might be less obvious for hybrid models considering that the phase boundaries are temperature dependent as well. These calculations show that the inclusion of thermal effects is generally important, while depending on the required precision an ansatz with an approximate thermal pressure component can work well at least for hadronic models.

\begin{figure*}
\centering
\includegraphics[width=0.98\columnwidth,trim=40 40 40 60,clip]{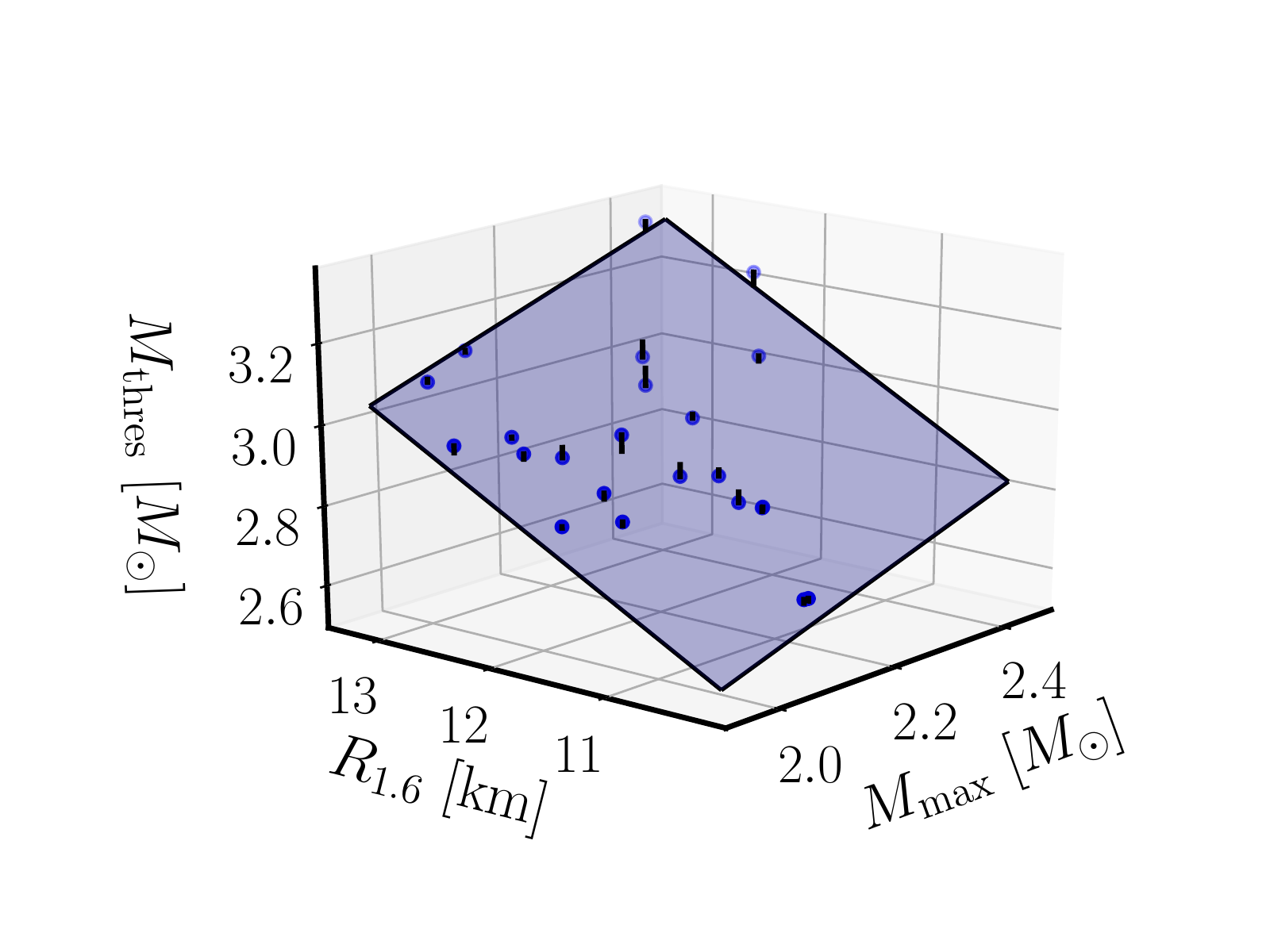}
\includegraphics[width=0.98\columnwidth,trim=40 40 40 60,clip]{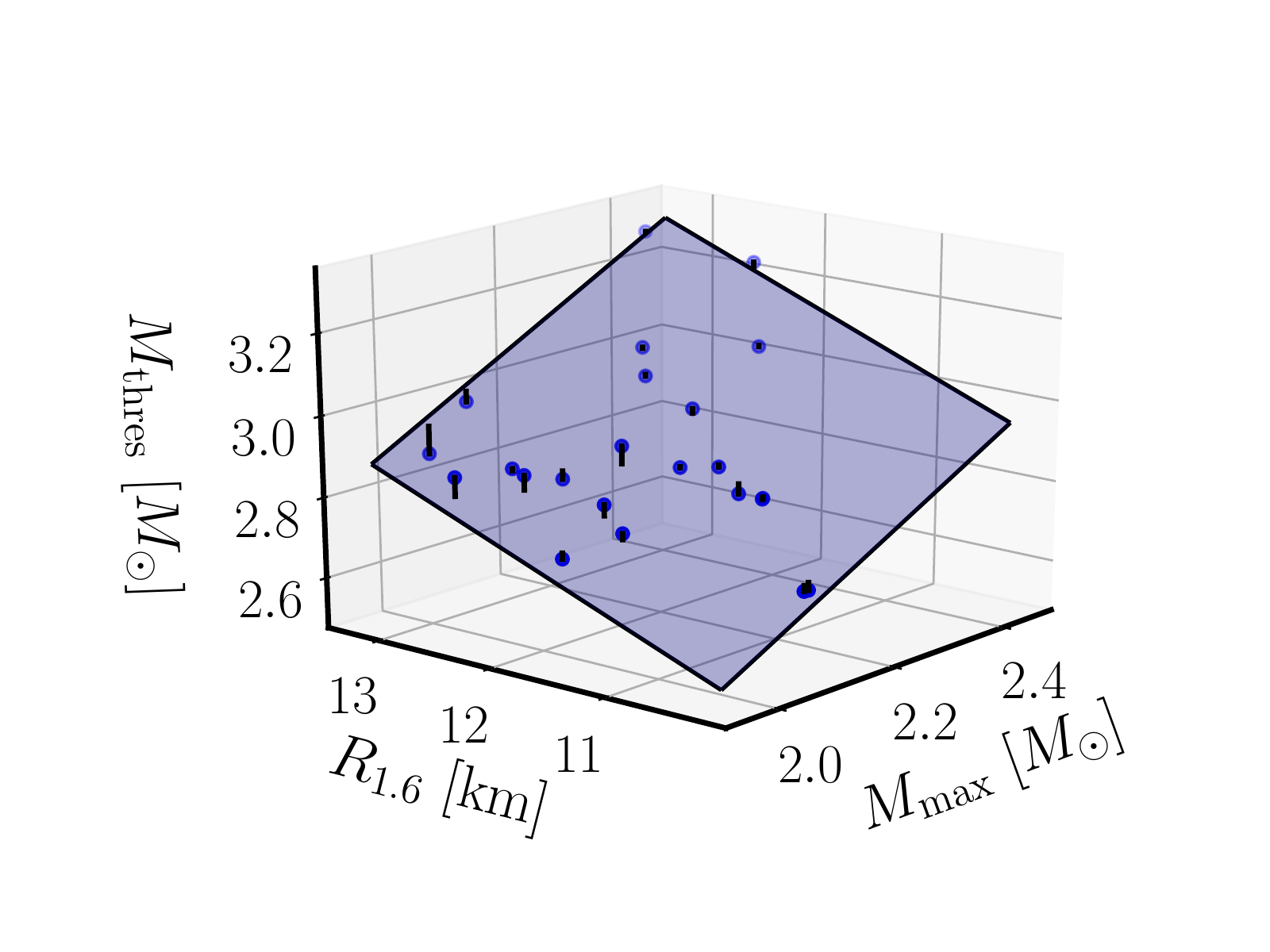}\\
\includegraphics[width=0.98\columnwidth,trim=40 40 40 60,clip]{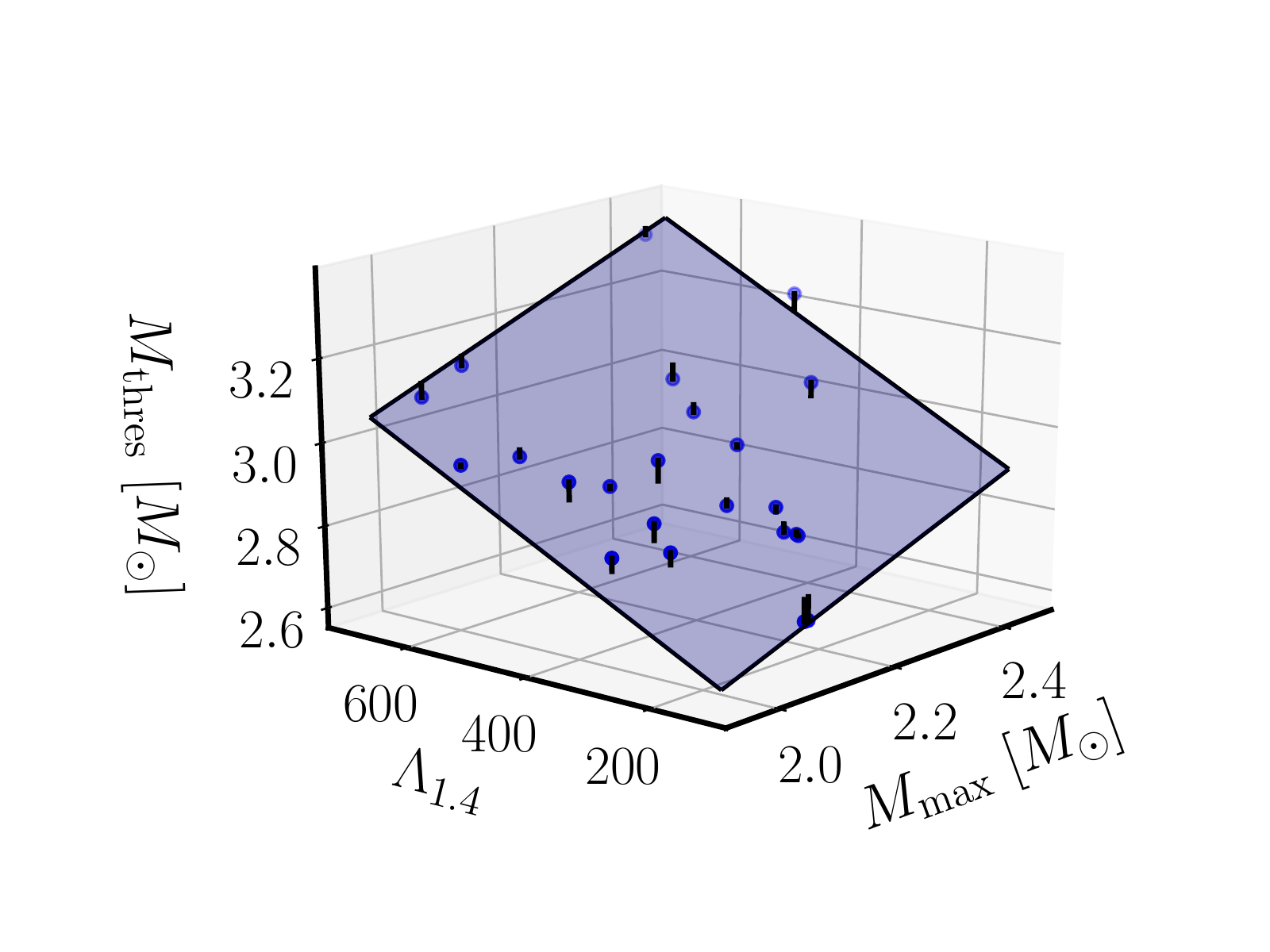}
\includegraphics[width=0.98\columnwidth,trim=40 40 40 60,clip]{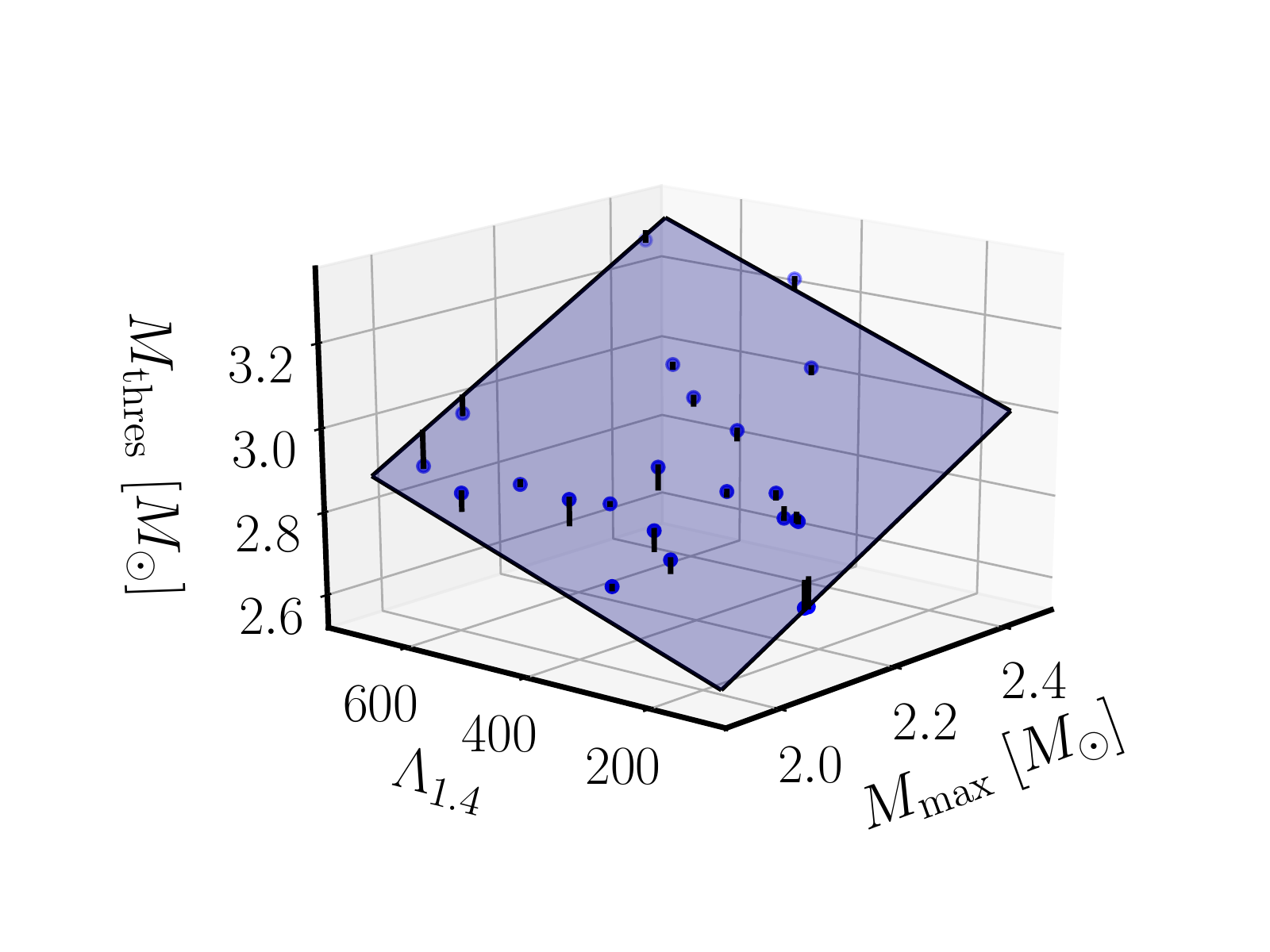}
\caption{Upper panels: Threshold binary mass $M_\mathrm{thres}$ for prompt BH formation as function of the maximum mass $M_\mathrm{max}$ of nonrotating NSs and the radius of a 1.6~$M_\odot$ NS for $q=M_1/M_2=1$ (left) and $q=0.7$ (right) with the base EoS sample. Blue plane is a bilinear fit (see Tab.~\ref{tab:mthr1}) to the data (blue points). Short black lines visualize deviations between fit and data. Lower panels: Same as the upper panels but with the tidal deformability $\Lambda_{1.4}$ of a 1.4~$M_\odot$ NS instead of $R_{1.6}$ ($q=1$ on the left, $q=0.7$ on the right; see Tab.~\ref{tab:mthr2}).}
\label{fig:mthr}
\end{figure*}

\begin{table*} 

\begin{tabular}{|l|l|c|c|c|c|c|c|c|c|c|}  \hline 
no. & fit $M_\mathrm{thres}(X,Y)$ & EoSs & $q$ & $a$ & $b$ & $c$ & max. & av. &  $\sum$ sq. res. & N \\ \hline 
1 & $a M_\mathrm{max} + b R_{1.6} + c$ & b & 1 & 
0.547 $\pm$ 0.036 & 0.1647 $\pm$ 0.006 & -0.198 $\pm$ 0.099 & 0.042 & 0.016 & 0.0096 & 23 \\ \hline % Mthres-R16-Mmax-q1-b-computed.pdf 0.77 
2 & $a M_\mathrm{max} + b R_{1.6} + c$ & b & 0.85 & 
0.629 $\pm$ 0.034 & 0.1486 $\pm$ 0.006 & -0.181 $\pm$ 0.093 & 0.041 & 0.016 & 0.0085 & 23 \\ \hline % Mthres-R16-Mmax-q085-b-computed.pdf 0.68 
3 & $a M_\mathrm{max} + b R_{1.6} + c$ & b & 0.7 & 
0.832 $\pm$ 0.042 & 0.1161 $\pm$ 0.007 & -0.276 $\pm$ 0.117 & 0.067 & 0.017 & 0.0134 & 23 \\ \hline % Mthres-R16-Mmax-q07-b-computed.pdf 1.07 
4 & $a M_\mathrm{max} + b R_{1.6} + c$ & b & 1, 0.85 & 
0.588 $\pm$ 0.026 & 0.1566 $\pm$ 0.004 & -0.189 $\pm$ 0.070 & 0.043 & 0.018 & 0.0209 & 46 \\ \hline % Mthres-R16-Mmax-q1085-b-computed.pdf 0.78 
5 & $a M_\mathrm{max} + b R_{1.6} + c$ & b & 1, 0.7 & 
0.689 $\pm$ 0.048 & 0.1404 $\pm$ 0.008 & -0.237 $\pm$ 0.132 & 0.137 & 0.028 & 0.0736 & 46 \\ \hline % Mthres-R16-Mmax-q107-b-computed.pdf 2.74 
5 * & $a M_\mathrm{max} + b R_{1.6} + c$ & b & 1, 0.7 & 
0.689 $\pm$ 0.048 & 0.1404 $\pm$ 0.008 & -0.237 $\pm$ 0.132 & 0.137 & 0.028 & 0.0736 & 46 \\ \hline % check085Mthres-R16-Mmax-q107-b-computed.pdf 2.74 
6 & $a M_\mathrm{max} + b R_{1.6} + c$ & b & 1, 0.85, 0.7 & 
0.669 $\pm$ 0.035 & 0.1431 $\pm$ 0.006 & -0.218 $\pm$ 0.097 & 0.151 & 0.027 & 0.0919 & 69 \\ \hline % Mthres-R16-Mmax-q108507-b-computed.pdf 2.23 
7 & $a M_\mathrm{max} + b R_{1.6} + c$ & b + h + e & 1 & 
0.610 $\pm$ 0.035 & 0.1641 $\pm$ 0.007 & -0.342 $\pm$ 0.086 & 0.107 & 0.031 & 0.0602 & 40 \\ \hline % Mthres-R16-Mmax-q1-bhe-computed.pdf 2.60 
8 & $a M_\mathrm{max} + b R_{1.6} + c$ & b + h + e & 0.85 & 
0.746 $\pm$ 0.039 & 0.1391 $\pm$ 0.008 & -0.341 $\pm$ 0.098 & 0.102 & 0.035 & 0.0782 & 40 \\ \hline % Mthres-R16-Mmax-q085-bhe-computed.pdf 3.38 
9 & $a M_\mathrm{max} + b R_{1.6} + c$ & b + h + e & 0.7 & 
0.914 $\pm$ 0.039 & 0.1121 $\pm$ 0.008 & -0.424 $\pm$ 0.098 & 0.117 & 0.033 & 0.0780 & 40 \\ \hline % Mthres-R16-Mmax-q07-bhe-computed.pdf 3.37 
10 & $a M_\mathrm{max} + b R_{1.6} + c$ & b + h + e & 1, 0.85 & 
0.678 $\pm$ 0.028 & 0.1516 $\pm$ 0.005 & -0.341 $\pm$ 0.069 & 0.117 & 0.034 & 0.1594 & 80 \\ \hline % Mthres-R16-Mmax-q1085-bhe-computed.pdf 3.31 
11 & $a M_\mathrm{max} + b R_{1.6} + c$ & b + h + e & 1, 0.7 & 
0.762 $\pm$ 0.039 & 0.1381 $\pm$ 0.008 & -0.383 $\pm$ 0.096 & 0.162 & 0.045 & 0.3115 & 80 \\ \hline % Mthres-R16-Mmax-q107-bhe-computed.pdf 6.47 
11 * & $a M_\mathrm{max} + b R_{1.6} + c$ & b + h + e & 1, 0.7 & 
0.762 $\pm$ 0.039 & 0.1381 $\pm$ 0.008 & -0.383 $\pm$ 0.096 & 0.162 & 0.045 & 0.3115 & 80 \\ \hline % check085Mthres-R16-Mmax-q107-bhe-computed.pdf 6.47 
12 & $a M_\mathrm{max} + b R_{1.6} + c$ & b + h + e & 1, 0.85, 0.7 & 
0.757 $\pm$ 0.029 & 0.1384 $\pm$ 0.006 & -0.369 $\pm$ 0.072 & 0.170 & 0.043 & 0.4003 & 120 \\ \hline % Mthres-R16-Mmax-q108507-bhe-computed.pdf 5.47 
13 & $a M_\mathrm{max} + b R_{1.6} + c$ & b + h & 1 & 
0.617 $\pm$ 0.061 & 0.1581 $\pm$ 0.011 & -0.287 $\pm$ 0.185 & 0.104 & 0.033 & 0.0529 & 32 \\ \hline % Mthres-R16-Mmax-q1-bh-computed.pdf 2.92 
14 & $a M_\mathrm{max} + b R_{1.6} + c$ & b + e & 1 & 
0.581 $\pm$ 0.022 & 0.1632 $\pm$ 0.004 & -0.254 $\pm$ 0.053 & 0.054 & 0.018 & 0.0161 & 31 \\ \hline % Mthres-R16-Mmax-q1-be-computed.pdf 0.92 
\hline 
15 & $a M_\mathrm{max} + b R_\mathrm{max} + c$ & b & 1 & 
0.450 $\pm$ 0.043 & 0.1891 $\pm$ 0.008 & -0.011 $\pm$ 0.114 & 0.059 & 0.018 & 0.0137 & 23 \\ \hline % Mthres-Rmax-Mmax-q1-b-computed.pdf 1.10 
16 & $a M_\mathrm{max} + b R_\mathrm{max} + c$ & b & 0.85 & 
0.544 $\pm$ 0.051 & 0.1687 $\pm$ 0.009 & 0.003 $\pm$ 0.133 & 0.071 & 0.022 & 0.0188 & 23 \\ \hline % Mthres-Rmax-Mmax-q085-b-computed.pdf 1.50 
17 & $a M_\mathrm{max} + b R_\mathrm{max} + c$ & b & 0.7 & 
0.764 $\pm$ 0.047 & 0.1331 $\pm$ 0.009 & -0.144 $\pm$ 0.123 & 0.054 & 0.021 & 0.0160 & 23 \\ \hline % Mthres-Rmax-Mmax-q07-b-computed.pdf 1.28 
18 & $a M_\mathrm{max} + b R_\mathrm{max} + c$ & b & 1, 0.85 & 
0.497 $\pm$ 0.034 & 0.1789 $\pm$ 0.006 & -0.004 $\pm$ 0.088 & 0.078 & 0.021 & 0.0357 & 46 \\ \hline % Mthres-Rmax-Mmax-q1085-b-computed.pdf 1.33 
19 & $a M_\mathrm{max} + b R_\mathrm{max} + c$ & b & 1, 0.7 & 
0.607 $\pm$ 0.051 & 0.1611 $\pm$ 0.009 & -0.077 $\pm$ 0.133 & 0.119 & 0.031 & 0.0802 & 46 \\ \hline % Mthres-Rmax-Mmax-q107-b-computed.pdf 2.98 
19 * & $a M_\mathrm{max} + b R_\mathrm{max} + c$ & b & 1, 0.7 & 
0.607 $\pm$ 0.051 & 0.1611 $\pm$ 0.009 & -0.077 $\pm$ 0.133 & 0.119 & 0.031 & 0.0802 & 46 \\ \hline % check085Mthres-Rmax-Mmax-q107-b-computed.pdf 2.98 
20 & $a M_\mathrm{max} + b R_\mathrm{max} + c$ & b & 1, 0.85, 0.7 & 
0.586 $\pm$ 0.039 & 0.1636 $\pm$ 0.007 & -0.051 $\pm$ 0.102 & 0.132 & 0.029 & 0.1085 & 69 \\ \hline % Mthres-Rmax-Mmax-q108507-b-computed.pdf 2.63 
21 & $a M_\mathrm{max} + b R_\mathrm{max} + c$ & b + h + e & 1 & 
0.507 $\pm$ 0.038 & 0.1885 $\pm$ 0.008 & -0.135 $\pm$ 0.085 & 0.126 & 0.030 & 0.0662 & 40 \\ \hline % Mthres-Rmax-Mmax-q1-bhe-computed.pdf 2.86 
22 & $a M_\mathrm{max} + b R_\mathrm{max} + c$ & b + h + e & 0.85 & 
0.654 $\pm$ 0.038 & 0.1617 $\pm$ 0.008 & -0.176 $\pm$ 0.085 & 0.100 & 0.032 & 0.0663 & 40 \\ \hline % Mthres-Rmax-Mmax-q085-bhe-computed.pdf 2.87 
23 & $a M_\mathrm{max} + b R_\mathrm{max} + c$ & b + h + e & 0.7 & 
0.833 $\pm$ 0.034 & 0.1330 $\pm$ 0.007 & -0.307 $\pm$ 0.075 & 0.079 & 0.030 & 0.0511 & 40 \\ \hline % Mthres-Rmax-Mmax-q07-bhe-computed.pdf 2.21 
24 & $a M_\mathrm{max} + b R_\mathrm{max} + c$ & b + h + e & 1, 0.85 & 
0.580 $\pm$ 0.028 & 0.1751 $\pm$ 0.006 & -0.156 $\pm$ 0.063 & 0.131 & 0.032 & 0.1520 & 80 \\ \hline % Mthres-Rmax-Mmax-q1085-bhe-computed.pdf 3.16 
25 & $a M_\mathrm{max} + b R_\mathrm{max} + c$ & b + h + e & 1, 0.7 & 
0.670 $\pm$ 0.039 & 0.1608 $\pm$ 0.008 & -0.221 $\pm$ 0.086 & 0.164 & 0.046 & 0.2838 & 80 \\ \hline % Mthres-Rmax-Mmax-q107-bhe-computed.pdf 5.90 
25 * & $a M_\mathrm{max} + b R_\mathrm{max} + c$ & b + h + e & 1, 0.7 & 
0.670 $\pm$ 0.039 & 0.1608 $\pm$ 0.008 & -0.221 $\pm$ 0.086 & 0.164 & 0.046 & 0.2838 & 80 \\ \hline % check085Mthres-Rmax-Mmax-q107-bhe-computed.pdf 5.90 
26 & $a M_\mathrm{max} + b R_\mathrm{max} + c$ & b + h + e & 1, 0.85, 0.7 & 
0.665 $\pm$ 0.029 & 0.1611 $\pm$ 0.006 & -0.206 $\pm$ 0.065 & 0.157 & 0.042 & 0.3606 & 120 \\ \hline % Mthres-Rmax-Mmax-q108507-bhe-computed.pdf 4.93 
27 & $a M_\mathrm{max} + b R_\mathrm{max} + c$ & b + h & 1 & 
0.550 $\pm$ 0.058 & 0.1689 $\pm$ 0.011 & -0.028 $\pm$ 0.161 & 0.092 & 0.029 & 0.0468 & 32 \\ \hline % Mthres-Rmax-Mmax-q1-bh-computed.pdf 2.58 
28 & $a M_\mathrm{max} + b R_\mathrm{max} + c$ & b + e & 1 & 
0.445 $\pm$ 0.026 & 0.1970 $\pm$ 0.005 & -0.079 $\pm$ 0.056 & 0.069 & 0.019 & 0.0195 & 31 \\ \hline % Mthres-Rmax-Mmax-q1-be-computed.pdf 1.11 
\end{tabular} 
\caption{Different bilinear fits (second column) describing the EoS dependence of the threshold binary mass $M_\mathrm{thres}$ for prompt BH formation (see main text). Third column specifies the set of EoSs used for the fit (``b'' $\equiv$ hadronic base sample (a), ``e'' $\equiv$ exlcuded hadronic sample (b), ``h'' $\equiv$ hybrid sample (c) as defined  in Sect.~\ref{sec:eos}). $q$ is the binary mass ratio of the underlying data. Fifth to seventh columns provide the fit parameters $a$, $b$ and $c$ and their respective variances with units such that masses are in $M_\odot$ and radii are in~km. The next two columns specify the maximum and average deviation between fit and the underlying data (in $M_\odot$). Last two columns give the sum of the squared residuals being minimized by the fit procedure and the number of data points included in the fit. For fits marked with an asterisk we compute deviations between fit and the data comparing additionally to the results for $q=0.85$ which are not employed for the fit.} 
\label{tab:mthr1} 
\end{table*}

\begin{table*} 
\begin{tabular}{|l|l|c|c|c|c|c|c|c|c|c|}  \hline 
no. & fit $M_\mathrm{thres}(X,Y)$ & EoSs & $q$ & $a$ & $b/10^{-4}$ & $c$ & max. & av. &  $\sum$ sq. res. & N \\ \hline 
29 & $a M_\mathrm{max} + b \Lambda_{1.4} + c$ & b & 1 & 
0.589 $\pm$ 0.052 & 7.973 $\pm$ 0.416 & 1.359 $\pm$ 0.112 & 0.056 & 0.025 & 0.0201 & 23 \\ \hline % Mthres-Lam14-Mmax-q1-b-computed.pdf 1.61 
30 & $a M_\mathrm{max} + b \Lambda_{1.4} + c$ & b & 0.85 & 
0.668 $\pm$ 0.055 & 7.126 $\pm$ 0.443 & 1.226 $\pm$ 0.119 & 0.060 & 0.027 & 0.0228 & 23 \\ \hline % Mthres-Lam14-Mmax-q085-b-computed.pdf 1.82 
31 & $a M_\mathrm{max} + b \Lambda_{1.4} + c$ & b & 0.7 & 
0.863 $\pm$ 0.061 & 5.469 $\pm$ 0.494 & 0.825 $\pm$ 0.133 & 0.081 & 0.027 & 0.0284 & 23 \\ \hline % Mthres-Lam14-Mmax-q07-b-computed.pdf 2.27 
32 & $a M_\mathrm{max} + b \Lambda_{1.4} + c$ & b & 1, 0.85 & 
0.629 $\pm$ 0.038 & 7.550 $\pm$ 0.303 & 1.293 $\pm$ 0.082 & 0.066 & 0.026 & 0.0460 & 46 \\ \hline % Mthres-Lam14-Mmax-q1085-b-computed.pdf 1.71 
33 & $a M_\mathrm{max} + b \Lambda_{1.4} + c$ & b & 1, 0.7 & 
0.726 $\pm$ 0.056 & 6.721 $\pm$ 0.449 & 1.092 $\pm$ 0.121 & 0.157 & 0.035 & 0.1007 & 46 \\ \hline % Mthres-Lam14-Mmax-q107-b-computed.pdf 3.75 
33 * & $a M_\mathrm{max} + b \Lambda_{1.4} + c$ & b & 1, 0.7 & 
0.726 $\pm$ 0.056 & 6.721 $\pm$ 0.449 & 1.092 $\pm$ 0.121 & 0.157 & 0.035 & 0.1007 & 46 \\ \hline % check085Mthres-Lam14-Mmax-q107-b-computed.pdf 3.75 
34 & $a M_\mathrm{max} + b \Lambda_{1.4} + c$ & b & 1, 0.85, 0.7 & 
0.707 $\pm$ 0.042 & 6.856 $\pm$ 0.340 & 1.137 $\pm$ 0.092 & 0.171 & 0.033 & 0.1333 & 69 \\ \hline % Mthres-Lam14-Mmax-q108507-b-computed.pdf 3.23 
35 & $a M_\mathrm{max} + b \Lambda_{1.4} + c$ & b + h + e & 1 & 
0.509 $\pm$ 0.047 & 6.026 $\pm$ 0.324 & 1.595 $\pm$ 0.094 & 0.095 & 0.041 & 0.0982 & 40 \\ \hline % Mthres-Lam14-Mmax-q1-bhe-computed.pdf 4.24 
36 & $a M_\mathrm{max} + b \Lambda_{1.4} + c$ & b + h + e & 0.85 & 
0.659 $\pm$ 0.048 & 5.122 $\pm$ 0.330 & 1.303 $\pm$ 0.096 & 0.088 & 0.045 & 0.1023 & 40 \\ \hline % Mthres-Lam14-Mmax-q085-bhe-computed.pdf 4.42 
37 & $a M_\mathrm{max} + b \Lambda_{1.4} + c$ & b + h + e & 0.7 & 
0.846 $\pm$ 0.047 & 4.109 $\pm$ 0.322 & 0.898 $\pm$ 0.094 & 0.106 & 0.042 & 0.0975 & 40 \\ \hline % Mthres-Lam14-Mmax-q07-bhe-computed.pdf 4.21 
38 & $a M_\mathrm{max} + b \Lambda_{1.4} + c$ & b + h + e & 1, 0.85 & 
0.584 $\pm$ 0.034 & 5.574 $\pm$ 0.238 & 1.449 $\pm$ 0.069 & 0.107 & 0.045 & 0.2208 & 80 \\ \hline % Mthres-Lam14-Mmax-q1085-bhe-computed.pdf 4.59 
39 & $a M_\mathrm{max} + b \Lambda_{1.4} + c$ & b + h + e & 1, 0.7 & 
0.677 $\pm$ 0.044 & 5.067 $\pm$ 0.307 & 1.246 $\pm$ 0.089 & 0.150 & 0.055 & 0.3675 & 80 \\ \hline % Mthres-Lam14-Mmax-q107-bhe-computed.pdf 7.64 
39 * & $a M_\mathrm{max} + b \Lambda_{1.4} + c$ & b + h + e & 1, 0.7 & 
0.677 $\pm$ 0.044 & 5.067 $\pm$ 0.307 & 1.246 $\pm$ 0.089 & 0.150 & 0.055 & 0.3675 & 80 \\ \hline % check085Mthres-Lam14-Mmax-q107-bhe-computed.pdf 7.64 
40 & $a M_\mathrm{max} + b \Lambda_{1.4} + c$ & b + h + e & 1, 0.85, 0.7 & 
0.671 $\pm$ 0.034 & 5.085 $\pm$ 0.232 & 1.265 $\pm$ 0.068 & 0.155 & 0.052 & 0.4804 & 120 \\ \hline % Mthres-Lam14-Mmax-q108507-bhe-computed.pdf 6.57 
41 & $a M_\mathrm{max} + b \Lambda_{1.4} + c$ & b + h & 1 & 
0.622 $\pm$ 0.063 & 7.837 $\pm$ 0.570 & 1.282 $\pm$ 0.135 & 0.097 & 0.034 & 0.0554 & 32 \\ \hline % Mthres-Lam14-Mmax-q1-bh-computed.pdf 3.05 
42 & $a M_\mathrm{max} + b \Lambda_{1.4} + c$ & b + e & 1 & 
0.492 $\pm$ 0.046 & 5.959 $\pm$ 0.303 & 1.645 $\pm$ 0.093 & 0.105 & 0.035 & 0.0643 & 31 \\ \hline % Mthres-Lam14-Mmax-q1-be-computed.pdf 3.67 
\hline 
43 & $a M_\mathrm{max} + b \tilde{\Lambda}_\mathrm{thres} + c$ & b & 1 & 
1.441 $\pm$ 0.101 & 27.52 $\pm$ 2.239 & -0.909 $\pm$ 0.261 & 0.085 & 0.037 & 0.0455 & 23 \\ \hline % Mthres-Lamthres-Mmax-q1-b-computed.pdf 3.64 
44 & $a M_\mathrm{max} + b \tilde{\Lambda}_\mathrm{thres} + c$ & b & 0.85 & 
1.409 $\pm$ 0.105 & 19.70 $\pm$ 1.902 & -0.641 $\pm$ 0.264 & 0.088 & 0.040 & 0.0500 & 23 \\ \hline % Mthres-Lamthres-Mmax-q085-b-computed.pdf 4.00 
45 & $a M_\mathrm{max} + b \tilde{\Lambda}_\mathrm{thres} + c$ & b & 0.7 & 
1.448 $\pm$ 0.109 & 9.484 $\pm$ 1.235 & -0.544 $\pm$ 0.264 & 0.112 & 0.037 & 0.0512 & 23 \\ \hline % Mthres-Lamthres-Mmax-q07-b-computed.pdf 4.10 
46 & $a M_\mathrm{max} + b \tilde{\Lambda}_\mathrm{thres} + c$ & b & 1, 0.85 & 
1.401 $\pm$ 0.080 & 22.46 $\pm$ 1.575 & -0.693 $\pm$ 0.202 & 0.109 & 0.044 & 0.1237 & 46 \\ \hline % Mthres-Lamthres-Mmax-q1085-b-computed.pdf 4.60 
47 & $a M_\mathrm{max} + b \tilde{\Lambda}_\mathrm{thres} + c$ & b & 1, 0.7 & 
1.130 $\pm$ 0.141 & 8.211 $\pm$ 1.875 & 0.235 $\pm$ 0.341 & 0.228 & 0.077 & 0.4328 & 46 \\ \hline % Mthres-Lamthres-Mmax-q107-b-computed.pdf 16.10 
47 * & $a M_\mathrm{max} + b \tilde{\Lambda}_\mathrm{thres} + c$ & b & 1, 0.7 & 
1.130 $\pm$ 0.141 & 8.211 $\pm$ 1.875 & 0.235 $\pm$ 0.341 & 0.228 & 0.077 & 0.4328 & 46 \\ \hline % check085Mthres-Lamthres-Mmax-q107-b-computed.pdf 16.10 
48 & $a M_\mathrm{max} + b \tilde{\Lambda}_\mathrm{thres} + c$ & b & 1, 0.85, 0.7 & 
1.152 $\pm$ 0.110 & 9.767 $\pm$ 1.572 & 0.157 $\pm$ 0.267 & 0.230 & 0.078 & 0.6012 & 69 \\ \hline % Mthres-Lamthres-Mmax-q108507-b-computed.pdf 14.57 
49 & $a M_\mathrm{max} + b \tilde{\Lambda}_\mathrm{thres} + c$ & b + h + e & 1 & 
1.389 $\pm$ 0.122 & 19.29 $\pm$ 3.467 & -0.577 $\pm$ 0.341 & 0.275 & 0.095 & 0.5545 & 40 \\ \hline % Mthres-Lamthres-Mmax-q1-bhe-computed.pdf 23.98 
50 & $a M_\mathrm{max} + b \tilde{\Lambda}_\mathrm{thres} + c$ & b + h + e & 0.85 & 
1.400 $\pm$ 0.104 & 11.91 $\pm$ 2.120 & -0.422 $\pm$ 0.273 & 0.209 & 0.087 & 0.4141 & 40 \\ \hline % Mthres-Lamthres-Mmax-q085-bhe-computed.pdf 17.91 
51 & $a M_\mathrm{max} + b \tilde{\Lambda}_\mathrm{thres} + c$ & b + h + e & 0.7 & 
1.458 $\pm$ 0.086 & 6.494 $\pm$ 1.138 & -0.479 $\pm$ 0.219 & 0.192 & 0.067 & 0.2795 & 40 \\ \hline % Mthres-Lamthres-Mmax-q07-bhe-computed.pdf 12.09 
52 & $a M_\mathrm{max} + b \tilde{\Lambda}_\mathrm{thres} + c$ & b + h + e & 1, 0.85 & 
1.352 $\pm$ 0.082 & 13.43 $\pm$ 1.904 & -0.344 $\pm$ 0.220 & 0.282 & 0.096 & 1.0914 & 80 \\ \hline % Mthres-Lamthres-Mmax-q1085-bhe-computed.pdf 22.68 
53 & $a M_\mathrm{max} + b \tilde{\Lambda}_\mathrm{thres} + c$ & b + h + e & 1, 0.7 & 
1.207 $\pm$ 0.090 & 4.317 $\pm$ 1.391 & 0.213 $\pm$ 0.226 & 0.367 & 0.105 & 1.4837 & 80 \\ \hline % Mthres-Lamthres-Mmax-q107-bhe-computed.pdf 30.83 
53 * & $a M_\mathrm{max} + b \tilde{\Lambda}_\mathrm{thres} + c$ & b + h + e & 1, 0.7 & 
1.207 $\pm$ 0.090 & 4.317 $\pm$ 1.391 & 0.213 $\pm$ 0.226 & 0.367 & 0.105 & 1.4837 & 80 \\ \hline % check085Mthres-Lamthres-Mmax-q107-bhe-computed.pdf 30.83 
54 & $a M_\mathrm{max} + b \tilde{\Lambda}_\mathrm{thres} + c$ & b + h + e & 1, 0.85, 0.7 & 
1.237 $\pm$ 0.071 & 5.585 $\pm$ 1.174 & 0.113 $\pm$ 0.179 & 0.355 & 0.106 & 2.0506 & 120 \\ \hline % Mthres-Lamthres-Mmax-q108507-bhe-computed.pdf 28.04 
55 & $a M_\mathrm{max} + b \tilde{\Lambda}_\mathrm{thres} + c$ & b + h & 1 & 
1.090 $\pm$ 0.194 & 12.09 $\pm$ 3.498 & 0.242 $\pm$ 0.490 & 0.184 & 0.081 & 0.2945 & 32 \\ \hline % Mthres-Lamthres-Mmax-q1-bh-computed.pdf 16.25 
56 & $a M_\mathrm{max} + b \tilde{\Lambda}_\mathrm{thres} + c$ & b + e & 1 & 
1.505 $\pm$ 0.067 & 30.20 $\pm$ 2.136 & -1.108 $\pm$ 0.190 & 0.165 & 0.047 & 0.1172 & 31 \\ \hline % Mthres-Lamthres-Mmax-q1-be-computed.pdf 6.70 
\end{tabular} 
\caption{Different bilinear fits (second column) describing the EoS dependence of the threshold binary mass $M_\mathrm{thres}$ for prompt BH formation (see main txet). Third column specifies the set of EoSs used for the fit (``b'' $\equiv$ hadronic base sample (a), ``e'' $\equiv$ exlcuded hadronic sample (b), ``h'' $\equiv$ hybrid sample (c) as defined  in Sect.~\ref{sec:eos}). $q$ is the binary mass ratio of the underlying data. Fifth to seventh columns provide the fit parameters $a$, $b$ and $c$ and their respective variances with units such that masses are in $M_\odot$. The next two columns specify the maximum and average deviation between fit and the underlying data (in $M_\odot$). Last two columns give the sum of the squared residuals being minimized by the fit procedure and the number of data points included in the fit. For fits marked with an asterisk we compute deviations between fit and the data comparing additionally to the results for $q=0.85$ which are not employed for the fit.}
\label{tab:mthr2} 
\end{table*}

\begin{table*} 
\begin{tabular}{|l|l|c|c|c|c|c|c|c|c|c|}  \hline 
no. & fit  $\mathcal{M}_\mathrm{thres}(X,Y)$ & EoSs & $q$ & $a$ & $b/10^{-2}$ & $c$ & max. & av. & $\sum$ sq. res. & N \\ \hline 
57 & $a M_\mathrm{max} + b R_{1.6} + c$ & b & 1 & 
0.238 $\pm$ 0.016 & 7.168 $\pm$ 0.255 & -0.086 $\pm$ 0.043 & 0.018 & 0.007 & 0.0018 & 23 \\ \hline % Mchthres-R16-Mmax-q1-b-computed.pdf 0.77 
58 & $a M_\mathrm{max} + b R_{1.6} + c$ & b & 0.85 & 
0.273 $\pm$ 0.015 & 6.443 $\pm$ 0.239 & -0.078 $\pm$ 0.040 & 0.018 & 0.007 & 0.0016 & 23 \\ \hline % Mchthres-R16-Mmax-q085-b-computed.pdf 0.68 
59 & $a M_\mathrm{max} + b R_{1.6} + c$ & b & 0.7 & 
0.355 $\pm$ 0.018 & 4.958 $\pm$ 0.296 & -0.118 $\pm$ 0.050 & 0.029 & 0.007 & 0.0024 & 23 \\ \hline % Mchthres-R16-Mmax-q07-b-computed.pdf 1.03 
60 & $a M_\mathrm{max} + b R_{1.6} + c$ & b & 1, 0.85 & 
0.255 $\pm$ 0.011 & 6.806 $\pm$ 0.186 & -0.082 $\pm$ 0.031 & 0.021 & 0.008 & 0.0042 & 46 \\ \hline % Mchthres-R16-Mmax-q1085-b-computed.pdf 0.82 
61 & $a M_\mathrm{max} + b R_{1.6} + c$ & b & 1, 0.7 & 
0.297 $\pm$ 0.031 & 6.063 $\pm$ 0.505 & -0.102 $\pm$ 0.085 & 0.071 & 0.021 & 0.0307 & 46 \\ \hline % Mchthres-R16-Mmax-q107-b-computed.pdf 6.02 
61 * & $a M_\mathrm{max} + b R_{1.6} + c$ & b & 1, 0.7 & 
0.297 $\pm$ 0.031 & 6.063 $\pm$ 0.505 & -0.102 $\pm$ 0.085 & 0.071 & 0.021 & 0.0307 & 46 \\ \hline % check085Mchthres-R16-Mmax-q107-b-computed.pdf 6.02 
62 & $a M_\mathrm{max} + b R_{1.6} + c$ & b & 1, 0.85, 0.7 & 
0.289 $\pm$ 0.022 & 6.190 $\pm$ 0.365 & -0.094 $\pm$ 0.062 & 0.079 & 0.019 & 0.0370 & 69 \\ \hline % Mchthres-R16-Mmax-q108507-b-computed.pdf 4.73 
63 & $a M_\mathrm{max} + b R_{1.6} + c$ & b + h + e & 1 & 
0.265 $\pm$ 0.015 & 7.143 $\pm$ 0.294 & -0.149 $\pm$ 0.037 & 0.047 & 0.013 & 0.0114 & 40 \\ \hline % Mchthres-R16-Mmax-q1-bhe-computed.pdf 2.60 
64 & $a M_\mathrm{max} + b R_{1.6} + c$ & b + h + e & 0.85 & 
0.323 $\pm$ 0.017 & 6.032 $\pm$ 0.334 & -0.148 $\pm$ 0.043 & 0.044 & 0.015 & 0.0147 & 40 \\ \hline % Mchthres-R16-Mmax-q085-bhe-computed.pdf 3.35 
65 & $a M_\mathrm{max} + b R_{1.6} + c$ & b + h + e & 0.7 & 
0.390 $\pm$ 0.017 & 4.788 $\pm$ 0.329 & -0.181 $\pm$ 0.042 & 0.050 & 0.014 & 0.0142 & 40 \\ \hline % Mchthres-R16-Mmax-q07-bhe-computed.pdf 3.25 
66 & $a M_\mathrm{max} + b R_{1.6} + c$ & b + h + e & 1, 0.85 & 
0.294 $\pm$ 0.012 & 6.587 $\pm$ 0.241 & -0.148 $\pm$ 0.031 & 0.053 & 0.015 & 0.0319 & 80 \\ \hline % Mchthres-R16-Mmax-q1085-bhe-computed.pdf 3.50 
67 & $a M_\mathrm{max} + b R_{1.6} + c$ & b + h + e & 1, 0.7 & 
0.328 $\pm$ 0.022 & 5.965 $\pm$ 0.424 & -0.165 $\pm$ 0.054 & 0.081 & 0.028 & 0.0985 & 80 \\ \hline % Mchthres-R16-Mmax-q107-bhe-computed.pdf 10.80 
67 * & $a M_\mathrm{max} + b R_{1.6} + c$ & b + h + e & 1, 0.7 & 
0.328 $\pm$ 0.022 & 5.965 $\pm$ 0.424 & -0.165 $\pm$ 0.054 & 0.081 & 0.028 & 0.0985 & 80 \\ \hline % check085Mchthres-R16-Mmax-q107-bhe-computed.pdf 10.80 
68 & $a M_\mathrm{max} + b R_{1.6} + c$ & b + h + e & 1, 0.85, 0.7 & 
0.326 $\pm$ 0.016 & 5.988 $\pm$ 0.309 & -0.159 $\pm$ 0.039 & 0.086 & 0.025 & 0.1196 & 120 \\ \hline % Mchthres-R16-Mmax-q108507-bhe-computed.pdf 8.63 
69 & $a M_\mathrm{max} + b R_{1.6} + c$ & b + h & 1 & 
0.269 $\pm$ 0.027 & 6.880 $\pm$ 0.488 & -0.125 $\pm$ 0.080 & 0.045 & 0.014 & 0.0100 & 32 \\ \hline % Mchthres-R16-Mmax-q1-bh-computed.pdf 2.92 
70 & $a M_\mathrm{max} + b R_{1.6} + c$ & b + e & 1 & 
0.253 $\pm$ 0.009 & 7.102 $\pm$ 0.176 & -0.111 $\pm$ 0.023 & 0.023 & 0.008 & 0.0030 & 31 \\ \hline % Mchthres-R16-Mmax-q1-be-computed.pdf 0.92 
\hline 
71 & $a M_\mathrm{max} + b R_\mathrm{max} + c$ & b & 1 & 
0.196 $\pm$ 0.019 & 8.229 $\pm$ 0.352 & -0.005 $\pm$ 0.049 & 0.026 & 0.008 & 0.0026 & 23 \\ \hline % Mchthres-Rmax-Mmax-q1-b-computed.pdf 1.10 
72 & $a M_\mathrm{max} + b R_\mathrm{max} + c$ & b & 0.85 & 
0.236 $\pm$ 0.022 & 7.313 $\pm$ 0.410 & 0.001 $\pm$ 0.058 & 0.031 & 0.010 & 0.0035 & 23 \\ \hline % Mchthres-Rmax-Mmax-q085-b-computed.pdf 1.49 
73 & $a M_\mathrm{max} + b R_\mathrm{max} + c$ & b & 0.7 & 
0.326 $\pm$ 0.020 & 5.684 $\pm$ 0.372 & -0.061 $\pm$ 0.052 & 0.023 & 0.009 & 0.0029 & 23 \\ \hline % Mchthres-Rmax-Mmax-q07-b-computed.pdf 1.23 
74 & $a M_\mathrm{max} + b R_\mathrm{max} + c$ & b & 1, 0.85 & 
0.216 $\pm$ 0.015 & 7.771 $\pm$ 0.277 & -0.002 $\pm$ 0.039 & 0.037 & 0.009 & 0.0070 & 46 \\ \hline % Mchthres-Rmax-Mmax-q1085-b-computed.pdf 1.37 
75 & $a M_\mathrm{max} + b R_\mathrm{max} + c$ & b & 1, 0.7 & 
0.261 $\pm$ 0.032 & 6.957 $\pm$ 0.594 & -0.033 $\pm$ 0.084 & 0.063 & 0.022 & 0.0319 & 46 \\ \hline % Mchthres-Rmax-Mmax-q107-b-computed.pdf 6.27 
75 * & $a M_\mathrm{max} + b R_\mathrm{max} + c$ & b & 1, 0.7 & 
0.261 $\pm$ 0.032 & 6.957 $\pm$ 0.594 & -0.033 $\pm$ 0.084 & 0.063 & 0.022 & 0.0319 & 46 \\ \hline % check085Mchthres-Rmax-Mmax-q107-b-computed.pdf 6.27 
76 & $a M_\mathrm{max} + b R_\mathrm{max} + c$ & b & 1, 0.85, 0.7 & 
0.253 $\pm$ 0.024 & 7.075 $\pm$ 0.439 & -0.022 $\pm$ 0.062 & 0.071 & 0.019 & 0.0401 & 69 \\ \hline % Mchthres-Rmax-Mmax-q108507-b-computed.pdf 5.13 
77 & $a M_\mathrm{max} + b R_\mathrm{max} + c$ & b + h + e & 1 & 
0.221 $\pm$ 0.017 & 8.206 $\pm$ 0.356 & -0.059 $\pm$ 0.037 & 0.055 & 0.013 & 0.0125 & 40 \\ \hline % Mchthres-Rmax-Mmax-q1-bhe-computed.pdf 2.86 
78 & $a M_\mathrm{max} + b R_\mathrm{max} + c$ & b + h + e & 0.85 & 
0.283 $\pm$ 0.017 & 7.012 $\pm$ 0.355 & -0.076 $\pm$ 0.037 & 0.043 & 0.014 & 0.0125 & 40 \\ \hline % Mchthres-Rmax-Mmax-q085-bhe-computed.pdf 2.84 
79 & $a M_\mathrm{max} + b R_\mathrm{max} + c$ & b + h + e & 0.7 & 
0.356 $\pm$ 0.014 & 5.682 $\pm$ 0.307 & -0.131 $\pm$ 0.032 & 0.034 & 0.013 & 0.0093 & 40 \\ \hline % Mchthres-Rmax-Mmax-q07-bhe-computed.pdf 2.13 
80 & $a M_\mathrm{max} + b R_\mathrm{max} + c$ & b + h + e & 1, 0.85 & 
0.252 $\pm$ 0.013 & 7.609 $\pm$ 0.272 & -0.068 $\pm$ 0.028 & 0.059 & 0.014 & 0.0305 & 80 \\ \hline % Mchthres-Rmax-Mmax-q1085-bhe-computed.pdf 3.35 
81 & $a M_\mathrm{max} + b R_\mathrm{max} + c$ & b + h + e & 1, 0.7 & 
0.288 $\pm$ 0.022 & 6.944 $\pm$ 0.476 & -0.095 $\pm$ 0.050 & 0.083 & 0.028 & 0.0934 & 80 \\ \hline % Mchthres-Rmax-Mmax-q107-bhe-computed.pdf 10.24 
81 * & $a M_\mathrm{max} + b R_\mathrm{max} + c$ & b + h + e & 1, 0.7 & 
0.288 $\pm$ 0.022 & 6.944 $\pm$ 0.476 & -0.095 $\pm$ 0.050 & 0.083 & 0.028 & 0.0934 & 80 \\ \hline % check085Mchthres-Rmax-Mmax-q107-bhe-computed.pdf 10.24 
82 & $a M_\mathrm{max} + b R_\mathrm{max} + c$ & b + h + e & 1, 085, 0.7 & 
0.287 $\pm$ 0.016 & 6.966 $\pm$ 0.345 & -0.089 $\pm$ 0.036 & 0.081 & 0.024 & 0.1122 & 120 \\ \hline % Mchthres-Rmax-Mmax-q108507-bhe-computed.pdf 8.10 
83 & $a M_\mathrm{max} + b R_\mathrm{max} + c$ & b + h & 1 & 
0.240 $\pm$ 0.025 & 7.351 $\pm$ 0.486 & -0.012 $\pm$ 0.070 & 0.040 & 0.013 & 0.0089 & 32 \\ \hline % Mchthres-Rmax-Mmax-q1-bh-computed.pdf 2.58 
84 & $a M_\mathrm{max} + b R_\mathrm{max} + c$ & b + e & 1 & 
0.194 $\pm$ 0.011 & 8.573 $\pm$ 0.234 & -0.034 $\pm$ 0.024 & 0.030 & 0.008 & 0.0037 & 31 \\ \hline % Mchthres-Rmax-Mmax-q1-be-computed.pdf 1.11 
\end{tabular} 
\caption{Different bilinear fits (second column) describing the EoS dependence of the  threshold chirp mass $\mathcal{M}_\mathrm{thres}$ for prompt BH formation (see main text). Third column specifies the set of EoSs used for the fit (``b'' $\equiv$ hadronic base sample (a), ``e'' $\equiv$ exlcuded hadronic sample (b), ``h'' $\equiv$ hybrid sample (c) as defined  in Sect.~\ref{sec:eos}). $q$ is the binary mass ratio of the underlying data. Fifth to seventh columns provide the fit parameters $a$, $b$ and $c$ and their respective variance with units such that masses are in $M_\odot$ and radii are in~km. The next two columns specify the maximum and average deviation between fit and the underlying data (in $M_\odot$). Last two columns give the sum of the squared residuals being minimized by the fit procedure and the number of data points included in the fit. For fits marked with an asterisk we compute deviations between fit and the data comparing additionally to the results for $q=0.85$ which are not employed for the fit.} 
\label{tab:mch1} 
\end{table*}

\begin{table*} 
\begin{tabular}{|l|l|c|c|c|c|c|c|c|c|c|}  \hline 
no. & fit  $\mathcal{M}_\mathrm{thres}(X,Y)$ & EoSs & $q$ & $a$ & $b/10^{-4}$ & $c$ & max. & av. &  $\sum$ sq. res. & N \\ \hline 
85 & $a M_\mathrm{max} + b \Lambda_{1.4} + c$ & b & 1 & 
0.257 $\pm$ 0.023 & $3.470 \pm 0.181$ & 0.592 $\pm$ 0.049 & 0.025 & 0.011 & 0.0038 & 23 \\ \hline % Mchthres-Lam14-Mmax-q1-b-computed.pdf 1.61 
86 & $a M_\mathrm{max} + b \Lambda_{1.4} + c$ & b & 0.85 & 
0.289 $\pm$ 0.024 & $3.090\pm 0.192$ & 0.531 $\pm$ 0.052 & 0.026 & 0.012 & 0.0043 & 23 \\ \hline % Mchthres-Lam14-Mmax-q085-b-computed.pdf 1.81 
87 & $a M_\mathrm{max} + b \Lambda_{1.4} + c$ & b & 0.7 & 
0.369 $\pm$ 0.026 & $2.336\pm 0.211$ & 0.352 $\pm$ 0.057 & 0.035 & 0.012 & 0.0052 & 23 \\ \hline % Mchthres-Lam14-Mmax-q07-b-computed.pdf 2.19 
88 & $a M_\mathrm{max} + b \Lambda_{1.4} + c$ & b & 1, 0.85 & 
0.273 $\pm$ 0.017 & $3.280\pm 0.134$ & 0.562 $\pm$ 0.036 & 0.031 & 0.011 & 0.0089 & 46 \\ \hline % Mchthres-Lam14-Mmax-q1085-b-computed.pdf 1.75 
89 & $a M_\mathrm{max} + b \Lambda_{1.4} + c$ & b & 1, 0.7 & 
0.313 $\pm$ 0.033 & $2.903\pm 0.267$ & 0.472 $\pm$ 0.072 & 0.080 & 0.023 & 0.0357 & 46 \\ \hline % Mchthres-Lam14-Mmax-q107-b-computed.pdf 7.01 
89 * & $a M_\mathrm{max} + b \Lambda_{1.4} + c$ & b & 1, 0.7 & 
0.313 $\pm$ 0.033 & $2.903\pm 0.267$ & 0.472 $\pm$ 0.072 & 0.080 & 0.023 & 0.0357 & 46 \\ \hline % check085Mchthres-Lam14-Mmax-q107-b-computed.pdf 7.01 
90 & $a M_\mathrm{max} + b \Lambda_{1.4} + c$ & b & 1, 0.85, 0.7 & 
0.305 $\pm$ 0.024 & $2.965\pm 0.197$ & 0.492 $\pm$ 0.053 & 0.088 & 0.020 & 0.0447 & 69 \\ \hline % Mchthres-Lam14-Mmax-q108507-b-computed.pdf 5.72 
91 & $a M_\mathrm{max} + b \Lambda_{1.4} + c$ & b + h + e & 1 & 
0.221 $\pm$ 0.020 & $2.623\pm0.141$ & 0.694 $\pm$ 0.041 & 0.041 & 0.018 & 0.0186 & 40 \\ \hline % Mchthres-Lam14-Mmax-q1-bhe-computed.pdf 4.24 
92 & $a M_\mathrm{max} + b \Lambda_{1.4} + c$ & b + h + e & 0.85 & 
0.286 $\pm$ 0.021 & $2.221\pm0.143$ & 0.565 $\pm$ 0.042 & 0.038 & 0.020 & 0.0192 & 40 \\ \hline % Mchthres-Lam14-Mmax-q085-bhe-computed.pdf 4.39 
93 & $a M_\mathrm{max} + b \Lambda_{1.4} + c$ & b + h + e & 0.7 & 
0.361 $\pm$ 0.020 & $1.755\pm0.138$ & 0.384 $\pm$ 0.040 & 0.045 & 0.018 & 0.0178 & 40 \\ \hline % Mchthres-Lam14-Mmax-q07-bhe-computed.pdf 4.06 
94 & $a M_\mathrm{max} + b \Lambda_{1.4} + c$ & b + h + e & 1, 0.85 & 
0.254 $\pm$ 0.015 & $2.422\pm0.106$ & 0.629 $\pm$ 0.031 & 0.049 & 0.020 & 0.0435 & 80 \\ \hline % Mchthres-Lam14-Mmax-q1085-bhe-computed.pdf 4.77 
95 & $a M_\mathrm{max} + b \Lambda_{1.4} + c$ & b + h + e & 1, 0.7 & 
0.291 $\pm$ 0.024 & $2.189\pm0.167$ & 0.539 $\pm$ 0.049 & 0.078 & 0.030 & 0.1089 & 80 \\ \hline % Mchthres-Lam14-Mmax-q107-bhe-computed.pdf 11.94 
95 * & $a M_\mathrm{max} + b \Lambda_{1.4} + c$ & b + h + e & 1, 0.7 & 
0.291 $\pm$ 0.024 & $2.189\pm0.167$ & 0.539 $\pm$ 0.049 & 0.078 & 0.030 & 0.1089 & 80 \\ \hline % check085Mchthres-Lam14-Mmax-q107-bhe-computed.pdf 11.94 
96 & $a M_\mathrm{max} + b \Lambda_{1.4} + c$ & b + h + e & 1, 0.85, 0.7 & 
0.289 $\pm$ 0.018 & $2.199\pm0.123$ & 0.548 $\pm$ 0.036 & 0.080 & 0.027 & 0.1346 & 120 \\ \hline % Mchthres-Lam14-Mmax-q108507-bhe-computed.pdf 9.71 
97 & $a M_\mathrm{max} + b \Lambda_{1.4} + c$ & b + h & 1 & 
0.271 $\pm$ 0.027 & $3.411\pm 0.248$ & 0.558 $\pm$ 0.059 & 0.042 & 0.015 & 0.0105 & 32 \\ \hline % Mchthres-Lam14-Mmax-q1-bh-computed.pdf 3.05 
98 & $a M_\mathrm{max} + b \Lambda_{1.4} + c$ & b + e & 1 & 
0.214 $\pm$ 0.020 & $2.594\pm 0.132$ & 0.716 $\pm$ 0.041 & 0.046 & 0.015 & 0.0122 & 31 \\ \hline % Mchthres-Lam14-Mmax-q1-be-computed.pdf 3.67 
\hline 
99 & $a M_\mathrm{max} + b \tilde{\Lambda}_\mathrm{thres} + c$ & b & 1 & 
0.627 $\pm$ 0.044 & $11.98\pm0.975$ & -0.396 $\pm$ 0.114 & 0.037 & 0.016 & 0.0086 & 23 \\ \hline % Mchthres-Lamthres-Mmax-q1-b-computed.pdf 3.64 
100 & $a M_\mathrm{max} + b \tilde{\Lambda}_\mathrm{thres} + c$ & b & 0.85 & 
0.611 $\pm$ 0.046 & $8.540\pm0.825$ & -0.278 $\pm$ 0.114 & 0.038 & 0.017 & 0.0094 & 23 \\ \hline % Mchthres-Lamthres-Mmax-q085-b-computed.pdf 3.97 
101 & $a M_\mathrm{max} + b \tilde{\Lambda}_\mathrm{thres} + c$ & b & 0.7 & 
0.618 $\pm$ 0.046 & $4.050\pm0.528$ & -0.232 $\pm$ 0.113 & 0.048 & 0.016 & 0.0093 & 23 \\ \hline % Mchthres-Lamthres-Mmax-q07-b-computed.pdf 3.95 
102 & $a M_\mathrm{max} + b \tilde{\Lambda}_\mathrm{thres} + c$ & b & 1, 0.85 & 
0.607 $\pm$ 0.035 & $9.720\pm0.701$ & -0.297 $\pm$ 0.090 & 0.050 & 0.019 & 0.0245 & 46 \\ \hline % Mchthres-Lamthres-Mmax-q1085-b-computed.pdf 4.81 
103 & $a M_\mathrm{max} + b \tilde{\Lambda}_\mathrm{thres} + c$ & b & 1, 0.7 & 
0.458 $\pm$ 0.071 & $2.867\pm0.946$ & 0.186 $\pm$ 0.172 & 0.112 & 0.040 & 0.1102 & 46 \\ \hline % Mchthres-Lamthres-Mmax-q107-b-computed.pdf 21.63 
103 * & $a M_\mathrm{max} + b \tilde{\Lambda}_\mathrm{thres} + c$ & b & 1, 0.7 & 
0.458 $\pm$ 0.071 & $2.867\pm0.946$ & 0.186 $\pm$ 0.172 & 0.112 & 0.040 & 0.1102 & 46 \\ \hline % check085Mchthres-Lamthres-Mmax-q107-b-computed.pdf 21.63 
104 & $a M_\mathrm{max} + b \tilde{\Lambda}_\mathrm{thres} + c$ & b & 1, 0.85, 0.7 & 
0.473 $\pm$ 0.055 & $3.619\pm0.784$ & 0.139 $\pm$ 0.133 & 0.101 & 0.040 & 0.1497 & 69 \\ \hline % Mchthres-Lamthres-Mmax-q108507-b-computed.pdf 19.16 
105 & $a M_\mathrm{max} + b \tilde{\Lambda}_\mathrm{thres} + c$ & b + h + e & 1 & 
0.604 $\pm$ 0.053 & $8.398\pm 1.509$ & -0.251 $\pm$ 0.148 & 0.120 & 0.041 & 0.1051 & 40 \\ \hline % Mchthres-Lamthres-Mmax-q1-bhe-computed.pdf 23.98 
106 & $a M_\mathrm{max} + b \tilde{\Lambda}_\mathrm{thres} + c$ & b + h + e & 0.85 & 
0.607 $\pm$ 0.045 & $5.162\pm0.919$ & -0.183 $\pm$ 0.118 & 0.091 & 0.038 & 0.0778 & 40 \\ \hline % Mchthres-Lamthres-Mmax-q085-bhe-computed.pdf 17.77 
107 & $a M_\mathrm{max} + b \tilde{\Lambda}_\mathrm{thres} + c$ & b + h + e & 0.7 & 
0.623 $\pm$ 0.037 & $2.774\pm0.486$ & -0.205 $\pm$ 0.093 & 0.082 & 0.029 & 0.0510 & 40 \\ \hline % Mchthres-Lamthres-Mmax-q07-bhe-computed.pdf 11.64 
108 & $a M_\mathrm{max} + b \tilde{\Lambda}_\mathrm{thres} + c$ & b + h + e & 1, 0.85 & 
0.586 $\pm$ 0.036 & $5.769\pm0.837$ & -0.144 $\pm$ 0.097 & 0.125 & 0.042 & 0.2108 & 80 \\ \hline % Mchthres-Lamthres-Mmax-q1085-bhe-computed.pdf 23.12 
109 & $a M_\mathrm{max} + b \tilde{\Lambda}_\mathrm{thres} + c$ & b + h + e & 1, 0.7 & 
0.502 $\pm$ 0.043 & $1.339\pm0.660$ & 0.150 $\pm$ 0.107 & 0.170 & 0.050 & 0.3340 & 80 \\ \hline % Mchthres-Lamthres-Mmax-q107-bhe-computed.pdf 36.63 
109 * & $a M_\mathrm{max} + b \tilde{\Lambda}_\mathrm{thres} + c$ & b + h + e & 1, 0.7 & 
0.502 $\pm$ 0.043 & $1.339\pm0.660$ & 0.150 $\pm$ 0.107 & 0.170 & 0.050 & 0.3340 & 80 \\ \hline % check085Mchthres-Lamthres-Mmax-q107-bhe-computed.pdf 36.63 
110 & $a M_\mathrm{max} + b \tilde{\Lambda}_\mathrm{thres} + c$ & b + h + e & 1, 0.85, 0.7 & 
0.519 $\pm$ 0.033 & $1.962\pm0.552$ & 0.097 $\pm$ 0.084 & 0.164 & 0.050 & 0.4535 & 120 \\ \hline % Mchthres-Lamthres-Mmax-q108507-bhe-computed.pdf 32.73 
111 & $a M_\mathrm{max} + b \tilde{\Lambda}_\mathrm{thres} + c$ & b + h & 1 & 
0.475 $\pm$ 0.085 & $5.264\pm1.522$ & 0.105 $\pm$ 0.213 & 0.080 & 0.035 & 0.0558 & 32 \\ \hline % Mchthres-Lamthres-Mmax-q1-bh-computed.pdf 16.25 
112 & $a M_\mathrm{max} + b \tilde{\Lambda}_\mathrm{thres} + c$ & b + e & 1 & 
0.655 $\pm$ 0.029 & $13.15\pm0.927$ & -0.482 $\pm$ 0.083 & 0.072 & 0.021 & 0.0222 & 31 \\ \hline % Mchthres-Lamthres-Mmax-q1-be-computed.pdf 6.70 
\end{tabular} 
\caption{Different bilinear fits (second column) describing the EoS dependence of the  threshold chirp mass $\mathcal{M}_\mathrm{thres}$ for prompt BH formation (see main text). Third column specifies the set of EoSs used for the fit (``b'' $\equiv$ hadronic base sample (a), ``e'' $\equiv$ exlcuded hadronic sample (b), ``h'' $\equiv$ hybrid sample (c) as defined  in Sect.~\ref{sec:eos}). $q$ is the binary mass ratio of the underlying data. Fifth to seventh columns provide the fit parameters $a$, $b$ and $c$ and their respective variance with units such that masses are in $M_\odot$. The next two columns specify the maximum and average deviation between fit and the underlying data (in $M_\odot$). Last two columns give the sum of the squared residuals being minimized by the fit procedure and the number of data points included in the fit. For fits marked with an asterisk we compute deviations between fit and the data comparing additionally to the results for $q=0.85$ which are not employed for the fit.} 
\label{tab:mch2} 
\end{table*}

\section{Relations for $M_\mathrm{thres}$} \label{sec:mthr}

In~\cite{Bauswein2020a} we discuss different relations that quantitatively describe the collapse behavior of NS mergers. To this end we construct bilinear fit formulae involving $M_\mathrm{thres}$ and certain stellar parameters, which characterize an EoS, for instance the maximum mass $M_\mathrm{max}$ of nonrotating NSs and NS radii or their tidal deformability (see Tab.~II in the Supplemental Material of~\cite{Bauswein2020a}). For most of these fits we choose $M_\mathrm{max}$ to be the dependent variable, but being linear the relations can easily be inverted to express for instance $M_\mathrm{thres}$ as function of $M_\mathrm{max}$ and another stellar property. We for instance consider dependencies on $\tilde{\Lambda}_\mathrm{thres}$, which is the combined tidal deformability of a binary with a mass $M_\mathrm{thres}$. This quantity can be inferred from future merger observations (see~\cite{Bauswein2020a} for a more detailed discussion).

Changes of the independent variables are useful depending on which of these quantities are actually measured. We also develop fits employing the different sets of EoSs as defined in Sect.~\ref{sec:eos}. The reasoning behind this is that one may make different assumptions about the general properties of the EoSs. Imposing for instance current astrophysical constraints one may exclude subset (b). One may also argue that indications of a phase transition may be revealed either by heavy-ion collisions, theoretical arguments or by other astronomical observations. For instance, we demonstrate in~\cite{Bauswein2020a} that a combined measurement of $M_\mathrm{thres}$ and $\tilde{\Lambda}_\mathrm{thres}$ may reveal signs of a phase transition (more details below). If there are no indications of a phase transition, one may want to focus on purely hadronic EoS models, i.e. the hadronic base sample possibly extended by the excluded hadronic sample.

Here we further extend the set of fits from~\cite{Bauswein2020a}. The additional relations are listed in Tab.~\ref{tab:mthr1} and~\ref{tab:mthr2}. We choose the threshold binary mass $M_\mathrm{thres}$ as dependent variable, i.e. we quantify the collapse behavior by
\begin{equation}
    M_\mathrm{thres}(X,Y)=a X + b Y +c
\end{equation}
with $X$ and $Y$ being stellar properties characterizing the EoS. The parameters $a$, $b$ and $c$ are obtained by least-squares fits to the data listed in Tab.~\ref{tab:data}. We provide the variances of the fit parameters from which confidence intervals can be drawn as appropriate. We assess the quality of these fits by providing the maximum residual and the average deviation between fit and the underlying data. Both are meaningful measures. Since the set of EoS models is not a statistical sample, the maximum residual may provide the best figure of merit to assess the quality of a fit. However, our sample may also include some rather extreme EoS models which may lead to larger deviations but might already be excluded by other measurements like for instance nuclear parameters derived from experiments. This is why also the average deviation may provide generally useful information about how well a certain relation describes the data as we avoid a possible bias by a single extreme model. In addition, we quote the sum of the squared residuals, which is the quantity which is actually minimized by the least-squares fit. Thus, as a cross-validation one can assess the normalized sum of the squared residuals and the reduced chi-square employing the number of data points and fit parameters and the precision to which we determine $M_\mathrm{thres}$. These figures of merit closely correlate with the average deviations and generally indicate good fits especially for the base sample\footnote{If the hybrid sample is included, the fits become systematically worse in comparison to relations for the base sample. This is understandable because a phase transition introduces an additional mechanism that affects the collapse behavior which may not be fully captured by the stellar parameters we are considering (see Sect.~\ref{sec:pt}). The larger reduced chi-square and the other figures of merit clearly indicate this deficiency of the modeling.} and relations with fixed $q$. We provide these numbers for all fits throughout the paper, but will focus the discussion on the more intuitive maximum and average deviations.

As in~\cite{Bauswein2020a} the relations in Tabs.~\ref{tab:mthr1} and~\ref{tab:mthr2} are built either for fixed binary mass ratios or for a range in $q$. Hence, the respective relations can be employed for equal-mass binaries, for asymmetric mergers or for observations which provide only a coarse constraint on the mass ratio (see also discussion below). As said, we also vary the underlying set of EoSs, since one may make different assumptions on which models are considered to be viable.

Finally, we employ different independent variables to describe the EoS dependence of $M_\mathrm{thres}$, but all relations involve $M_\mathrm{max}$ (see Sect.~\ref{sec:add} for univariate relations without $M_\mathrm{max}$). This includes the radius $R_{1.6}$ of a 1.6~$M_\odot$ NS, the radius $R_\mathrm{max}$ of the nonrotating maximum-mass NS, the tidal deformability of a 1.4~$M_\odot$ NS and the combined tidal deformability $\tilde{\Lambda}_\mathrm{thres}$ of the binary at the threshold mass, which generally depends on $q$. We use different independent variables in the relations because different applications of these relations may require different quantities. One should also bear in mind that these quantities may be obtained with different accuracy. In particular, we point out that the relations with $R_{1.6}$ or $\Lambda_{1.4}$ may be the most useful since these parameters may be obtained with high precision in the near future, for instance from GW measurements of the inspiral phase~\cite{Chatziioannou2020} or the postmerger phase~\cite{Bauswein2016}, or from X-ray timing~\cite{Miller2019,Riley2019,Raaijmakers2019}.

Considering the different relations in Tab.~\ref{tab:mthr1} and~\ref{tab:mthr2}, we make the following general observations. (i) For $q=1$ and the base sample, $R_{1.6}$ yields the tightest relations, whereas $\tilde{\Lambda}_\mathrm{thres}$ leads to the largest deviations. This may be explained by the fact that $R_{1.6}$ is a quantity that characterizes the density regime of the EoS which determines the dynamics of systems close to the collapse (see Fig.~1 in the Supplemental Material of~\cite{Bauswein2020a}) (ii) Relations for fixed mass ratio are tighter than those with a range in $q$. The one for $q=1$ is slightly better for most fits. (iii) Not unexpected, the relations are tighter if we only consider the base sample of EoS models. Including hybrid EoSs with phase transition leads to significantly worse relations. This can be seen in particular for relations with $q=1$ involving NS radii. Adding the excluded sample to the base sample only slightly increases the deviations compared to the base sample alone. Combining the base sample and the hybrid sample instead leads to larger deviations (compare for instance fits 1, 13 and 14).

(iv) Without explicitly listing deviations, we remark that the fits for $q=1$ also describe the results for $q=0.85$ relatively well and vice versa, since the threshold mass for $0.85\leq q \leq 1$ do not differ very much (cf. Sect.~\ref{sec:q}). (v) Considering the maximum residuals, fits based on data for $q=1$ and $q=0.7$ are generally somewhat tighter than fits which employ $q=1$, $q=0.85$ and $q=0.7$ (cf. fits 5 and 6). We explain this by the fact that $M_\mathrm{thres}$ is very similar for $q=1$ and $q=0.85$. Thus, the inclusion of $q=0.85$ results shifts the fit towards the $M_\mathrm{thres}(q=1)$, which implies a slightly worse description of the data for $q=0.7$. We consider both types of fits to be useful for describing $M_\mathrm{thres}$ in the range $0.7\leq q \leq1$. Relations based on only $q=1$ and $q=0.7$ may be more accurate and thus preferable if one wishes to minimize the maximum deviations and binaries with small $q$ are likely or possible to occur. (For the table entries with an asterisk we determine the deviations by comparing additionally to the $q=0.85$ data which are not employed for constructing the fit. The $q=0.85$ deviations do not increase the overall deviations which means that the $q=0.85$ data is well captured by those fits.) Fits employing $q=1$, $q=0.85$ and $q=0.7$, i.e. with a bias towards the results of equal-mass mergers, may provide on average a better description by taking into account that for uniformly distributed $q$ the threshold mass is more likely to be closer to $M_\mathrm{thres}(q=1)$. See also Sect.~\ref{sec:q}.

Examples for $M_\mathrm{thres}$ relations are shown in Fig.~\ref{fig:mthr} for fixed $q$ with the base EoS sample. In the upper panels $M_\mathrm{thres}$ is given as function of $M_\mathrm{max}$ and $R_{1.6}$, whereas in the lower panels we use $M_\mathrm{max}$ and the tidal deformability of a 1.4~$M_\odot$ NS as independent variables. For all of these cases with fixed mass ratio and the base sample, we find the maximum residual to be significantly better than 0.1~$M_\odot$ with an average deviation of less than 0.03~$M_\odot$ (see Tabs.~\ref{tab:mthr1} and~\ref{tab:mthr2}).

In addition, we construct another set of fits describing the collapse behavior by the threshold chirp mass $\mathcal{M}_\mathrm{thres}$ for prompt BH formation. We provide these relations in Tabs.~\ref{tab:mch1} and~\ref{tab:mch2}. We recall that the chirp mass is measured with high precision during the GW inspiral, whereas the total mass is computed from the chirp mass and constraints on the mass ratio $q$. Since $\mathcal{M}$ is the quantity which is directly measured, the relations in Tabs.~\ref{tab:mch1} and~\ref{tab:mch2} may be more useful for many applications.

For fixed mass ratio, the conversion between $M_\mathrm{tot}$ and $\mathcal{M}$ is trivial ($M_\mathrm{tot}=\mathcal{M} q^{0.6} (1+1/q)^{1.2}$), and thus the conversion between $M_\mathrm{thres}$ and $\mathcal{M}_\mathrm{thres}$. Here we are mostly interested in the relations for a range in $q$ to check whether $\mathcal{M}_\mathrm{thres}$ or $M_\mathrm{thres}$ yields tighter relations. Although the absolute deviations are smaller for $\mathcal{M}_\mathrm{thres}$, the relative deviations are larger. Recall that the chirp mass is smaller than the total binary mass. The reason for the larger deviations is that the chirp mass decreases with binary mass asymmetry for constant total mass: $\mathcal{M} =M_\mathrm{tot} q^{-0.6} (1+1/q)^{-1.2}$. Hence, if for a specific EoS the threshold binary mass was the same for $q=1$ and $q=0.7$, $\mathcal{M}_\mathrm{thres}$ for this EoS differs between symmetric and asymmetric binaries. This additional $q$ dependence leads to the stronger relative deviations in the $\mathcal{M}_\mathrm{thres}$ relations if the fit includes data for different $q$. {In other words, because of the definition of the chirp mass, the relative differences between $\mathcal{M}_\mathrm{thres}(q=1)$ and $\mathcal{M}_\mathrm{thres}(q=0.7)$ are larger than those for $M_\mathrm{thres}$. Consequenctly combining $q=1$ and $q=0.7$ data in a single relation leads to larger scatter.}  %close(); execfile("checkrelativedifferenmthrvsmchthrMchthres-R16-Mmax-q107-base.py")
We also refer to the discussion below providing tight fits for $\mathcal{M}_\mathrm{thres}$ and $M_\mathrm{thres}$ including an explicit $q$ dependence. For fixed mass ratio the relative deviations between fit and data are in fact identical comparing $\mathcal{M}_\mathrm{thres}(X,Y)$ and $M_\mathrm{thres}(X,Y)$. 

We finally comment on the choice of the independent variables in the relations for $M_\mathrm{thres}$ and $\mathcal{M}_\mathrm{thres}$ in Tabs.~\ref{tab:mthr1},~\ref{tab:mthr2},~\ref{tab:mch1} and~\ref{tab:mch2}. These stellar parameters express the EoS dependence of the collapse behavior and characterize the EoS in the density regime most relevant for prompt BH formation. The maximum mass $M_\mathrm{max}$ of nonrotating NSs is a natural choice since it determines the threshold to BH formation for nonrotating stars. However, it is clear from the different mass-radius relations of nonrotating NSs and the simulation results for $M_\mathrm{thres}$ that also other EoS parameters affect the collapse behavior (two EoSs with the same $M_\mathrm{max}$ can yield very different $M_\mathrm{thres}$). In Tabs.~\ref{tab:mthr1},~\ref{tab:mthr2},~\ref{tab:mch1} and~\ref{tab:mch2} we have chosen for instance the radii or the closely related tidal deformability of NSs with masses in the range 1.4 to 1.6~$M_\odot$, which characterize the EoS at intermediate to high densities. This is motivated by the fact that fit formulae involving these parameters yield the tightest relations, which suggests that these quantities express the fundamental dependencies. We discuss these findings in Appendix~\ref{sec:opt} and point out that fit formulae for $M_\mathrm{thres}$ can also be constructed for radii or tidal deformablities of NSs with other fiducial masses in the range 1.1 to 2.0~$M_\odot$. These relations are also relatively tight, the maximum deviations roughly double if one employs the stellar properties of very light or very massive stars in comparison to 1.4 to 1.6~$M_\odot$ NSs (see Appendix~\ref{sec:opt} for additional details).

We also refer to some subtle aspects of the inversion of $M_\mathrm{thres}$ and $\mathcal{M}_\mathrm{thres}$ relations. It is trivial to invert a bilinear relation, but it is at least in principle not obvious that the inverted relation represents the optimal description of the data if dependent and independent variables are exchanged. This is because a least-squares fit is constructed by minimizing the residuals with respect to the dependent variable. Thus, changing the dependent variable and minizing the residuals with respect to this variable yields another fit, which generally will (slightly) differ from the inverted relation derived from a fit to another dependent variable. By comparing inverted relations and fits for different dependent variables and by orthogonal regression (total least squares) we show that these subtleties are in practice irrelevant for the relations discussed here\footnote{We assess these aspects by considering the $q=1$ data of the base sample and relations between $M_\mathrm{thres}$, $M_\mathrm{max}$ and $R_{1.6}$ (fit~1). We compute a least-squares fit for each variable $M_\mathrm{thres}$, $M_\mathrm{max}$ and $R_{1.6}$ and invert the respective relation. Moreover, we model the data by a total least-squares fit (orthogonal regression). For every fit and every inverted relation we compute the maximum and average deviation between relation and data. Comparing these deviations for every dependent variable, we find that they coincide to within at least a few per cent (typically less than 1\%). Hence, we conclude that inverting relations is as good as performing directly a fit to data with another dependent variable. We expect this to hold for all fits in this paper considering the similarity between these relations.}. In essence, the inversion of the relations given in Tabs.~\ref{tab:mthr1} to~\ref{tab:mch2} represent a very good description of the data.

\subsection{Current and future EoS constraints from $M_\mathrm{thres}$}

The advantages of bilinear relations for $M_\mathrm{thres}$ is the fact that they are readily invertible. One can thus obtain expressions for NS radii or the tidal deformability as function of $M_\mathrm{thres}$ and $M_\mathrm{max}$. In the spirit of~\cite{Bauswein2017,Bauswein2019b,Bauswein2019c} one can employ additional relations between $M_\mathrm{max}$ and the radius $R$ or the tidal deformability $\Lambda$ to obtain a lower or upper bound on $M_\mathrm{max}$ for a given $R$ or $\Lambda$. Adopting such a relation to constrain $M_\mathrm{max}(R)$ or $M_\mathrm{max}(\Lambda)$, one can then directly read off a bound on $R$ or $\Lambda$ for a given $M_\mathrm{thres}$ constraint.

\begin{figure}% r16-limit-prospects.py
\centering
\includegraphics[width=\columnwidth]{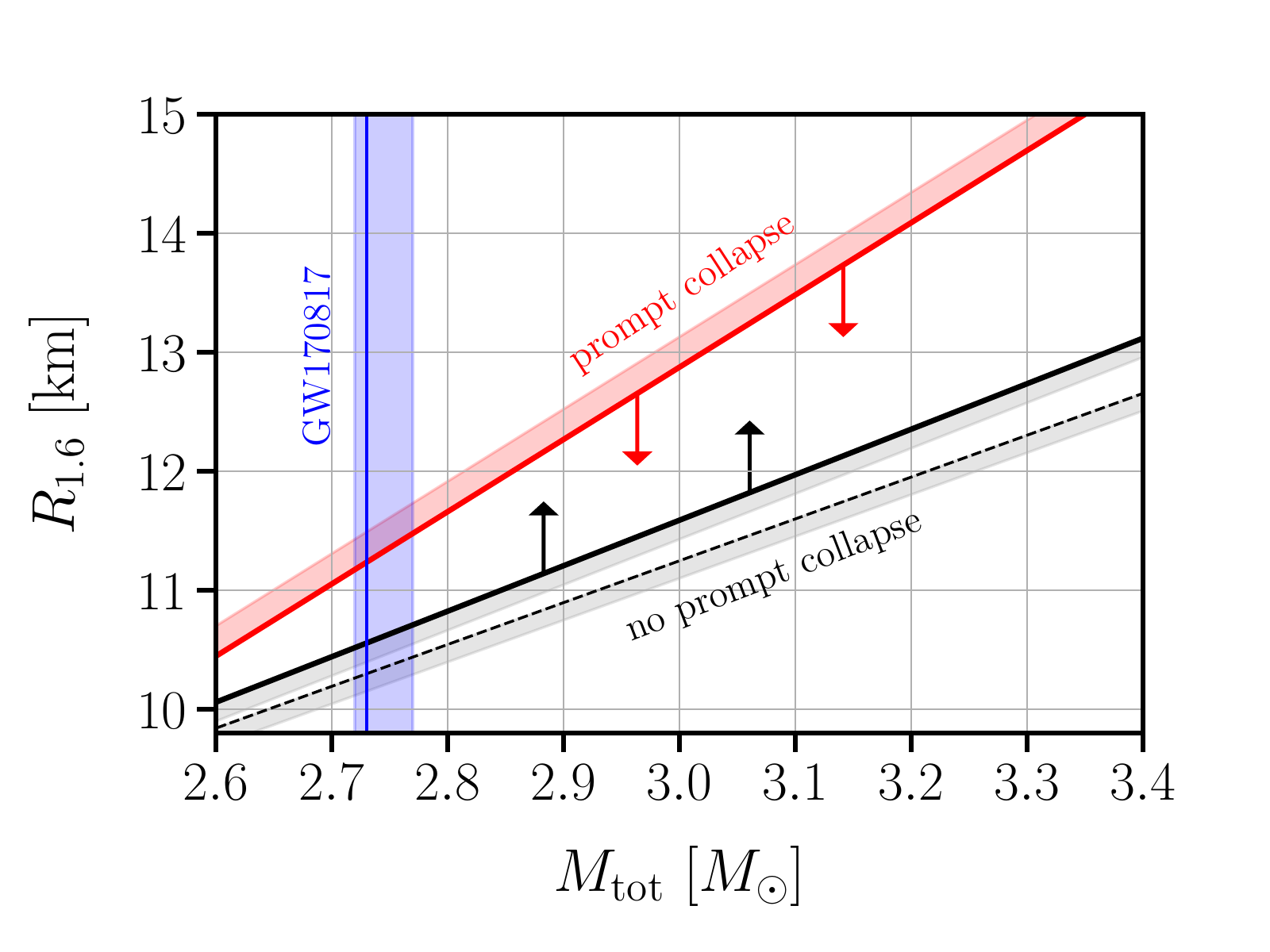}
\caption{Bounds on NS radius $R_{1.6}$ implied by GW detections with total binary mass $M_\mathrm{tot}$. Black lines show a lower limit on $R_{1.6}$ if an observation provides evidence for no direct BH formation. Red line displays an upper limit on $R_{1.6}$ inferred from events with indications for a prompt collapse of the merger remnant. The shaded areas indicate the maximum scatter of the underlying empirical relations for $M_\mathrm{thres}$; the dashed black line employs a very conservative assumption about the maximum stiffness of NS matter at higher densities (see text). The blue vertical line provides the total binary mass of GW170817 with the error bar indicated by the blue shaded area.}
\label{fig:rlimit}
\end{figure}

We exemplify the constraints for an event which provides evidence for no direct BH formation, i.e. with $M_\mathrm{tot}<M_\mathrm{thres}$, as it was arguably the case for GW170817~\cite{Margalit2017,Bauswein2017,Radice2018}. Additionally, we employ an upper bound on $M_\mathrm{max}$, which may be expressed as $M_\mathrm{max}<M_\mathrm{max}^\mathrm{up}=w_1 R + w_2$ with some coefficients $w_1$ and $w_2$ to be specified. Such relations can be inferred by causality arguments, which limit the maximum stiffness of any EoS at higher densities by requiring that the speed of sound should not exceed the speed of light. See~\cite{Bauswein2019b,Bauswein2019c} for details and for the case $R=R_{1.6}$, which yields $M_\mathrm{max}^\mathrm{up}=\frac{1}{3.1}\frac{c^2}{G}R_{1.6}$, hence $w_1=0.219$ and $w_2=0$ for $R$ in km. Alternatively, we consider the 40 EoS models of this study and find an upper limit $M_\mathrm{max}(R_{1.6})<M_\mathrm{max}^\mathrm{up}(R_{1.6})=0.177 R_{1.6} + 0.306$. Following these arguments we then obtain
\begin{equation}\label{eq:ansatzr}
    \begin{split}
        M_\mathrm{tot} & < a M_\mathrm{max} + b R + c \\
        & <  a M_\mathrm{max}^\mathrm{up} + b R + c \\
        & < a w_1 R + a w_2 + b R + c
    \end{split}
\end{equation} 
with the parameters $a$, $b$ and $c$ taken from Tab.~\ref{tab:mthr1} {or~\ref{tab:mthr2}} as appropriate. From this results the limit
\begin{equation}
    R > \frac{M_\mathrm{tot} - c - a w_2}{a w_1 + b}.
\end{equation}
This limit is displayed as black solid line in Fig.~\ref{fig:rlimit}, where we employ the parameters from fit~1 from Tab.~\ref{tab:mthr1}. (It is trivial to obtain similar limits for the tidal deformability by replacing $R$ by $\Lambda$ in the derivation. We also refer to the following section and the discussion in Sect.~\ref{sec:add} for a justification to use fit~1, i.e. a $M_\mathrm{thres}$ relation for $q=1$. Below we describe that for most EoS models the threshold mass of asymmetric mergers is approximately equal or smaller than the one of equal-mass systems. This implies that using $M_\mathrm{thres}(q=1)$, being usually equal or larger than $M_\mathrm{thres}(q<1)$, is a safe choice in Eq.~\eqref{eq:ansatzr} independent of the mass ratio $q$, which may not be well known.) A very conservative limit is given by the dashed black curve, for which we employ the above mentioned causality argument resulting in the very strict limit $M_\mathrm{max}^\mathrm{up}(R_{1.6})=\frac{1}{3.1}\frac{c^2}{G}R_{1.6}$ (see~\cite{Bauswein2017,Bauswein2019c}). The dashed curve is likely very restrictive since we assume here the extreme case that the speed of sound equals the speed of light above a certain density to obtain an upper bound on $M_\mathrm{max}$. Microphysical EoS models are typically not as stiff, which is why the solid curve represents a more appropriate constraint. The relations for $M_\mathrm{thres}$ have a certain scatter $\delta M$ (see Tab.~\ref{tab:mthr1} and~\ref{tab:mthr2}), which can be easily taken into account by shifting the relation and setting $M_\mathrm{tot}<M_\mathrm{thres} -\delta M$. This results in
\begin{equation}\label{eq:rlimit}
    R > \frac{M_\mathrm{tot} -\delta M - c - a w_2}{a w_1 + b}
\end{equation}
and is visualized by the shaded area, where we adopted the maximum deviation from Tab.~\ref{tab:mthr1}.

For the case of GW170817 with $M_\mathrm{tot}^\mathrm{GW170817}=2.73_{-0.01}^{+0.04}~M_\odot$~\cite{Abbott2019} we find $R_{1.6}> 10.56_{-0.04}^{+0.15}$~km, which is fully in line with our previous constraint in~\cite{Bauswein2017} based an older relation for $M_\mathrm{thres}$~\cite{Bauswein2013} and slightly different assumptions\footnote{As in~\cite{Bauswein2017} we do not include the maximum deviation in the error analysis ($\delta M =0$) since our assumptions are generally conservative. In~\cite{Bauswein2017} we employed the very conservative $M_\mathrm{max}^\mathrm{up}$ relation based on causality arguments but assumed that the total binary mass was 0.1~$M_\odot$ below the threshold mass, which we do not adopt here. Note that the binary mass of GW170817 has been slightly revised in~\cite{Abbott2019} compared to~\cite{Abbott2017}, which was employed in~\cite{Bauswein2017}. Adopting the same assumptions and binary mass as in~\cite{Bauswein2017} we in fact find $R_{1.6}>10.68$~km for the new relation (fit 1), i.e. the limit as in~\cite{Bauswein2017}.}. The total binary mass of GW170817 is depicted by the blue line in Fig.~\ref{fig:rlimit}.

It is straightforward to apply Eq.~\eqref{eq:rlimit} to any other new GW event which indicates that no prompt collapse took place. An updated limit can be easily read off from Fig.~\ref{fig:rlimit} for any total binary mass of a new measurement. 

Following a very similar reasoning one can infer upper limits on $R$ from GW events with evidence for a prompt collapse, i.e. $M_\mathrm{tot}>M_\mathrm{thres}$. In this case one may simply insert the current lower limit $M_\mathrm{max}^\mathrm{low,obs}$ from pulsar observations~\cite{Antoniadis2013,Arzoumanian2018a,Cromartie2019}. This leads to an upper limit
\begin{equation}\label{eq:rup}
    R < \frac{M_\mathrm{tot}+\delta M -a M_\mathrm{max}^\mathrm{low,obs} -c}{b}.
\end{equation}
This upper bound is also visualized in Fig.~\ref{fig:rlimit} by the red line\footnote{The shown limit is drawn for equal-mass mergers. For some EoS models $M_\mathrm{thres}$ of asymmetric mergers is smaller than that of equal-mass binaries. Hence, if the mass ratio in a detection deviates from unity or if only a range of $q$ is given, a corresponding line with the measured $q$ or the lower bound of the inferred range of $q$, respectively, should be computed. This can be readily done by $M_\mathrm{thres}$ relations for asymmetric mergers, see in particular Sect.~\ref{sec:q}}.  From this figure one can easily read off which constraint on NS properties are implied by new detections with given $M_\mathrm{tot}$. It is apparent from Fig.~\ref{fig:rlimit} how the constraints tighten as better limits on $M_\mathrm{thres}$ become available. In this context we also refer to~\cite{Bauswein2019c}, which sketches an observing strategy to efficiently determine $M_\mathrm{thres}$ from electromagnetic counterparts assuming that binary mass estimates would be circulated with a GW trigger (see also~\cite{Margalit2019}).

Obviously Eqs.~\eqref{eq:rlimit} and~\eqref{eq:rup} yield limits on any radius $R$ or tidal deformability $\Lambda$ by employing the corresponding $M_\mathrm{thres}(M_\mathrm{max},R)$ or $M_\mathrm{thres}(M_\mathrm{max},\Lambda)$ relation and the fit parameters $a$, $b$ and $c$ (see Tabs.~\ref{tab:mthr1}~and~\ref{tab:mthr2} and in particular Tabs.~\ref{tab:optr} and~\ref{tab:optlam} from Appendix~\ref{sec:opt} for stellar parameters of different NS masses).

Considering GW170817, for instance, fit 29 implies $\Lambda_{1.4}>150$ (with $w_1=0.001124$ and $w_2=1.9646$ determined empirically from the base sample). This limit is very small, also in comparison to previous constraints~\cite{Radice2018,Radice2019}, and arguably conservative. But in fact the $M_\mathrm{thres}$ data shows that there potentially exist EoS models with $\Lambda_{1.4}$ somewhat above 150 which would not lead to a prompt collapse and thus be compatible with GW170817 (see also~\cite{Kiuchi2019}). 

The exact lower limit is sensitive to the adopted upper limit on $M_\mathrm{max}$ for a given $\Lambda_{1.4}$, i.e. $w_1$ and $w_2$, which we determine empirically based on the available EoS sample. This sample contains models which are rather stiff like for instance WFF1 and WFF2 and which become acausal at higher densities\footnote{For those two EoS models, for example, the speed of sound exceeds $c$ for rest-mass densities above $\rho=1.45\times 10^{15}\mathrm{g/cm^3}$ and $\rho=1.40\times 10^{15}\mathrm{g/cm^3}$, respectively, which corresponds to the central densities of NSs with 1.9~$M_\odot$ and 2.0~$M_\odot$.}. This may yield a slightly too high $M_\mathrm{max}^\mathrm{up}$ as function of $R_{1.6}$ or $\Lambda_{1.4}$. Omitting such extreme EoS models, we find somewhat higher limits of $\Lambda_{1.4}>180$ and $R_{1.6}>10.7$~km.

See also Fig.~\ref{fig:pt} to read off a limit on $\tilde{\Lambda}_\mathrm{thres}$, which similarly constrains $\Lambda_{1.37}\equiv \Lambda(1.37~M_\odot)$ (cf. discussion in~\cite{Bauswein2020a,Bauswein2019a,Bauswein2019b}). These are the limits implied by the currently available data and it is likely that those will be further strengthened by future events (Fig.~\ref{fig:rlimit}). Ref.~\cite{Bauswein2017} provides a deeper discussion of the particular advantages and prospects of this type of EoS constraints, which for example rely only on the mass inference of GW measurements and are thus rather insensitive to the waveform models used in the analysis~\cite{Abbott2019}. We also refer to~\cite{Bauswein2020a} for a description how $M_\mathrm{thres}$ relations can be employed to determine $M_\mathrm{max}$. This requires the measurement or constraint of one additional NS parameter, which could be for instance $R_{1.6}$, $\Lambda_{1.4}$ or $\Lambda_\mathrm{thres}$ (see fits 25 to 32). Employing $\Lambda_\mathrm{thres}$ has the advantage that it can be directly determined from the inspiral GW signal of the same merger events which yield the $M_\mathrm{thres}$ estimate.

\begin{figure}% Dmthres-r16-mmax-computed.py
\centering
\includegraphics[width=\columnwidth,trim=40 40 40 60,clip]{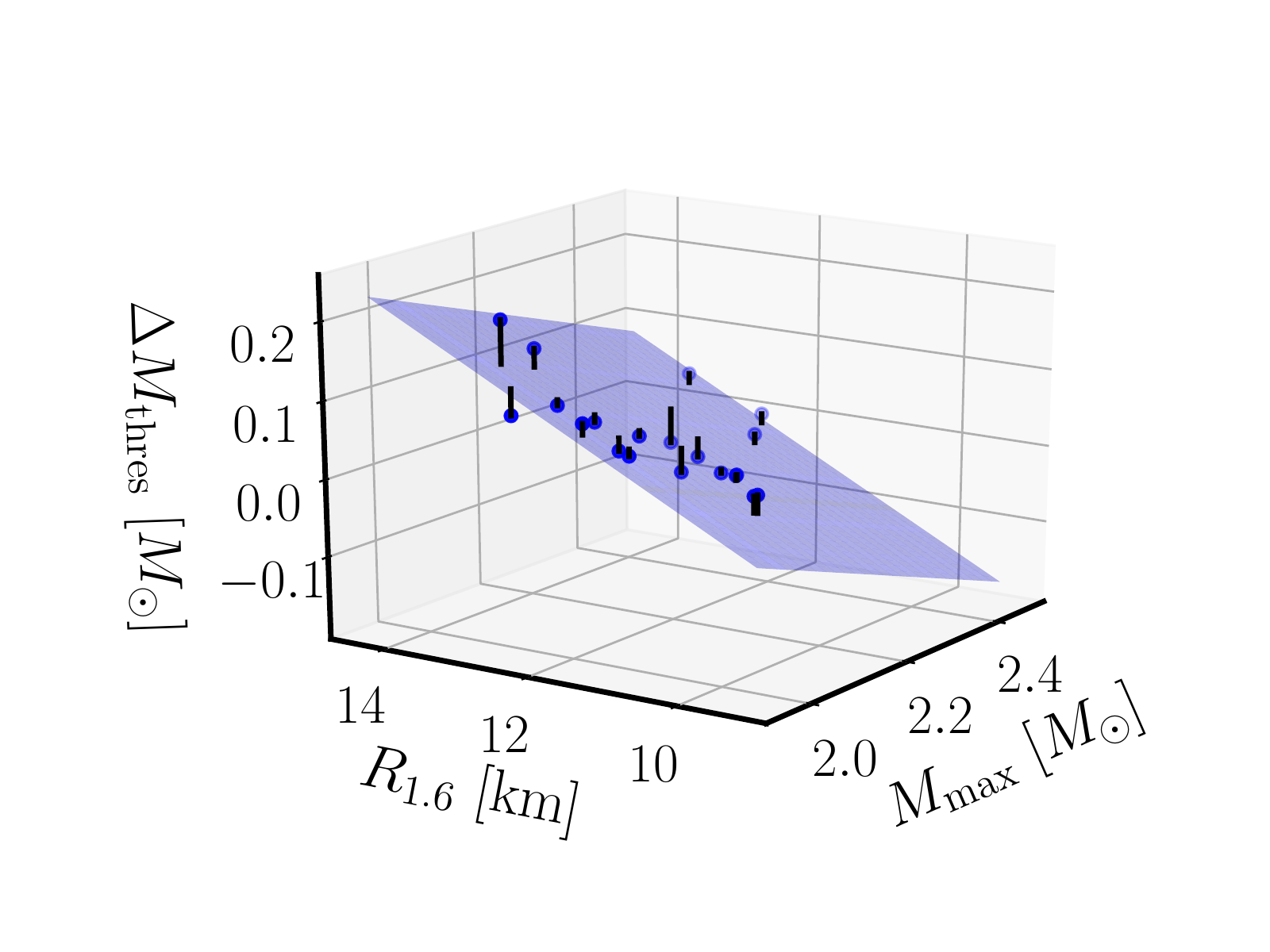}
\caption{Difference $\Delta M_\mathrm{thres}$ between the threshold mass of equal-mass mergers and the threshold mass of asymmetric binaries with mass ratio $q=0.7$ as function of $M_\mathrm{max}$ and $R_{1.6}$ (blue dots). The blue plane displays a bilinear fit (see Tab.~\ref{tab:Dmthr}). Deviations between the fit and the underlying data are illustrated by black lines. This Figure is identical to Fig.~3 in the Supplemental Material of~\cite{Bauswein2020a}.}
\label{fig:DMthr}
\end{figure}

\section{Binary mass ratio effects on $M_\mathrm{thres}$} \label{sec:q}

In~\cite{Bauswein2020a} we already mentioned that the binary mass ratio $q$ affects the threshold mass for prompt BH formation in a systematic way, i.e. the difference between $M_\mathrm{thres}$ of equal-mass mergers and of asymmetric binaries depends in a particular way on the EoS (see fit 12 in Tab.~II in the Supplemental Material of~\cite{Bauswein2020a}). We define $\Delta M_\mathrm{thres}=M_\mathrm{thres}(q=1)-M_\mathrm{thres}(q=0.7)$. For the purely hadronic base sample Fig.~\ref{fig:DMthr} shows $\Delta M_\mathrm{thres}$ as function of $M_\mathrm{max}$ and $R_{1.6}$ revealing clearly that the mass ratio effect itself is EoS dependent. The relation is well described by a bilinear fit. We find a maximum residual of 0.06~$M_\odot$ and an average deviation between the bilinear fit and the data of 0.02~$M_\odot$. Interestingly, if one removes the DD2-Hyp, BHBLP and DD2 EoSs, the fit gets even tighter with a maximum residual of about 0.028~$M_\odot$ and an average deviation of only 0.012~$M_\odot$. %0.0116 
These three EoS models are at most marginally compatible with the constraints from GW170817~\cite{Abbott2019}.

This demonstrates that the difference in $M_\mathrm{thres}$ between symmetric and asymmetric binaries increases for larger radii $R_{1.6}$ but decreases with increasing maximum mass. One may thus summarize that stiff EoS at moderate NS densities enlarge the difference $\Delta M_\mathrm{thres}$, while the difference is reduced if the EoS remains very stiff at the highest densities. We remark that possibly the different collapse behavior for different $q$, which is expressed through $\Delta M_\mathrm{thres}$, may also inform about details of the high-density EoS. In principle, $\Delta M_\mathrm{thres}$ is measurable from several different detections, i.e. from measurements of $M_\mathrm{thres}$ with different $q$.

The dependence in Fig.~\ref{fig:DMthr} also explains why in~\cite{Kiuchi2019} the threshold mass for asymmetric binaries seems to  increase for very soft EoSs\footnote{In~\cite{Kiuchi2019} the threshold mass was not directly determined, but the collapse behavior of binaries with the same total mass but different $q$ suggest an increase of the remnant stability and $M_\mathrm{thres}$ correspondingly with the system's asymmetry.}, whereas our earlier results~\cite{Bauswein2013,Bauswein2017} have shown the opposite behavior for moderately stiff EoSs.

\begin{table*} 
\begin{tabular}{|l|c|c|c|c|c|c|c|c|}  \hline 
fit $=\Delta M_\mathrm{thres}(X,Y)$ & EoS sample & $a$ & $b$ & $c$ & max. & av. & $\sum$ sq. res. & N \\ \hline 

%execfile("Dmthres-r16-mmax-computed.py") eos sample to be adpated in script by hand

$a M_\mathrm{max} +b R_{1.6} +c$ & base           &  -0.285 $\pm$ 0.041 & $ (4.859 \pm 0.672 )\times 10^{-2}$ & 0.079 $\pm$ 0.114 & 0.061 & 0.019 & 0.0127 & 23  \\ \hline %   1.01
$a M_\mathrm{max} +b R_{1.6} +c$ & base + hyb     &  -0.296 $\pm$ 0.031 & $ (4.964 \pm 0.576 )\times 10^{-2}$ & 0.093 $\pm$ 0.095 & 0.056 & 0.017 & 0.0140 & 32  \\ \hline %   0.77
$a M_\mathrm{max} +b R_{1.6} +c$ & base + ex      &  -0.302 $\pm$ 0.021 & $ (5.218 \pm 0.388 )\times 10^{-2}$ & 0.075 $\pm$ 0.051 & 0.052 & 0.018 & 0.0148 & 31  \\ \hline %   0.85
$a M_\mathrm{max} +b R_{1.6} +c$ & base + hyb +ex &  -0.304 $\pm$ 0.018 & $ (5.198 \pm 0.347 )\times 10^{-2}$ & 0.083 $\pm$ 0.044 & 0.051 & 0.016 & 0.0159 & 40  \\ \hline %   0.69

% execfile("Dmthres-rmax-mmax-computed.py")  eos sample to be adpated in script by hand

$a M_\mathrm{max} +b R_\mathrm{max} +c$ & base            & -0.314 $\pm$ 0.042 & $ (5.596 \pm 0.780 )\times 10^{-2}$ & 0.132 $\pm$ 0.110 & 0.067 & 0.018 & 0.0128 & 23  \\ \hline %   1.02
$a M_\mathrm{max} +b R_\mathrm{max} +c$ & base + hyb      & -0.313 $\pm$ 0.041 & $ (4.525 \pm 0.788 )\times 10^{-2}$ & 0.247 $\pm$ 0.114 & 0.073 & 0.021 & 0.0233 & 32  \\ \hline %   1.29
$a M_\mathrm{max} +b R_\mathrm{max} +c$ & base + ex       & -0.344 $\pm$ 0.024 & $ (6.231 \pm 0.505 )\times 10^{-2}$ & 0.135 $\pm$ 0.053 & 0.066 & 0.018 & 0.0172 & 31  \\ \hline %   0.98
$a M_\mathrm{max} +b R_\mathrm{max} +c$ & base + hyb + ex & -0.327 $\pm$ 0.026 & $ (5.548 \pm 0.546 )\times 10^{-2}$ & 0.172 $\pm$ 0.057 & 0.069 & 0.021 & 0.0295 & 40  \\ \hline %   1.28

% execfile("Dmthres-lam14-mmax-computed.py")  eos sample to be adpated in script by hand

$a M_\mathrm{max} +b \Lambda_{1.4} +c$ & base            & -0.274 $\pm$ 0.035 & $ (2.504 \pm 0.283 )\times 10^{-4}$ & 0.535 $\pm$ 0.076 & 0.049 & 0.015 & 0.0093 & 23  \\ \hline %   0.74
$a M_\mathrm{max} +b \Lambda_{1.4} +c$ & base + hyb      & -0.295 $\pm$ 0.029 & $ (2.553 \pm 0.260 )\times 10^{-4}$ & 0.582 $\pm$ 0.062 & 0.045 & 0.015 & 0.0115 & 32  \\ \hline %   0.64
$a M_\mathrm{max} +b \Lambda_{1.4} +c$ & base + ex       & -0.333 $\pm$ 0.023 & $ (1.934 \pm 0.156 )\times 10^{-4}$ & 0.685 $\pm$ 0.048 & 0.058 & 0.019 & 0.0170 & 31  \\ \hline %   0.97
$a M_\mathrm{max} +b \Lambda_{1.4} +c$ & base + hyb + ex & -0.337 $\pm$ 0.021 & $ (1.917 \pm 0.142 )\times 10^{-4}$ & 0.697 $\pm$ 0.041 & 0.055 & 0.018 & 0.0188 & 40  \\ \hline %   0.81

\end{tabular} 
\caption{Fits describing $\Delta M_\mathrm{thres}$ as the difference between the threshold mass for prompt collapse of equal-mass mergers and of asymmetric binaries with $q=0.7$ as function of different stellar parameters (first column). Second column specifies the set of EoSs used for the fit (``base'' $\equiv$ hadronic base sample (a), ``ex'' $\equiv$ exlcuded hadronic sample (b), ``hyb'' $\equiv$ hybrid sample (c) as defined in Sect.~\ref{sec:eos}). Third to fifth columns provide the fit parameters $a$, $b$ and $c$ and their respective variances with units such that masses are in $M_\odot$ and radii are in~km. The next two columns specify the maximum and average deviation between fit and the underlying data (in $M_\odot$). Last two columns give the sum of the squared residuals being minimized by the fit procedure and the number of data points included in the fit. } 
\label{tab:Dmthr} 
\end{table*}

Not unexpectedly we find that $\Delta M_\mathrm{thres}$ can also be described by relations involving other independent variables like $R_\mathrm{max}$ or $\Lambda_{1.4}$ and that similar fits exist for enlarged EoS samples. We summarize these findings in Tab.~\ref{tab:Dmthr}, where we assess the quality of the fits by the maximum residual and the average deviation between fit and data. The relation $\Delta M_\mathrm{thres}(M_\mathrm{max},R_{1.6})$ becomes slightly tighter (according to sum of the squared residuals, the maximum and average deviations in Tab.~\ref{tab:Dmthr}) if we include all the EoS models in the fit\footnote{This behavior seems somewhat peculiar, we note, however, that removing the very stiff EoS models DD2, DD2+Hyp and BHBLP (which are at most marginally compatible with the tidal deformability of GW170817) from the base sample leads to the tightest fit with a maximum residual of 0.028~$M_\odot$ and an average deviation of 0.012~$M_\odot$. (In this case the fit parameters are $a-0.263$,  $b=3.050\times 10^{-02}$ and $c=0.239$.) We thus suspect that the inclusion of additional excluded stiff models shifts the fit towards these models and yields a better description of these three outliers, which ``spoil'' the fit of the base sample.}. 

We remark that we observe a qualitatively similar behavior if we consider the difference $M_\mathrm{thres}(q=1)-M_\mathrm{thres}(q=0.85)$, which can be similarly described by fits as in Tab.~\ref{tab:Dmthr}. A few more details are provided in Appendix~\ref{app:data} and Fig.~\ref{fig:Dmthr08507}. The absolute differences $M_\mathrm{thres}(q=1)-M_\mathrm{thres}(q=0.85)$, however, are smaller compared to the comparison between $q=1$ and $q=0.7$, i.e. the threshold mass does not change strongly in the range $0.85\leq q \leq 1$. The deviations from a fit are in relation to the absolute values more significant, and thus the overall trend is less pronounced but still clearly visible. For six particularly soft EoS models $M_\mathrm{thres}(q=0.85)$ is marginally larger than the threshold mass of the corresponding equal-mass merger (by at most 0.05~$M_\odot$; see Tab.~\ref{tab:data}). Also the very stiff NL3 leads to a slightly larger $M_\mathrm{thres}(q=0.85)$. This is fully in line with the behavior shown in Fig.~\ref{fig:DMthr}, where models with relatively small $R_{1.6}$ in relation to $M_\mathrm{max}$ have $\Delta M_\mathrm{thres}\approx 0$ (cf. also panels in Fig.~\ref{fig:Dmthr08507}).

\begin{figure*}
\centering
\includegraphics[width=\columnwidth]{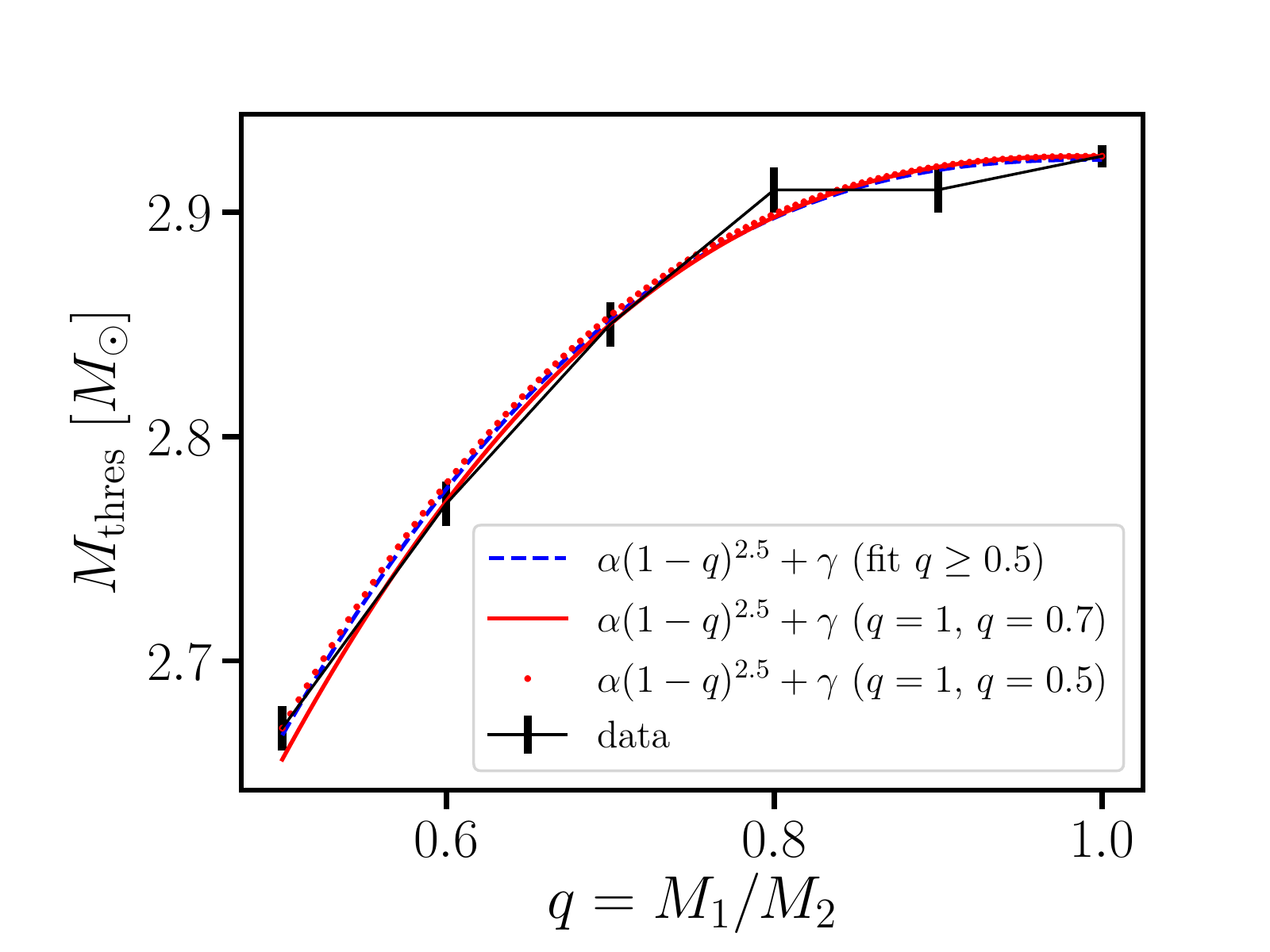}
\includegraphics[width=\columnwidth]{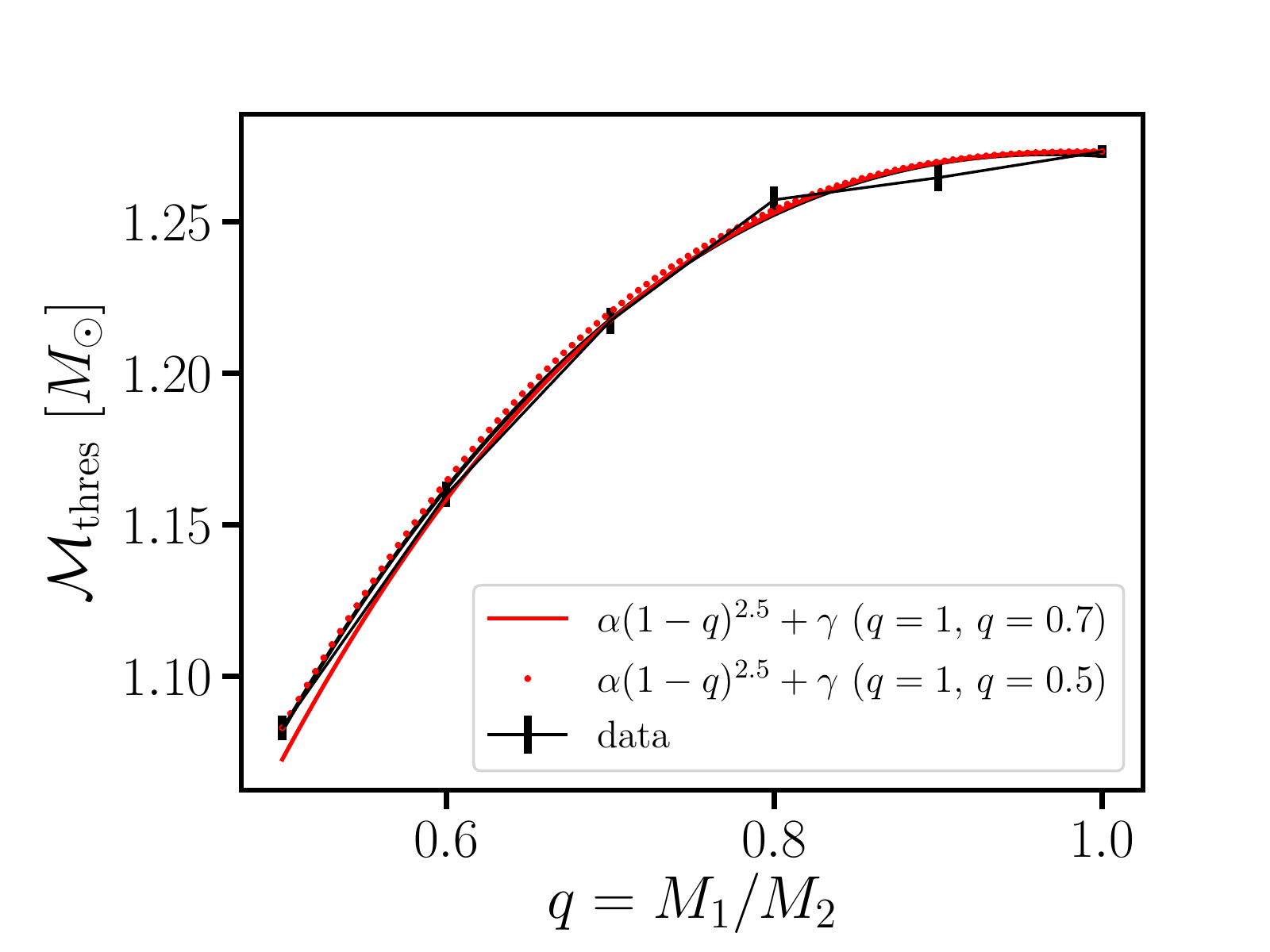}
\caption{Left panel: Threshold mass $M_\mathrm{thres}$ for prompt BH formation as function of mass ratio $q$ for the DD2F EoS. Black error bars indicate finite precision to which $M_\mathrm{thres}$ is given from merger simulations for fixed $q$. Blue dashed curve displays a fit $M_\mathrm{thres}(q)=\alpha(1-q)^{2.5}+\gamma$  to the data. Red curves are directly computed from specific data points assuming a function $M_\mathrm{thres}(q)=\alpha(1-q)^{2.5}+\gamma$ (see main text). Right panel: Threshold chirp mass $\mathcal{M}_\mathrm{thres}$ for prompt BH formation as function of the mass ratio $q$ for the DD2F EoS. Black error bars indicate finite precision to which $\mathcal{M}_\mathrm{thres}$ is given from merger simulations. Red curves are directly computed from specific data points assuming a function $\mathcal{M}_\mathrm{thres}(q)=\alpha(1-q)^{2.5}+\gamma$ (see main text).}
\label{fig:dd2f}
\end{figure*}

\subsection{Combining fits}

Considering the above dependencies of $\Delta M_\mathrm{thres}$, we devise various extensions to include mass ratio effects. By this, we obtain particularly tight relations describing the collapse behavior for different $q$. An obvious choice is to combine two independent fits for samples with different fixed $q$ through some interpolation, e.g. fit~1 and fit~2 from Tab.~\ref{tab:mthr1} describing the range $0.85\leq q \leq 1$ or fit~2 and fit~3 to cover the range $0.7\leq q \leq 0.85$ (or any other combination of fits from Tabs.~\ref{tab:mthr1},~\ref{tab:mthr1},~\ref{tab:mch1} and~\ref{tab:mch2} or Tab.~II from the Supplemental Material in~\cite{Bauswein2020a}). We remind that all these relations are bilinear implying that they can be easily inverted to employ any quantity as the dependent one.

A very first ansatz could be a linear interpolation. Schematically, we then express the full q-dependent relation as
\begin{equation}\label{eq:q}%new w085
    \mathrm{fit}(q)=(\mathrm{fit}^{q=1}-\mathrm{fit}^{q=0.85}) (q-1)/0.15  +\mathrm{fit}^{q=1}
\end{equation}
to be valid in the whole range $0.85\leq q \leq 1$. This allows a simple and straightforward implementation of mass ratio effects for any two fits regardless of what are the dependent and independent variables. A larger range in $q$ can be covered either by employing fits for $q=1$ and $q=0.7$ in Eq.~\eqref{eq:q} or by a piecewise linear ansatz describing separately the ranges $q=[1,0.85]$ and $q=[0.85,0.7]$.

For a larger range in $q$, a simple linear dependence on $q$ may not provide the most accurate description of $M_\mathrm{thres}(q)$ (see also Fig.~2 in the Supplemental Material of~\cite{Bauswein2020a}). Since mass ratio effects become stronger with asymmetry, i.e. decreasing $q$, one may for instance replace Eq.~\eqref{eq:q} by a higher-order polynomial such that the derivative with respect to $q$ is zero at $q=1$. We assess this point by investigating $M_\mathrm{thres}(q)$ for selected EoS models. Fig.~\ref{fig:dd2f} (left panel) shows $M_\mathrm{thres}$ for the DD2F EoS as function of the mass ratio in the range $0.5\le q\le 1$. The black error bars indicate the precision to which $M_\mathrm{thres}(q)$ was determined from simulations for this EoS. One can clearly recognize that mass ratio effects on the threshold mass become stronger with mass asymmetry.

We take different approaches to quantify the behavior in Fig.~\ref{fig:dd2f}. Generally, we find that higher-order polynomials provide a good description of the data. The blue curve in Fig.~\ref{fig:dd2f} displays a least-squares fit of the form $M_\mathrm{thres}(q)=\alpha (1-q)^{2.5}+\gamma$ (resulting in $\alpha=-1.452 M_\odot$ and $\gamma=2.923 M_\odot$)\footnote{In the figure and the fit we intentionally do not include the data for $q=0.85$ because a uniform sampling in $q$ should provide a more representative description. In any case, the additional data point is consistent with the different curves and leaves the fit parameters basically unchanged if included. A least-squares fit with the $q=0.85$ result yields $\alpha=-1.475 M_\odot$ and $\gamma=2.927 M_\odot$, which lies virtually on top of the dashed blue line.}. A second approach is more suited to the idea behind Eq.~\eqref{eq:q}. We simply compute the coefficients in  $M_\mathrm{thres}(q)=\alpha (1-q)^{2.5}+\gamma$ by requiring that the function should reproduce the results for $q=1$ and $q=0.7$ (red solid curve; $\alpha=-1.521 M_\odot$ and $\gamma=2.925 M_\odot$). The red dotted curve uses $M_\mathrm{thres}(q=0.5)$ instead of $M_\mathrm{thres}(q=0.7)$ (yielding $\alpha=-1.443 M_\odot$ and $\gamma=2.925 M_\odot$). It is apparent that this description is comparable to an actual fit, and we conclude that this procedure describes the data sufficiently well even for very asymmetric binaries. Note that the solid red curve also reproduces the data for $q=0.5$, i.e. by an extrapolation beyond the region $0.7\le q\le 1$. The deviations are smaller than the maximum deviations in Tabs.~\ref{tab:mthr1}~and~\ref{tab:mthr2}. We emphasize that Fig.~\ref{fig:dd2f} shows an example for an EoS which leads to relatively strong mass ratio effects, i.e. a large $\Delta M_\mathrm{thres}$. As discussed above, other EoSs result in much smaller differences in $M_\mathrm{thres}$ for $q=1$ and $q=0.7$.

We also remark that functions like $\alpha(1-q)^2+\gamma$ or $\alpha(1-q)^3+\gamma$ similarly reproduce the data as the ansatz with the power 2.5. Fitting the data $M_\mathrm{thres}(q)$ for DD2F to a function $\alpha(1-q)^n+\gamma$, $n$ is found to be 2.47. For $M_\mathrm{thres}(q)$ with the SFHX EoS in the range $0.6\le q\le 1$ we find $n=3.53$; % 3.45 for 0.5<=q<=1
for the SAPR and DD2 EoSs $M_\mathrm{thres}(q)$ is best described by higher-order polynomials with $n\sim 5$. This is because in the range $0.8 \leq q \leq 1$ $M_\mathrm{thres}(q)$ changes only slightly, but mass ratio effects on the threshold mass become continuously stronger around $q=0.7$ and below. Generally, the power will depend on the EoS, the range of $q$ which is considered for a fit, and also on the accuracy to which $M_\mathrm{thres}$ is determined. Corroborating the precise functional dependence on $q$ will require to study the behavior of $M_\mathrm{thres}(q)$ for many more EoSs, which we leave for future work. Apparently, there is no universal power $n$ valid for all EoS and a larger range in $q$. We find that $n=3$ yields a good description of all data. See discussion below and in Appendix~\ref{app:q}, which shows that a simple linear $q$ dependence does not describe the data very well. Higher powers cause a very strong reduction of $M_\mathrm{thres}$ at smaller $q$, which may even become unphysical as $M_\mathrm{thres}$ approaches $M_\mathrm{max}$. We thus adopt below a dependence $M_\mathrm{thres}=\alpha(1-q)^n+\gamma$ with $n=3$. 

Ref.~\cite{Bernuzzi2020} speculated about a dependence $M_\mathrm{thres}(q)=M_\mathrm{thres}^{q=1}\left( \frac{q}{(1+q)^2}\right)^{0.6}$, which would lead to a constant difference $\Delta M_\mathrm{thres}=M_\mathrm{thres}^{q=1}-M_\mathrm{thres}^{q=0.7}$ for all EoSs. This function does not lead to a very accurate description of mass ratio effects on the threshold mass because it does not take into account that the influence of $q$ is different for different EoSs. See Fig.~\ref{fig:DMthr}; recall that $\Delta M_\mathrm{thres}\approx 0$ for some EoSs, i.e. $M_\mathrm{thres}(q=1)\approx M_\mathrm{thres}(q=0.7)$. We also note that for the range of binary mass ratios considered in this study, $M_\mathrm{thres}(q)$ changes smoothly with $q$ and we do not observe an abrupt change of the collapse behavior that could indicate the occurence of a qualitatively new mechanism related to enhanced tidal disruption 
triggering the collapse in very asymmetric binaries~\cite{Bernuzzi2020}. We observe a smooth transition with asymmetric merger dynamics and tidal disruption already for slightly asymmetric systems, which continuously becomes more pronounced for more extreme mass ratios (see also discussion in Sect.~\ref{sec:toy}).

Finally, we note that also the threshold for prompt BH formation expressed by the chirp mass $\mathcal{M}_\mathrm{thres}$ is well captured by an ansatz $\mathcal{M}_\mathrm{thres}(q)=\alpha(1-q)^{2.5}+\gamma$ with the power 2.5 (see right panel of Fig.~\ref{fig:dd2f}). For the red solid curve we employ $M_\mathrm{thres}(q=1)$ and $M_\mathrm{thres}(q=0.7)$ to determine the coefficients $\alpha=1.135 M_\odot$ and $\gamma=1.273 M_\odot$. The red dotted curve is given by $\alpha=-1.076 M_\odot$ and $\gamma=1.273 M_\odot$ based on the data points $M_\mathrm{thres}(q=1)$ and $M_\mathrm{thres}(q=0.5)$.

Based on these observations for an individual EoS, we devise a general recipe to describe $q$-dependent relations for $M_\mathrm{thres}$ using fits for fixed mass ratio as in Tabs.~\ref{tab:mthr1} and~\ref{tab:mthr2}. It reads 
\begin{equation}\label{eq:Mqgeneral}
    M_\mathrm{thres}(q)=\alpha (1-q)^{3}+M_\mathrm{thres}(q=1)
\end{equation}
where the coefficient $\alpha$ is given by
\begin{equation}
\alpha=\frac{M_\mathrm{thres}(q=1-\Delta q)-M_\mathrm{thres}(q=1)}{\Delta q^{3}}.
\end{equation}

In this study we describe fits for $M_\mathrm{thres}$ with fixed mass ratios of $q=1$, $q=0.85$ and $q=0.7$. Since the high-order polynomial apparently fits the data well in the full range $0.7\leq q \leq 1$, we set $\Delta q=0.3$ and $\alpha$ is given by $\frac{M_\mathrm{thres}(0.7)-M_\mathrm{thres}(1)}{0.3^{3}}$ with fit functions for $M_\mathrm{thres}(0.7)$ and $M_\mathrm{thres}(1)$ listed in Tabs.~\ref{tab:mthr1} and~\ref{tab:mthr2}. The range of validity is $0.7\le q\le 1$ but we conclude from Fig.~\ref{fig:dd2f} that one can expect that this ansatz will also work reasonably well for some range below $q=0.7$. Also, considering the behavior in Fig.~\ref{fig:dd2f} for the DD2F EoS (and a few other EoS models for which we obtained $M_\mathrm{thres}(q)$ with finer spacing in the range $0.5\le q\le 1$, see also Fig.~2 in the Supplemental Material of~\cite{Bauswein2020a}), we suspect that q-dependent relations given by Eq.~\eqref{eq:Mqgeneral} show a tightness, which is roughly comparable to the one found for fits with fixed mass ratio and which is significantly better than that of fits which combine $q=0.7$, $q=0.85$ and $q=1$ data without explicit q-dependence (like fits 4-6, 10-12, 18-20, 24-26, 32-34, 38-40, 46-48 and 52-54 in Tabs.~\ref{tab:mthr1} and~\ref{tab:mthr2}).

It is also clear from the right panel in Fig.~\ref{fig:dd2f} that the ansatz in Eq.~\eqref{eq:Mqgeneral} also works for improved descriptions of $\mathcal{M}_\mathrm{thres}(q)$.

\subsection{General $q$-dependent fit formula}

Based on the considerations above, we directly develop $q$-dependent fit formulae $M_\mathrm{thres}(q,X,Y)$ as alternative to Eq.~\eqref{eq:Mqgeneral}. First, we note that in our case the parameter $\alpha$ in Eq.~\eqref{eq:Mqgeneral} is given by $\alpha=-\Delta M_\mathrm{thres} /0.3^{3}$ for the range $0.7\leq q \leq 1$. Hence, Eq.~\eqref{eq:Mqgeneral} reads
\begin{equation}\label{eq:fullq}
M_\mathrm{thres}(q)=-\frac{\Delta M_\mathrm{thres}}{0.3^{3}} \delta q^{3}+M_\mathrm{thres}(q=1)
\end{equation}
with $\delta q\equiv 1-q$. As discussed above, $\Delta M_\mathrm{thres}$ can also be described by fit formulae (with the same independent variables $X$ and $Y$ as a fit for $M_\mathrm{thres}(q=1)$). In combination this suggests the ansatz
\begin{equation}\label{eq:fullqfit}
\begin{split}
    &M_\mathrm{thres}(q,M_\mathrm{max},R_{1.6})=c_1 M_\mathrm{max} + c_2 R_{1.6} +c_3 \\
    &+ c_4 \delta q^{3} M_\mathrm{max}+ c_5 \delta q^{3} R_{1.6} + c_6 \delta q^{3}.
\end{split}    
\end{equation}

Using this non-linear ansatz in a least-squares fit to the $M_\mathrm{thres}$ data for $q=0.7$, $q=0.85$ and $q=1$, we realize that the last term does not yield a significant improvement regarding the tightness of the relation. Thus, we omit the last term in  Eq.~\eqref{eq:fullqfit} and set $c_6=0$. Fitting the data of the base sample we then determine $c_1=0.578$, $c_2=0.161$, $c_3=-0.218$, $c_4=8.987$ $c_5=-1.767$ (with $c_6=0$ by construction). Remarkably, the maximum residual is only 0.066~$M_\odot$ (with an average deviation between fit and data of 0.017~$M_\odot$). Hence, the higher-dimensional, more general $q$-dependent relation given by Eq.~\eqref{eq:fullqfit} is approximately as tight as the relations for fixed $q$ (see fits~1 to~3 in Tab.~\ref{tab:mthr1}). This compares to a maximum residual of 0.151~$M_\odot$ (0.027~$M_\odot$ average deviation) for a fit combining the data for all mass ratios without an explicit $q$ dependence (fit~6 in Tab.~\ref{tab:mthr1}). For a more detailed comparison we individually check the deviations of the data sets with fixed $q$ against the higher-dimensional formula. We find maximum and average deviations of 0.047~$M_\odot$ and 0.016~$M_\odot$ for $q=1$, 0.044~$M_\odot$ and 0.017~$M_\odot$ for $q=0.85$, and 0.066~$M_\odot$ and 0.018~$M_\odot$ for $q=0.7$. Comparing these numbers to those for fits 1, 2 and 3 in Tab.~\ref{tab:mthr1}, shows that the more general formula with explicit q-dependence is nearly as accurate as the fits for fixed $q$, although it is valid for the full range $0.7 \leq q \leq 1$. In line with these numbers we find that the reduced chi-square is 0.84 for the relation with explicit mass ratio dependence (similar to the reduced chi-square of fits~1 to~3 for fixed $q$), whereas fit~6 from Tab.~\ref{tab:mthr1}, i.e. the relation without q dependence, yields about 2.23.

\begin{table*} 
\begin{tabular}{|l|c|c|c|c|c|c|c|c|c|}  \hline 
%fit / EoSs  & $c_1$ & $c_2$ & $c_3$ & $c_4$ & $c_5$ & max. & av. & $\sum$ sq. res. & N \\ \hline 

%

\multicolumn{10}{|l|}{$M_\mathrm{thres}(q,M_\mathrm{max},R_{1.6})=c_1 M_\mathrm{max} + c_2 R_{1.6} +c_3 + c_4 \delta q^3 M_\mathrm{max}+ c_5 \delta q^3 R_{1.6}$}  \\ \hline %execfile("Mthres-r16-fullq.py") but adat!!

% base  & 0.578 & 0.161 & -0.218 & 8.987 & -1.767 & 0.066 & 0.017 \\ \hline
% base + hyb  & 0.663 & 0.154 & -0.329 & 9.058 & -1.784 & 0.116 & 0.035 \\ \hline
% base + ex  & 0.633 & 0.156 & -0.274 & 9.120 & -1.783 & 0.108 & 0.022 \\ \hline
% base + hyb + ex  & 0.664 & 0.157 & -0.369 & 9.145 & -1.792 & 0.118 & 0.033 \\ \hline
% 

% 
% b  & 0.578 $\pm$ 0.025 & 0.161 $\pm$ 0.004 & -0.218 $\pm$ 0.060 & 8.987 $\pm$ 1.268 & -1.767 $\pm$ 0.229 & 0.066 & 0.017 & 0.0335 & 69 \\ \hline    %  0.838225448368
% b + h  & 0.663 $\pm$ 0.045 & 0.154 $\pm$ 0.008 & -0.329 $\pm$ 0.116 & 9.058 $\pm$ 2.321 & -1.784 $\pm$ 0.412 & 0.116 & 0.035 & 0.1950 & 96 \\ \hline    %  3.4285837241
% b + e  & 0.633 $\pm$ 0.019 & 0.156 $\pm$ 0.003 & -0.274 $\pm$ 0.037 & 9.120 $\pm$ 1.087 & -1.783 $\pm$ 0.192 & 0.108 & 0.022 & 0.0738 & 93 \\ \hline    %  1.3411195402
% b + h + e  & 0.664 $\pm$ 0.028 & 0.157 $\pm$ 0.005 & -0.369 $\pm$ 0.055 & 9.145 $\pm$ 1.628 & -1.792 $\pm$ 0.284 & 0.118 & 0.033 & 0.2281 & 120 \\ \hline    %  3.17412949385

EoSs  & $c_1$ & $c_2/\times 10^{-1}$ & $c_3$ & $c_4$ & $c_5$ & max. & av. & $\sum$ sq. res. & N \\ \hline 

b  & 0.578 $\pm$ 0.025 & $1.610\pm0.042$ & -0.218 $\pm$ 0.060 & 8.987 $\pm$ 1.268 & -1.767 $\pm$ 0.229 & 0.066 & 0.017 & 0.0335 & 69 \\ \hline    %  0.838225448368
b + h  & 0.663 $\pm$ 0.045 & $1.535\pm 0.082$ & -0.329 $\pm$ 0.116 & 9.058 $\pm$ 2.321 & -1.784 $\pm$ 0.412 & 0.116 & 0.035 & 0.1950 & 96 \\ \hline    %  3.4285837241
b + e  & 0.633 $\pm$ 0.019 & $1.555 \pm 0.034$ & -0.274 $\pm$ 0.037 & 9.120 $\pm$ 1.087 & -1.783 $\pm$ 0.192 & 0.108 & 0.022 & 0.0738 & 93 \\ \hline    %  1.3411195402
b + h + e  & 0.664 $\pm$ 0.028 & $1.566 \pm 0.052$ & -0.369 $\pm$ 0.055 & 9.145 $\pm$ 1.628 & -1.792 $\pm$ 0.284 & 0.118 & 0.033 & 0.2281 & 120 \\ \hline    %  3.17412949385

\hline
\hline

\multicolumn{10}{|l|}{$M_\mathrm{thres}(q,M_\mathrm{max},R_\mathrm{max})=c_1 M_\mathrm{max} + c_2 R_\mathrm{max} +c_3 + c_4 \delta q^3 M_\mathrm{max}+ c_5 \delta q^3 R_\mathrm{max}$}  \\ \hline

% b  & 0.491 & 0.185 & -0.051 & 9.382 & -2.088 & 0.071 & 0.021 \\ \hline
% b + h  & 0.613 & 0.166 & -0.135 & 8.209 & -1.859 & 0.102 & 0.032 \\ \hline
% b + e  & 0.507 & 0.189 & -0.129 & 10.074 & -2.218 & 0.074 & 0.021 \\ \hline
% b + h + e  & 0.571 & 0.182 & -0.206 & 9.249 & -2.061 & 0.121 & 0.031 \\ \hline

% b  & 0.491 $\pm$ 0.032 & 0.185 $\pm$ 0.006 & -0.051 $\pm$ 0.071 & 9.382 $\pm$ 1.656 & -2.088 $\pm$ 0.340 & 0.071 & 0.021 & 0.0512 & 69 \\ \hline    %  1.27880933885
% b + h  & 0.613 $\pm$ 0.040 & 0.166 $\pm$ 0.008 & -0.135 $\pm$ 0.096 & 8.209 $\pm$ 2.074 & -1.859 $\pm$ 0.419 & 0.102 & 0.032 & 0.1551 & 96 \\ \hline    %  2.72695810713
% b + e  & 0.507 $\pm$ 0.020 & 0.189 $\pm$ 0.004 & -0.129 $\pm$ 0.036 & 10.074 $\pm$ 1.205 & -2.218 $\pm$ 0.242 & 0.074 & 0.021 & 0.0733 & 93 \\ \hline    %  1.33314929289
% b + h + e  & 0.571 $\pm$ 0.027 & 0.182 $\pm$ 0.006 & -0.206 $\pm$ 0.048 & 9.249 $\pm$ 1.623 & -2.061 $\pm$ 0.323 & 0.121 & 0.031 & 0.1974 & 120 \\ \hline    %  2.74603516996

EoSs  & $c_1$ & $c_2/\times 10^{-1}$ & $c_3$ & $c_4$ & $c_5$ & max. & av. & $\sum$ sq. res. & N \\ \hline 

b  & 0.491 $\pm$ 0.032 & $1.847 \pm 0.061$ & -0.051 $\pm$ 0.071 & 9.382 $\pm$ 1.656 & -2.088 $\pm$ 0.340 & 0.071 & 0.021 & 0.0512 & 69 \\ \hline    %  1.27880933885
b + h  & 0.613 $\pm$ 0.040 & $1.665 \pm 0.079$ & -0.135 $\pm$ 0.096 & 8.209 $\pm$ 2.074 & -1.859 $\pm$ 0.419 & 0.102 & 0.032 & 0.1551 & 96 \\ \hline    %  2.72695810713
b + e  & 0.507 $\pm$ 0.020 & $1.888 \pm 0.042$ & -0.129 $\pm$ 0.036 & 10.074 $\pm$ 1.205 & -2.218 $\pm$ 0.242 & 0.074 & 0.021 & 0.0733 & 93 \\ \hline    %  1.33314929289
b + h + e  & 0.571 $\pm$ 0.027 & $1.820 \pm 0.057$ & -0.206 $\pm$ 0.048 & 9.249 $\pm$ 1.623 & -2.061 $\pm$ 0.323 & 0.121 & 0.031 & 0.1974 & 120 \\ \hline    %  2.74603516996

\hline
\hline

\multicolumn{10}{|l|}{$M_\mathrm{thres}(q,M_\mathrm{max},\Lambda_{1.4})=c_1 M_\mathrm{max} + c_2 \Lambda_{1.4} +c_3 + c_4 \delta q^3 M_\mathrm{max}+ c_5 \delta q^3 \Lambda_{1.4}$}  \\ \hline

% b  & 0.698 & 0.001 & 1.137 & 0.900 & -0.009 & 0.096 & 0.029 \\ \hline
% b + h  & 0.750 & 0.001 & 1.016 & 0.878 & -0.010 & 0.111 & 0.040 \\ \hline
% b + e  & 0.651 & 0.001 & 1.325 & -0.026 & -0.004 & 0.114 & 0.043 \\ \hline
% b + h + e  & 0.674 & 0.001 & 1.265 & -0.292 & -0.003 & 0.123 & 0.048 \\ \hline

% 
% b  & 0.698 $\pm$ 0.035 & 0.001 $\pm$ 0.000 & 1.137 $\pm$ 0.074 & 0.900 $\pm$ 0.455 & -0.009 $\pm$ 0.002 & 0.096 & 0.029 & 0.0854 & 69 \\ \hline    %  2.13590525948
% b + h  & 0.750 $\pm$ 0.041 & 0.001 $\pm$ 0.000 & 1.016 $\pm$ 0.088 & 0.878 $\pm$ 0.616 & -0.010 $\pm$ 0.003 & 0.111 & 0.040 & 0.2242 & 96 \\ \hline    %  3.94120403473
% b + e  & 0.651 $\pm$ 0.029 & 0.001 $\pm$ 0.000 & 1.325 $\pm$ 0.059 & -0.026 $\pm$ 0.435 & -0.004 $\pm$ 0.001 & 0.114 & 0.043 & 0.2404 & 93 \\ \hline    %  4.37139109437
% b + h + e  & 0.674 $\pm$ 0.030 & 0.001 $\pm$ 0.000 & 1.265 $\pm$ 0.060 & -0.292 $\pm$ 0.444 & -0.003 $\pm$ 0.002 & 0.123 & 0.048 & 0.3746 & 120 \\ \hline    %  5.21152232584
% 
EoSs  & $c_1$ & $c_2/\times 10^{-4}$ & $c_3$ & $c_4$ & $c_5/\times 10^{-3}$ & max. & av. & $\sum$ sq. res. & N \\ \hline 

b  & 0.698 $\pm$ 0.035 & $7.772 \pm 0.361$ & 1.137 $\pm$ 0.074 & 0.900 $\pm$ 0.455 & $-9.050 \pm 2.295$ & 0.096 & 0.029 & 0.0854 & 69 \\ \hline    %  2.13590525948
b + h  & 0.750 $\pm$ 0.041 & $7.670 \pm 0.487$ & 1.016 $\pm$ 0.088 & 0.878 $\pm$ 0.616 & $-9.763 \pm 3.078$ & 0.111 & 0.040 & 0.2242 & 96 \\ \hline    %  3.94120403473
b + e  & 0.651 $\pm$ 0.029 & $5.370 \pm 0.243$ & 1.325 $\pm$ 0.059 & -0.026 $\pm$ 0.435 & $-3.541 \pm 1.483$ & 0.114 & 0.043 & 0.2404 & 93 \\ \hline    %  4.37139109437
b + h + e  & 0.674 $\pm$ 0.030 & $5.402 \pm 0.263$ & 1.265 $\pm$ 0.060 & -0.292 $\pm$ 0.444 & $-3.121 \pm 1.595$ & 0.123 & 0.048 & 0.3746 & 120 \\ \hline    %  5.21152232584

\hline
\hline

\multicolumn{10}{|l|}{$M_\mathrm{thres}(q,M_\mathrm{max},\tilde{\Lambda}_\mathrm{thres})=c_1 M_\mathrm{max} + c_2 \tilde{\Lambda}_\mathrm{thres} +c_3 + c_4 \delta q^3 M_\mathrm{max}+ c_5 \delta q^3 \tilde{\Lambda}_\mathrm{thres}$}  \\ \hline

% b  & 1.380 & 0.002 & -0.669 & 4.464 & -0.051 & 0.114 & 0.040 \\ \hline
% b + h  & 1.107 & 0.001 & 0.263 & 1.914 & -0.024 & 0.187 & 0.078 \\ \hline
% b + e  & 1.408 & 0.002 & -0.652 & 3.854 & -0.047 & 0.274 & 0.056 \\ \hline
% b + h + e  & 1.374 & 0.001 & -0.428 & 2.384 & -0.032 & 0.268 & 0.086 \\ \hline

% b  & 1.380 $\pm$ 0.061 & 0.002 $\pm$ 0.000 & -0.669 $\pm$ 0.154 & 4.464 $\pm$ 0.755 & -0.051 $\pm$ 0.005 & 0.114 & 0.040 & 0.1617 & 69 \\ \hline    %  4.04356000058
% b + h  & 1.107 $\pm$ 0.105 & 0.001 $\pm$ 0.000 & 0.263 $\pm$ 0.259 & 1.914 $\pm$ 1.157 & -0.024 $\pm$ 0.007 & 0.187 & 0.078 & 0.7930 & 96 \\ \hline    %  13.9437036652
% b + e  & 1.408 $\pm$ 0.044 & 0.002 $\pm$ 0.000 & -0.652 $\pm$ 0.116 & 3.854 $\pm$ 0.758 & -0.047 $\pm$ 0.005 & 0.274 & 0.056 & 0.5000 & 93 \\ \hline    %  9.09162165794
% b + h + e  & 1.374 $\pm$ 0.061 & 0.001 $\pm$ 0.000 & -0.428 $\pm$ 0.161 & 2.384 $\pm$ 1.015 & -0.032 $\pm$ 0.006 & 0.268 & 0.086 & 1.3246 & 120 \\ \hline    %  18.4289563276
% 

EoSs  & $c_1$ & $c_2/\times 10^{-3}$ & $c_3$ & $c_4$ & $c_5/\times 10^{-3}$ & max. & av. & $\sum$ sq. res. & N \\ \hline

b  & 1.380 $\pm$ 0.061 & $2.352 \pm 0.141$ & -0.669 $\pm$ 0.154 & 4.464 $\pm$ 0.755 & $-5.076\pm0.530$ & 0.114 & 0.040 & 0.1617 & 69 \\ \hline    %  4.04356000058
b + h  & 1.107 $\pm$ 0.105 & $0.998 \pm 0.191$ & 0.263 $\pm$ 0.259 & 1.914 $\pm$ 1.157 & $-2.416\pm0.707$ & 0.187 & 0.078 & 0.7930 & 96 \\ \hline    %  13.9437036652
b + e  & 1.408 $\pm$ 0.044 & $2.112 \pm 0.131$ & -0.652 $\pm$ 0.116 & 3.854 $\pm$ 0.758 & $-4.681\pm0.496$ & 0.274 & 0.056 & 0.5000 & 93 \\ \hline    %  9.09162165794
b + h + e  & 1.374 $\pm$ 0.061 & $1.490 \pm 0.164$ & -0.428 $\pm$ 0.161 & 2.384 $\pm$ 1.015 & $-3.206\pm0.611$ & 0.268 & 0.086 & 1.3246 & 120 \\ \hline    %  18.4289563276

\hline

\end{tabular} 
\caption{Different fits describing the EoS dependence of the threshold binary mass $M_\mathrm{thres}$ for prompt BH formation including an explicit dependence on the binary mass ratio $q$ through $\delta q=1-q$ (see main text). First column specifies the set of EoSs used for the fit (``b'' $\equiv$ hadronic base sample (a), ``e'' $\equiv$ exlcuded hadronic sample (b), ``h'' $\equiv$ hybrid sample (c) as defined  in Sect.~\ref{sec:eos}). Fit parameters $c_i$ and their respective variances are given in second to sixth columns. The units of the fit parameters $c_i$ are such that masses are in $M_\odot$, radii in km and tidal deformabilities dimensionless. These relations are obtained by least-square fits to the $M_\mathrm{thres}$ data for $q=1$, $q=0.85$ and $q=0.7$. Seventh and eighth columns specify the maximum and average deviation between fit and the underlying data. Last two columns give the sum of the squared residuals being minimized by the fit procedure and the number of data points included in the fit.} 
\label{tab:generalq} 
\end{table*}

With the ansatz in Eq.~\eqref{eq:fullqfit} setting $c_6=0$, we obtain various fit formulae for the different EoS samples. The fit parameters are given in Tab.~\ref{tab:generalq} together with the maximum residual and the average deviation between fit and data to quantify the accuracy of the relation. Table~\ref{tab:generalq} includes also relations with other independent variables, where we replace $R_{1.6}$ by $R_\mathrm{max}$, $\Lambda_{1.4}$ or $\tilde{\Lambda}_\mathrm{thres}$. As explained, these choices are motivated by what is assumed to be known about the EoS and which quantities are measured. Also for these relations we find that the $q$-dependent fit is approximately as accurate as the fits for fixed mass ratio in Tabs.~\ref{tab:mthr1} and~\ref{tab:mthr2}. For completeness we provide a table with the fit parameters for the full ansatz in Eq.~\eqref{eq:fullqfit} with all six terms in Appendix~\ref{app:q}. There, we also include fits with the ansatz $M_\mathrm{thres}(q,M_\mathrm{max},R_{1.6})=c_1 M_\mathrm{max} + c_2 R_{1.6} +c_3 + c_4 \delta q^{3} R_{1.6}$, i.e. we keep only the dominant EoS dependence of $\Delta M_\mathrm{thres}$ on $R_{1.6}$ (Tab.~\ref{tab:generalqfourterms}). These relations are not particularly tight compared to fits without explicit $q$ dependence.

\begin{figure}% execfile("gformula-plot.py") %updated for w085
\centering
\includegraphics[width=\columnwidth]{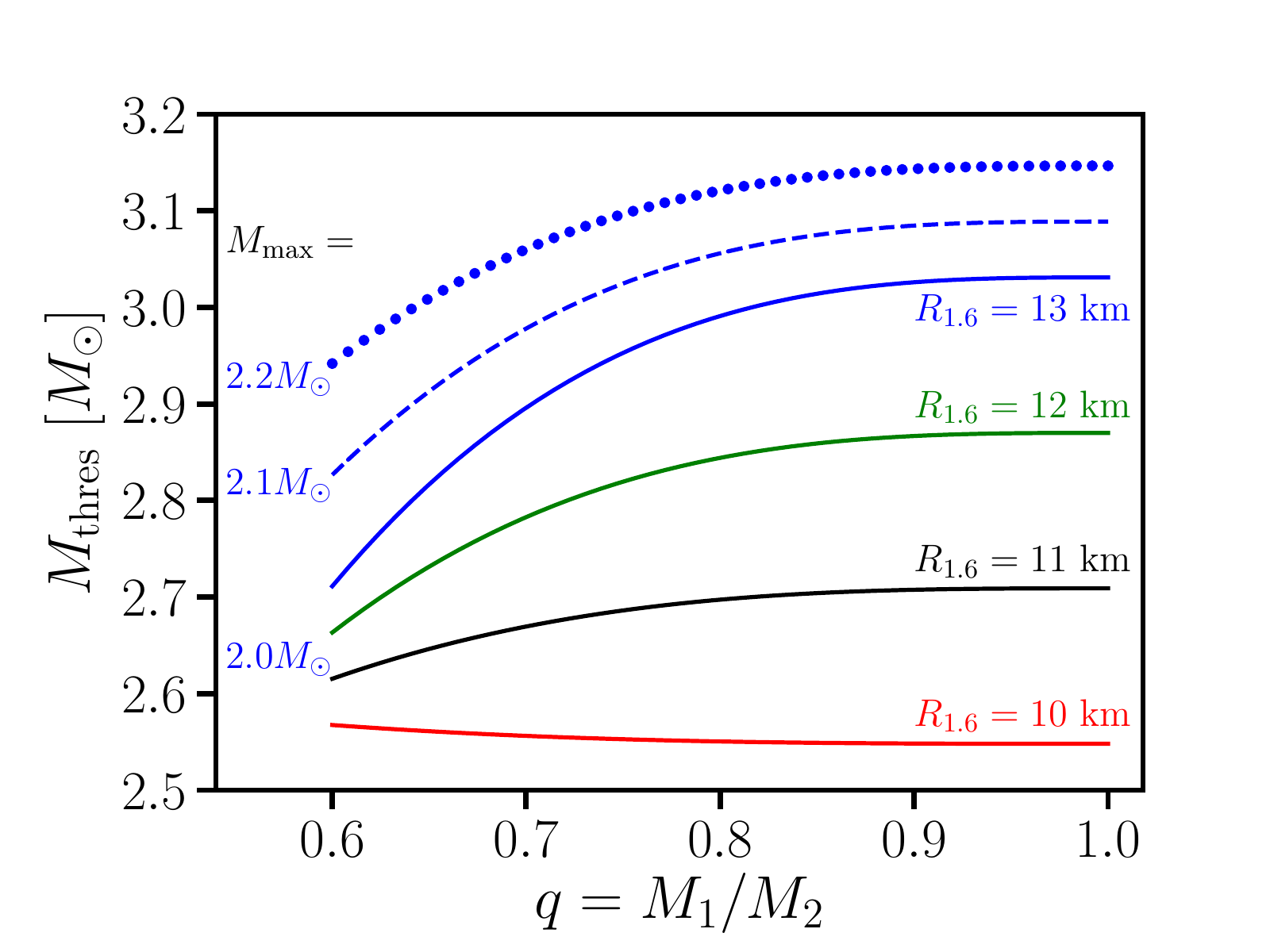}
\caption{Threshold binary mass for prompt collapse as function of binary mass ratio $q$ for different EoS properties based Eq.~\eqref{eq:fullqfit}. Solid curves assume a fixed maximum mass of $M_\mathrm{max}=2.0M_\odot$ but different NS radii. Blue curves show $M_\mathrm{thres}(q)$ for a fixed radius $R_{1.6}=13$~km but with $M_\mathrm{max}$ being 2.0~$M_\odot$ (solid), 2.1~$M_\odot$ (dashed) and 2.2~$M_\odot$ (dotted).}
\label{fig:gformula}
\end{figure}

To illustrate the influence of different EoS properties on $M_\mathrm{thres}(q)$, we plot Eq.~\eqref{eq:fullqfit} for fixed $M_\mathrm{max}$ and $R_{1.6}$ in Fig.~\ref{fig:gformula} with $c_i$ as given above. One can clearly recognize that $M_\mathrm{thres}$ increases with $M_\mathrm{max}$ and $R_{1.6}$, but that the impact of the binary mass ratio is very different depending on $M_\mathrm{max}$ and $R_{1.6}$.

We add some more remarks. Generating fits with a dependence on $\delta q$ to some other power $n$, e.g. 2, 2.5 or 4, leads to similarly tight fits. In Appendix~\ref{app:q} Tab.~\ref{tab:power} provides a comparison between fits like Eq.~\eqref{eq:fullqfit} adopting different powers $n$. Considering the deviations between fits and data we conclude that $n=3$ is the best choice to capture the higher-order behavior of $M_\mathrm{thres}(q)$ also indicated in Fig.~\ref{fig:dd2f}. The accuracy of the relations is not particularly sensitive to the exact value of $n$ (as long as $n$ is larger than abut 2 and smaller than about 4; see Appendix~\ref{app:q} for a brief discussion).

Finally, based on the findings for a single EoS in Fig.~\ref{fig:dd2f} (and a few other selected EoS models, i.e. SAPR, DD2, and SFHX in addition to DD2F), we suspect that the general $q$-dependent relations for $M_\mathrm{thres}$  (Eq.~\eqref{eq:fullqfit} and Tab.~\ref{tab:generalq}) hold also for mass ratios somewhat smaller than 0.7.

\begin{table*} 
\begin{tabular}{|l|c|c|c|c|c|c|c|c|c|}  \hline 
%fit / EoS sample  & $c_1$ & $c_2$ & $c_3$ & $c_4$ & $c_5$ & max. dev. & av. dev. & $\sum$ sq. res. & N \\ \hline 

\multicolumn{10}{|l|}{$\mathcal{M}_\mathrm{thres}(q,M_\mathrm{max},R_{1.6})=c_1 M_\mathrm{max} + c_2 R_{1.6} +c_3 + c_4 \delta q^3 M_\mathrm{max}+ c_5 \delta q^3 R_{1.6}$}  \\ \hline

% b  & 0.251 & 0.070 & -0.094 & 3.685 & -0.799 & 0.028 & 0.007 \\ \hline
% bh  & 0.288 & 0.067 & -0.142 & 3.696 & -0.803 & 0.049 & 0.015 \\ \hline
% be  & 0.275 & 0.068 & -0.118 & 3.727 & -0.803 & 0.048 & 0.009 \\ \hline
% bhe  & 0.289 & 0.068 & -0.159 & 3.732 & -0.806 & 0.050 & 0.015 \\ \hline

% b  & 0.251 & 0.070 & -0.094 & 3.685 & -0.799 & 0.028 & 0.007 & 0.0061 & 69 \\ \hline    %  0.809856527695
% b + h  & 0.288 & 0.067 & -0.142 & 3.696 & -0.803 & 0.049 & 0.015 & 0.0364 & 96 \\ \hline    %  3.37858339668
% b + e  & 0.275 & 0.068 & -0.118 & 3.727 & -0.803 & 0.048 & 0.009 & 0.0140 & 93 \\ \hline    %  1.33922647126
% b + h + e  & 0.289 & 0.068 & -0.159 & 3.732 & -0.806 & 0.050 & 0.015 & 0.0429 & 120 \\ \hline    %  3.14996318952
% 

EoSs  & $c_1$ & $c_2/\times 10^{-2}$ & $c_3/\times 10^{-1}$ & $c_4$ & $c_5/\times 10^{-1}$ & max. dev. & av. dev. & $\sum$ sq. res. & N \\ \hline 

b  & 0.251 $\pm$ 0.011 & $6.999 \pm 0.181$ & $-0.941 \pm 0.255$  & 3.685 $\pm$ 0.542 & $-7.992 \pm 0.980$  & 0.028 & 0.007 & 0.0061 & 69 \\ \hline    %  0.809856527695
b + h  & 0.288 $\pm$ 0.019 & $6.672 \pm 0.352$ & $-1.422 \pm 0.450$  & 3.696 $\pm$ 1.003 & $-8.027 \pm 1.778$  & 0.049 & 0.015 & 0.0364 & 96 \\ \hline    %  3.37858339668
b + e  & 0.275 $\pm$ 0.008 & $6.759 \pm 0.149$ & $-1.184 \pm 0.162$  & 3.727 $\pm$ 0.473 & $-8.035 \pm 0.833$  & 0.048 & 0.009 & 0.0140 & 93 \\ \hline    %  1.33922647126
b + h + e  & 0.289 $\pm$ 0.012 & $6.804 \pm 0.225$ & $-1.593 \pm 0.238$  & 3.732 $\pm$ 0.706 & $-8.061 \pm 1.233$  & 0.050 & 0.015 & 0.0429 & 120 \\ \hline    %  3.14996318952

\hline
\hline

\multicolumn{10}{|l|}{$\mathcal{M}_\mathrm{thres}(q,M_\mathrm{max},R_\mathrm{max})=c_1 M_\mathrm{max} + c_2 R_\mathrm{max} +c_3 + c_4 \delta q^3 M_\mathrm{max}+ c_5 \delta q^3 R_\mathrm{max}$}  \\ \hline

% b  & 0.214 & 0.080 & -0.022 & 3.866 & -0.945 & 0.031 & 0.009 \\ \hline
% bh  & 0.266 & 0.072 & -0.058 & 3.336 & -0.841 & 0.043 & 0.014 \\ \hline
% be  & 0.221 & 0.082 & -0.055 & 4.162 & -1.001 & 0.033 & 0.009 \\ \hline
% bhe  & 0.248 & 0.079 & -0.089 & 3.796 & -0.930 & 0.054 & 0.013 \\ \hline
% b  & 0.214 & 0.080 & -0.022 & 3.866 & -0.945 & 0.031 & 0.009 & 0.0094 & 69 \\ \hline    %  1.24621807899
% bh  & 0.266 & 0.072 & -0.058 & 3.336 & -0.841 & 0.043 & 0.014 & 0.0290 & 96 \\ \hline    %  2.69487280223
% be  & 0.221 & 0.082 & -0.055 & 4.162 & -1.001 & 0.033 & 0.009 & 0.0139 & 93 \\ \hline    %  1.33296464895
% bhe  & 0.248 & 0.079 & -0.089 & 3.796 & -0.930 & 0.054 & 0.013 & 0.0373 & 120 \\ \hline    %  2.73547647031

EoSs  & $c_1$ & $c_2/\times 10^{-2}$ & $c_3/\times 10^{-2}$ & $c_4$ & $c_5/\times 10^{-1}$ & max. dev. & av. dev. & $\sum$ sq. res. & N \\ \hline 

b  & 0.214 $\pm$ 0.014 & $8.032 \pm 0.262$ & $-2.162 \pm 3.044$  & 3.866 $\pm$ 0.711 & $-9.448 \pm 1.459$  & 0.031 & 0.009 & 0.0094 & 69 \\ \hline    %  1.24621807899
b + h  & 0.266 $\pm$ 0.017 & $7.236\pm 0.340$ & $-5.781 \pm 4.144$  & 3.336 $\pm$ 0.898 & $-8.409 \pm 1.812$  & 0.043 & 0.014 & 0.0290 & 96 \\ \hline    %  2.69487280223
b + e  & 0.221 $\pm$ 0.009 & $8.208 \pm 0.182$ & $-5.537 \pm 1.546$  & 4.162 $\pm$ 0.525 & $-10.01 \pm 1.052$  & 0.033 & 0.009 & 0.0139 & 93 \\ \hline    %  1.33296464895
b + h + e  & 0.248 $\pm$ 0.012 & $7.908 \pm 0.246$ & $-8.881 \pm 2.092$  & 3.796 $\pm$ 0.705 & $-9.304 \pm 1.403$  & 0.054 & 0.013 & 0.0373 & 120 \\ \hline    %  2.73547647031

\hline
\hline

\multicolumn{10}{|l|}{$\mathcal{M}_\mathrm{thres}(q,M_\mathrm{max},\Lambda_{1.4})=c_1 M_\mathrm{max} + c_2 \Lambda_{1.4} +c_3 + c_4 \delta q^3 M_\mathrm{max}+ c_5 \delta q^3 \Lambda_{1.4}$}  \\ \hline

% b  & 0.305 & 0.000 & 0.492 & 0.029 & -0.004 & 0.041 & 0.012 \\ \hline
% bh  & 0.327 & 0.000 & 0.440 & 0.021 & -0.004 & 0.048 & 0.017 \\ \hline
% be  & 0.285 & 0.000 & 0.573 & -0.387 & -0.002 & 0.049 & 0.019 \\ \hline
% bhe  & 0.295 & 0.000 & 0.548 & -0.503 & -0.001 & 0.053 & 0.021 \\ \hline

% b  & 0.305 & 0.000 & 0.492 & 0.029 & -0.004 & 0.041 & 0.012 & 0.0160 & 69 \\ \hline    %  2.10899878888
% bh  & 0.327 & 0.000 & 0.440 & 0.021 & -0.004 & 0.048 & 0.017 & 0.0420 & 96 \\ \hline    %  3.90088628518
% be  & 0.285 & 0.000 & 0.573 & -0.387 & -0.002 & 0.049 & 0.019 & 0.0459 & 93 \\ \hline    %  4.40574018031
% bhe  & 0.295 & 0.000 & 0.548 & -0.503 & -0.001 & 0.053 & 0.021 & 0.0711 & 120 \\ \hline    %  5.22463791404
% 

EoSs  & $c_1$ & $c_2/\times 10^{-4}$ & $c_3/\times 10^{-1}$ & $c_4$ & $c_5/\times 10^{-3}$ & max. dev. & av. dev. & $\sum$ sq. res. & N \\ \hline 

b  & 0.305 $\pm$ 0.015 & $3.381 \pm 0.156$ & $4.918 \pm 0.322$  & 0.029 $\pm$ 0.197 & $-4.102 \pm 0.993$  & 0.041 & 0.012 & 0.0160 & 69 \\ \hline    %  2.10899878888
b + h  & 0.327 $\pm$ 0.018 & $3.337 \pm 0.211$ & $4.399 \pm 0.383$  & 0.021 $\pm$ 0.267 & $-4.412 \pm 1.333$  & 0.048 & 0.017 & 0.0420 & 96 \\ \hline    %  3.90088628518
b + e  & 0.285 $\pm$ 0.013 & $2.332 \pm 0.106$ & $5.732 \pm 0.257$  & -0.387 $\pm$ 0.190 & $-1.625 \pm 0.648$  & 0.049 & 0.019 & 0.0459 & 93 \\ \hline    %  4.40574018031
b + h + e  & 0.295 $\pm$ 0.013 & $2.346 \pm 0.114$ & $5.475 \pm 0.262$  & -0.503 $\pm$ 0.193 & $-1.443 \pm 0.695$  & 0.053 & 0.021 & 0.0711 & 120 \\ \hline    %  5.22463791404

\hline
\hline

\multicolumn{10}{|l|}{$\mathcal{M}_\mathrm{thres}(q,M_\mathrm{max},\tilde{\Lambda}_\mathrm{thres})=c_1 M_\mathrm{max} + c_2 \tilde{\Lambda}_\mathrm{thres} +c_3 + c_4 \delta q^3 M_\mathrm{max}+ c_5 \delta q^3 \tilde{\Lambda}_\mathrm{thres}$}  \\ \hline

% b  & 0.600 & 0.001 & -0.289 & 1.572 & -0.022 & 0.049 & 0.017 \\ \hline
% bh  & 0.481 & 0.000 & 0.115 & 0.451 & -0.011 & 0.082 & 0.034 \\ \hline
% be  & 0.612 & 0.001 & -0.281 & 1.295 & -0.020 & 0.120 & 0.024 \\ \hline
% bhe  & 0.597 & 0.001 & -0.184 & 0.651 & -0.014 & 0.118 & 0.037 \\ \hline

% b  & 0.600 & 0.001 & -0.289 & 1.572 & -0.022 & 0.049 & 0.017 & 0.0303 & 69 \\ \hline    %  4.0039822891
% bh  & 0.481 & 0.000 & 0.115 & 0.451 & -0.011 & 0.082 & 0.034 & 0.1489 & 96 \\ \hline    %  13.8147862488
% be  & 0.612 & 0.001 & -0.281 & 1.295 & -0.020 & 0.120 & 0.024 & 0.0951 & 93 \\ \hline    %  9.12584203825
% bhe  & 0.597 & 0.001 & -0.184 & 0.651 & -0.014 & 0.118 & 0.037 & 0.2498 & 120 \\ \hline    %  18.3415834126

EoSs  & $c_1$ & $c_2/\times 10^{-4}$ & $c_3/\times 10^{-1}$ & $c_4$ & $c_5/\times 10^{-2}$ & max. dev. & av. dev. & $\sum$ sq. res. & N \\ \hline 

b  & 0.600 $\pm$ 0.026 & $10.20 \pm 0.610$ & $-2.889 \pm 0.667$  & 1.572 $\pm$ 0.327 & $-2.221 \pm 0.230$  & 0.049 & 0.017 & 0.0303 & 69 \\ \hline    %  4.0039822891
b + h  & 0.481 $\pm$ 0.045 & $4.322 \pm 0.826$ & $1.154 \pm 1.123$  & 0.451 $\pm$ 0.501 & $-1.055 \pm 0.306$  & 0.082 & 0.034 & 0.1489 & 96 \\ \hline    %  13.8147862488
b + e  & 0.612 $\pm$ 0.019 & $9.151 \pm 0.571$ & $-2.811 \pm 0.507$  & 1.295 $\pm$ 0.331 & $-2.043 \pm 0.216$  & 0.120 & 0.024 & 0.0951 & 93 \\ \hline    %  9.12584203825
b + h + e  & 0.597 $\pm$ 0.026 & $6.445 \pm 0.713$ & $-1.836 \pm 0.698$  & 0.651 $\pm$ 0.441 & $-1.397 \pm 0.265$  & 0.118 & 0.037 & 0.2498 & 120 \\ \hline    %  18.3415834126

\hline

\end{tabular} 
\caption{Different fits describing the EoS dependence of the threshold chirp mass $\mathcal{M}_\mathrm{thres}$ for prompt BH formation including an explicit dependence on the binary mass ratio $q$ through $\delta q=1-q$ (see main text). First column specifies the set of EoSs used for the fit (``b'' $\equiv$ hadronic base sample (a), ``e'' $\equiv$ exlcuded hadronic sample (b), ``h'' $\equiv$ hybrid sample (c) as defined  in Sect.~\ref{sec:eos}). Fit parameters $c_i$ and their respective variances are given in second to sixth columns. The units of the fit parameters $c_i$ are such that masses are in $M_\odot$, radii in km and tidal deformabilities dimensionless. These relations are obtained by least-square fits to the $\mathcal{M}_\mathrm{thres}$ data for $q=1$, $q=0.85$ and $q=0.7$. Seventh and eighth columns specify the maximum and average deviation between fit and the underlying data. Last two columns give the sum of the squared residuals being minimized by the fit procedure and the number of data points included in the fit.} 
\label{tab:generalqchirp} 
\end{table*}

We describe two possibilities to derive $q$-dependent fits for the chirp mass threshold $\mathcal{M}_\mathrm{thres}(q)$. We can employ the fits for $M_\mathrm{thres}$ (Eq.~\eqref{eq:fullqfit} and Tab.~\ref{tab:generalq}) and use the general relation $\mathcal{M}=M_\mathrm{tot}q^{0.6}(1+q)^{-1.2}$. This leads to
\begin{equation}
    \mathcal{M}_\mathrm{thres}(q)=M_\mathrm{thres}(q)q^{0.6}(1+q)^{-1.2}
\end{equation}
with $M_\mathrm{thres}(q)$ taken form Tab.~\ref{tab:generalq}. Alternatively, we directly fit the data for $\mathcal{M}_\mathrm{thres}(q)$ using an ansatz
\begin{equation}\label{eq:mchfullqfit}
\begin{split}
    &\mathcal{M}_\mathrm{thres}(q,M_\mathrm{max},R_{1.6})=c_1 M_\mathrm{max} + c_2 R_{1.6} +c_3 \\
    &+ c_4 \delta q^3 M_\mathrm{max}+ c_5 \delta q^3 R_{1.6}.% + c_6 \delta q^3.
\end{split}
\end{equation}
The resulting fit parameters are given in Tab.~\ref{tab:generalqchirp} for different sets of EoSs. The table also provides similar fits for other independent quantities like $R_\mathrm{max}$, $\Lambda_{1.4}$ and $\tilde{\Lambda}_\mathrm{thres}$. Both approaches lead to the same tightness of the final relations.

\begin{figure}% ang.py
\centering
\includegraphics[width=\columnwidth]{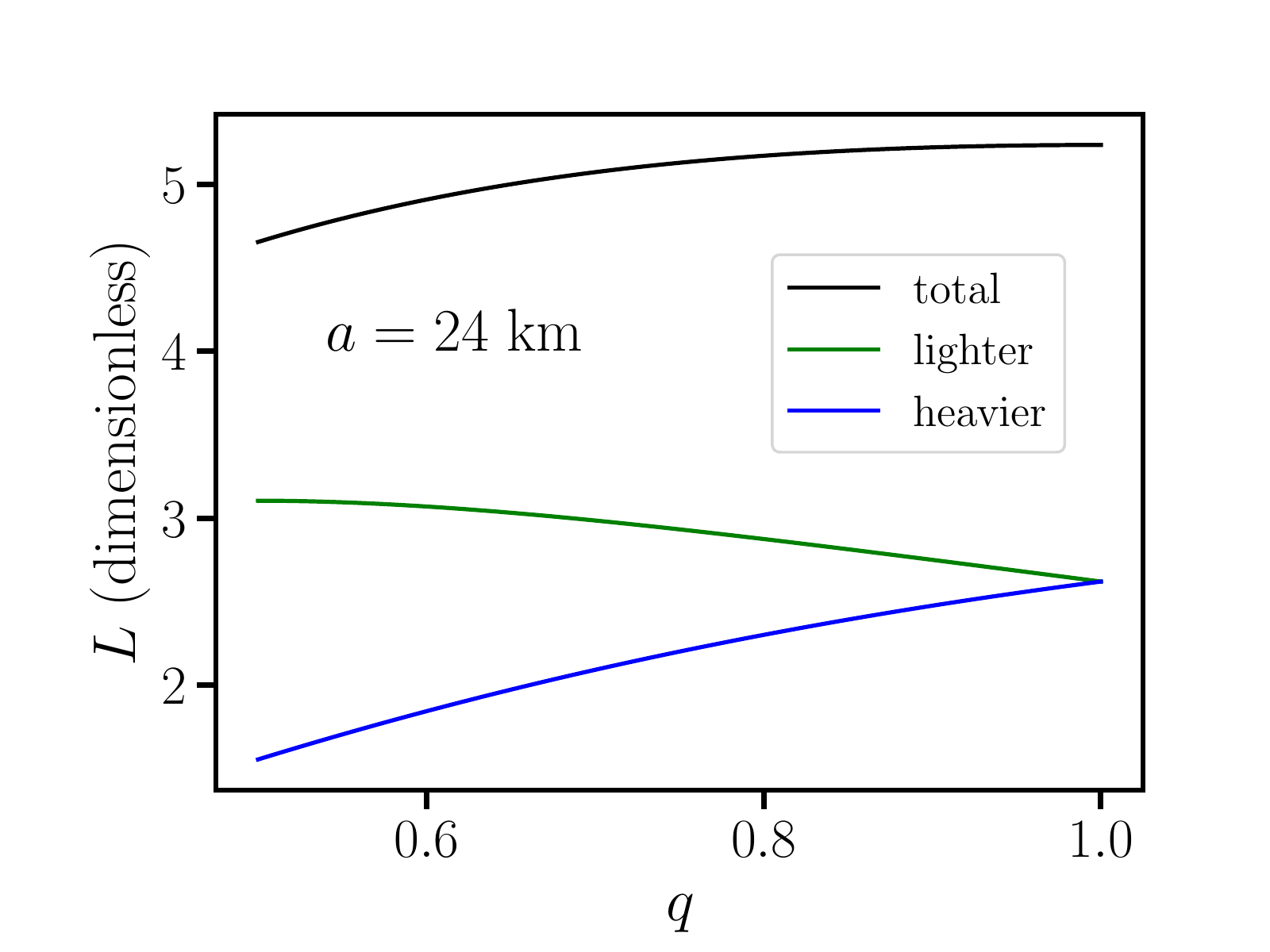}
\caption{Angular momentum of a binary with a total mass of 3~$M_\odot$ as function of the mass ratio $q$ within a Newtonian point-particle approximation at an orbital distance of 24~km. Within this model the total angular momentum decreases with binary mass asymmetry (black). The contribution from the lighter binary component grows (green curve), whereas the angular momentum of the more massive binary component decreases even stronger (blue curve).}
\label{fig:ang}
\end{figure}
\begin{figure}% RM-asymJ.py
\centering
\includegraphics[width=\columnwidth]{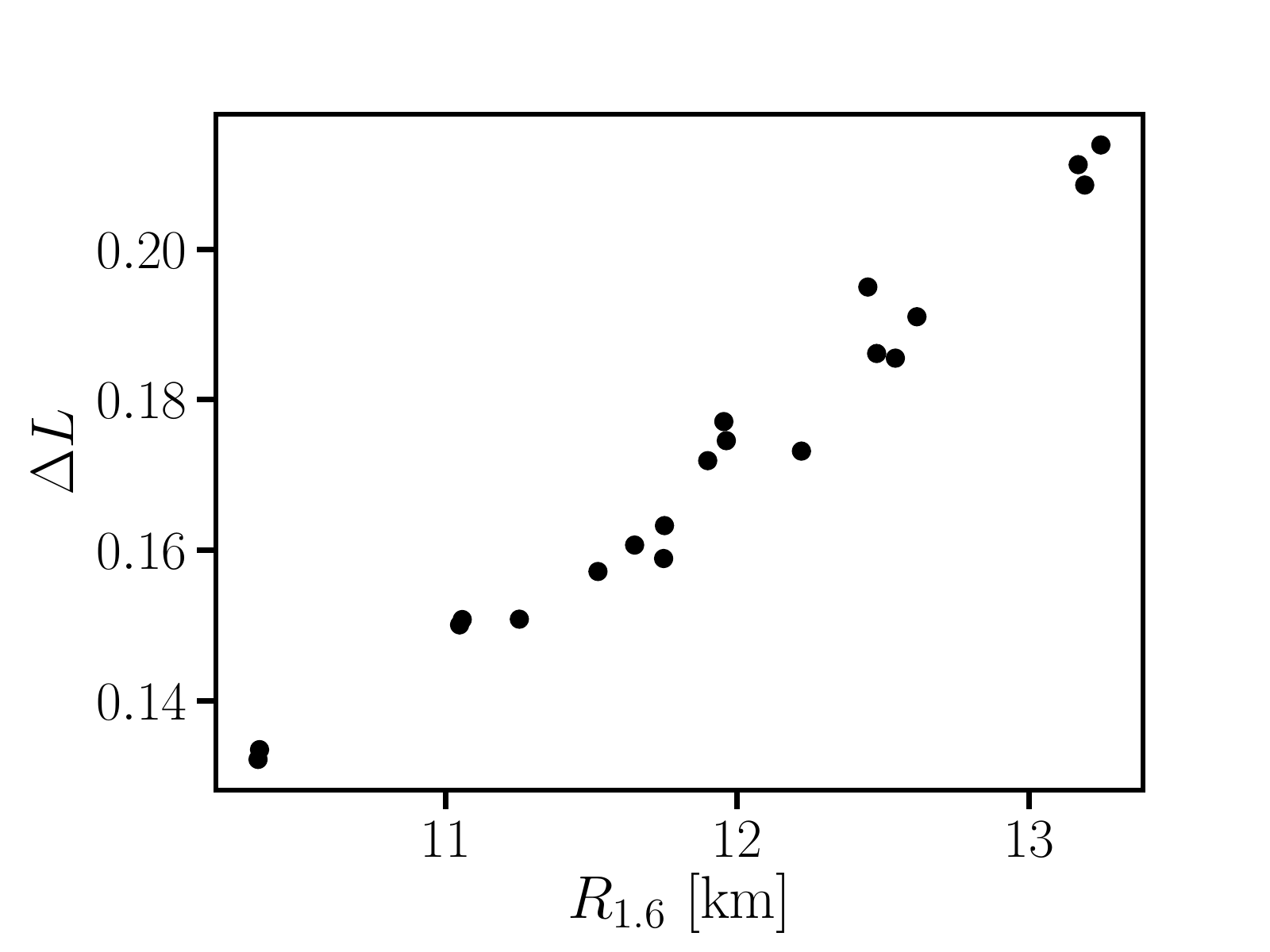}
\caption{Toy model estimate of the difference in the remnant's angular momentum between symmetric ($q=1$) and asymmetric mergers ($q=0.7$) as function of the radius $R_{1.6}$ of a 1.6~$M_\odot$ NS for different EoSs of the base sample. See text for more explanations.}
\label{fig:DL}
\end{figure}

\section{Semi-analytic model for mass ratio effects on $M_\mathrm{thres}$} \label{sec:toy}

One can work out a tentative explanation for mass ratio effects on $M_\mathrm{thres}$, i.e. Fig.~\ref{fig:DMthr}, by Newtonian point-particle dynamics. In a first step we assume that symmetric and asymmetric binaries merge at the same orbital distance $a$ and that the angular momentum $L$ of the remnant can be estimated by the orbital angular momentum at this distance. For the same orbital separation an asymmetric binary has less angular momentum than an equal-mass binary of the same total mass. Using Kepler's law, the total angular momentum is given by
\begin{equation}\label{eq:ang}
    L(q)=L_1+L_2=\frac{q a^2}{(1+q)^3}M_\mathrm{tot} \omega + \frac{q^2 a^2}{(1+q)^3}M_\mathrm{tot} \omega
\end{equation}
with $\omega=\sqrt{M_\mathrm{tot}/a^3}$ (gravitational constant $G$ suppressed for clarity). $L(q)$ is an increasing function of $q$, i.e. the angular momentum and thus the stability of the remnant decreases with the binary mass asymmetry (see Fig.~\ref{fig:ang}). This illustrates that generally asymmetric mergers may have smaller $M_\mathrm{thres}$. Here, we assume that the threshold mass is determined by a competition between the destabilizing effect of mass and the centrifugal support with increasing angular momentum (see~\cite{Bauswein2017a} for a semi-analytic model employing relativistic stationary stellar models).

Additionally, the angular momentum $L_1$ of the lighter binary component increases with asymmetry, while the more massive star carries less angular momentum ($L_2$) for smaller $q$ (see Fig.~\ref{fig:ang}). Considering the dynamics of asymmetric mergers where the less massive component is disrupted and wrapped around the more massive star, this may imply that there is less angular momentum in the very center of the remnant, which is formed by the more massive star. Hence, not only the total angular momentum is reduced for asymmetric mergers, but also its distribution may be such that the central core has less centrifugal support (see Fig.~12 in~\cite{Ciolfi2017} for stable remnants at late times considering systems of constant rest mass).

These simple considerations, however, do not explain the particular EoS dependence of the difference $\Delta M_\mathrm{thres}=M_\mathrm{thres}(q=1)-M_\mathrm{thres}(q=0.7)$ (or similarly the difference between the threshold mass for $q=1$ and $q=0.85$, see Fig.~\ref{fig:Dmthr08507}), which dominantly depends on $R_{1.6}$ (see Fig.~\ref{fig:DMthr}). Thus, we refine the model in a second step by assuming that the orbital distance at merging, i.e. the distance which determines the remnant's angular momentum, is given by $a=R_1+R_2=R(M_1)+R(M_2)$. Now, the model depends explicitly on the EoS through the different mass-radius relations. This estimate takes into account that the merging may occur at different orbital distances depending on the binary masses and the EoS. We compute the angular momentum of the remnant through Eq.~\eqref{eq:ang} by fixing $M_\mathrm{tot}=M_\mathrm{thres}(q=1)$ for every EoS of the base sample. We compare the angular momentum of equal-mass and asymmetric mergers of the same total mass by defining $\Delta L=L(q=1)-L(q=0.7)$. Fig.~\ref{fig:DL} displays $\Delta L$ as function of $R_{1.6}$ for all EoSs of the base sample. Note first that $\Delta L$ is positive, as already suspected from the arguments above. Moreover, the toy model qualitatively reproduces the EoS dependence by showing that larger NSs may lead to a stronger difference in the injected angular momentum in symmetric and asymmetric mergers. The model does not predict a slight increase of $M_\mathrm{thres}$ for small deviations from $q=1$ as it is indicated by our data for some relatively soft EoSs. Obviously, one should not overrate this simplistic model as the merging is a highly dynamical process and the angular momentum is a time-dependent quantity. Also, the exact distribution of angular momentum may be crucial, and mass ejection and disk formation in asymmetric systems may be enhanced. 

%%%%%%%%%%%%%%%%%%%%%%%%%%%%%%%%%%%%%%%%%%%%%%%%%%%%%%%%%%%%%%%%%%%%%%%%%%%%%%%%%%%%%%%%%

\section{Signature of a phase transition} \label{sec:pt}
\begin{figure}% lamthres-mthres-add-computed.py
\centering
\includegraphics[width=\columnwidth]{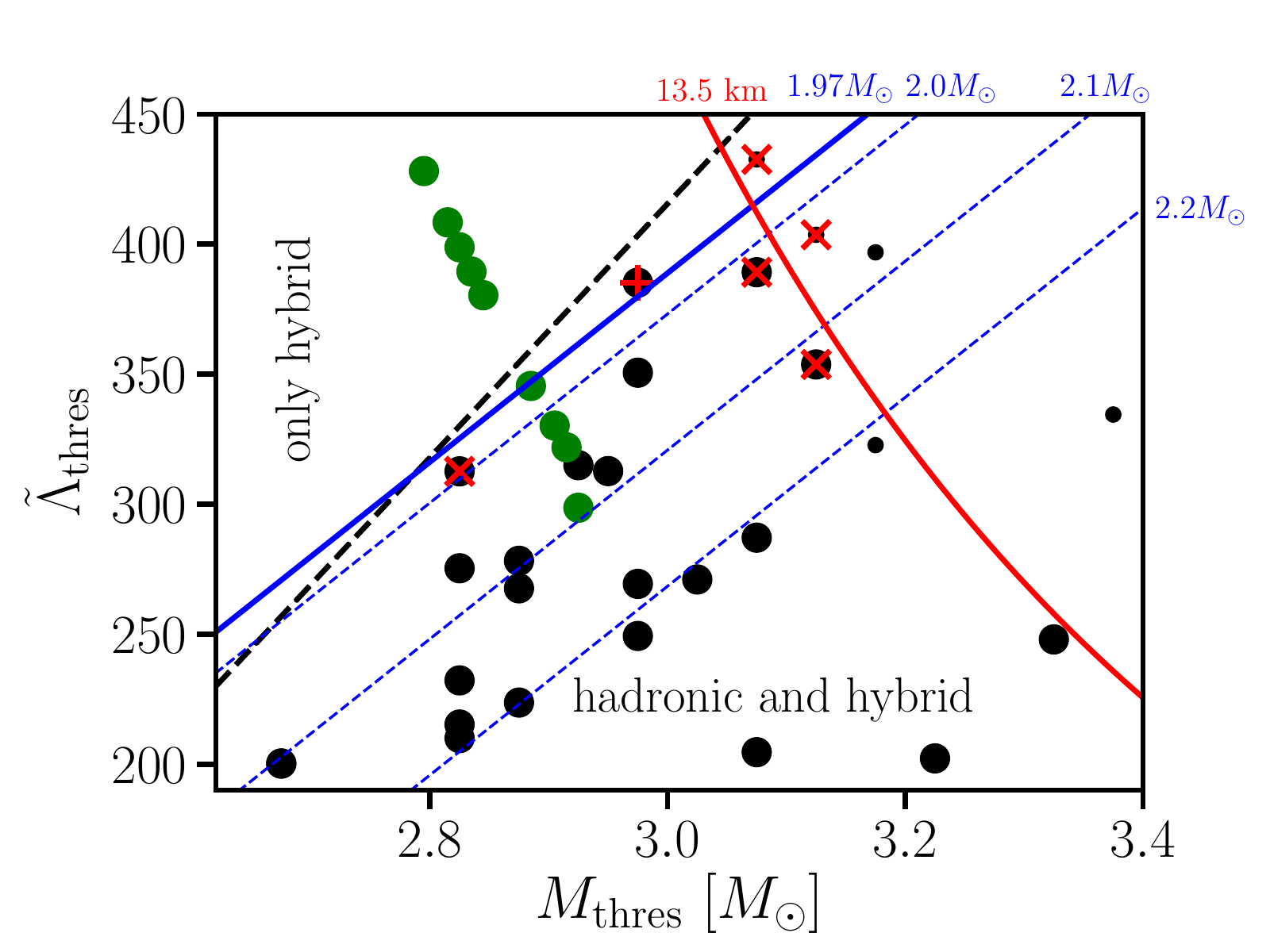}
\caption{Threshold mass $M_\mathrm{thres}$ and combined tidal deformability $\tilde{\Lambda}_\mathrm{thres}$ of binary systems with $M_\mathrm{tot}=M_\mathrm{thres}$ for different EoSs and $q=1$. EoS models are classified as in Sect.~\ref{sec:eos}: large black dots show the purely hadronic base sample, small black dots display the excluded hadronic sample, i.e. models which are incompatible with GW170817, and green symbols indicate hybrid models. (Two excluded models with $M_\mathrm{thres}>3.4~M_\odot$ lie outside the plot limits.) Overplotted crosses mark hyperonic EoSs. The red plus sign corresponds to the ALF2 EoS, where quark matter resembles properties of hadronic matter~\cite{Alford2005,Read2009a}. The black dashed line separates an area where only hybrid models occur, from a ``mixed'' regime with both hybrid and purely hadronic EoS. Blue lines display curves of constant $M_\mathrm{max}$, which are given by fit 43 in Tab.~\ref{tab:mthr2} for the hadronic base sample. Since viable hadronic models should have $M_\mathrm{max}\geq 1.97~M_\odot$, no hadronic EoS (large black dots) occurs significantly to the  left the blue solid line. Note that all except for one hybrid model (green dots) fulfill $M_\mathrm{max}\geq 1.97~M_\odot$, although some models occur on the left of the blue solid line. The red curve marks configurations which have approximately a NS radius of 13.5~km (with a mass of $M_\mathrm{thres}/2$ ) as an approximate upper bound on NS radii from GW170817. See main text for more explanations. A very similar figure can be found in~\cite{Bauswein2020a}.} 
\label{fig:pt}
\end{figure}

\subsection{General idea and equal-mass mergers}

In Sect.~\ref{sec:eos} we introduce different sets of EoS models, and we distinguish in particular purely hadronic EoSs and models with a phase transition to deconfined quark matter. The main purpose of this classification is to understand the impact of a phase transition on the collapse behavior (cf. Ref.~\cite{Bauswein2020a}).

Specifically, the combination of $M_\mathrm{thres}$ and $\tilde{\Lambda}_\mathrm{thres}$ may be indicative of a phase transition to deconfined quark matter. EoSs which undergo a phase transition tend to have relatively high $\tilde{\Lambda}_\mathrm{thres}$ in comparison to $M_\mathrm{thres}$. This is shown in Fig.~\ref{fig:pt}, which is nearly identical to Fig.~2 in~\cite{Bauswein2020a}. Hybrid models occur in the upper left corner of the diagram, which is not reached by hadronic models. Hence, measuring a combination $(\tilde{\Lambda}_\mathrm{thres},M_\mathrm{thres})$ in the regime above the dashed line provides evidence for a phase transition.

This particular impact of a phase transition on the collapse behavior is understandable. The gravitational collapse of the merger remnant, i.e. $M_\mathrm{thres}$, is determined by the very high density regime of the EoS. The tidal deformability $\tilde{\Lambda}_\mathrm{thres}$ is measured during the inspiral before merging and it thus contains only information about the EoS at moderate densities. Recall that for equal-mass mergers $\tilde{\Lambda}_\mathrm{thres}=\tilde{\Lambda}(M_\mathrm{thres}/2)$.

For most of the tested hybrid EoSs, we find that $M_\mathrm{thres}/2<M_\mathrm{onset}$ with $M_\mathrm{onset}$ being the mass of the lightest NS where quark matter occurs. Hence, the merging stars are purely hadronic and the phase transition takes place at densities, which are only reached after merging. This implies that $\tilde{\Lambda}_\mathrm{thres}$ is only sensitive to the hadronic regime of the EoS, whereas $M_\mathrm{thres}$ is strongly affected by the presence of quark matter.  Concretely, $M_\mathrm{thres}$ is significantly reduced compared to the purely hadronic reference model because the phase transition leads to a sudden softening of the EoS at higher densities. The reduction of $M_\mathrm{thres}$ implies an increase of $\tilde{\Lambda}_\mathrm{thres}=\tilde{\Lambda}(M_\mathrm{thres}/2)$ since the tidal deformability is larger for smaller masses. This explains why the regime beyond the dashed curve in Fig.~\ref{fig:pt} is only reached by hybrid EoSs which can lead to the combination of large $\tilde{\Lambda}_\mathrm{thres}$ and small $M_\mathrm{thres}$. The dashed line as criterion for the identification of a phase transition is drawn by hand and given by $\tilde{\Lambda}_\mathrm{thres}^\mathrm{hybrid}= 488 (M_\mathrm{thres}/M_\odot)-1050$. We remark that a similar figure can be obtained for $M_\mathrm{thres}$ and $\Lambda_{1.4}$. In this case the signature is less pronounced as the hybrid models are less distinguishable from the purely hadronic EoSs.

As is apparent from Fig.~\ref{fig:pt}, not all hybrid models are located at $\tilde{\Lambda}_\mathrm{thres}>\tilde{\Lambda}_\mathrm{thres}^\mathrm{hybrid}$. These are models with an EoS of the quark phase, which is relatively stiff. This stabilizes the remnant and yields a threshold mass that is comparable to that of the purely hadronic reference model. We thus conclude that a measured combination $(\tilde{\Lambda}_\mathrm{thres},M_\mathrm{thres})$ with $\tilde{\Lambda}_\mathrm{thres}<\tilde{\Lambda}_\mathrm{thres}^\mathrm{hybrid}(M_\mathrm{thres})$ is uninformative about the presence of a phase transition. In other words, depending on the properties of quark matter, a hybrid model can but does not need to be located in the regime labeled as ``only hybrid'' in Fig.~\ref{fig:pt}.

In~\cite{Bauswein2020a} we point out several advantages of the procedure to identify the presence of a phase transition through the comparison between $\tilde{\Lambda}_\mathrm{thres}$ and $\tilde{\Lambda}_\mathrm{thres}^\mathrm{hybrid}(M_\mathrm{thres})$. In particular, $\tilde{\Lambda}_\mathrm{thres}$ is larger and thus easier to measure compared to the combined tidal deformability of binaries composed of hybrid stars because $\tilde{\Lambda}_\mathrm{thres}=\tilde{\Lambda}(M_\mathrm{thres}/2)>\tilde{\Lambda}(M_\mathrm{onset})$ for most hybrid models. Also, a measurement of $\tilde{\Lambda}_\mathrm{thres}$ does not need to be very precise, whereas the identification of a phase transition through a kink in the relation $\Lambda(M)$ at higher masses around $M_\mathrm{onset}$ requires very accurate observations of $\Lambda$ (e.g. Fig.~3 in~\cite{Bauswein2019a}) or a larger number of events~\cite{Chatziioannou2020a,Chen2019}. Alternatively, measuring an increased dominant GW frequency of the postmerger phase would indicate the occurrence of a phase transition~\cite{Bauswein2019,Breschi2019,Weih2019,Bauswein2020hybrid}. See~\cite{Bauswein2020a} for some more advantageous aspects of the $\tilde{\Lambda}_\mathrm{thres}^\mathrm{hybrid}-M_\mathrm{thres}$ comparison.

For these reasons the new signature of a phase transition through a constraint on $(\tilde{\Lambda}_\mathrm{thres},M_\mathrm{thres})$ is very promising. However, one point is critical, namely the fact that no purely hadronic EoS is located above the dashed line in Fig.~\ref{fig:pt}. Our sample of hadronic models is relatively large, and thus we basically test the full range of viable models with a finite number of candidate EoSs. It is reasonable to expect that any other hadronic EoS is sufficiently similar to one of our tested models that it would lead to a very similar combination of $\tilde{\Lambda}_\mathrm{thres}$ and $M_\mathrm{thres}$. Similarly, we anticipate based on the physical understanding described above that also other hybrid models which are not considered here, may occur in the ``only hybrid'' regime. We remark that our hybrid models employ the same hadronic model (DD2F) at densities below the onset density of the phase transition. Because $M_\mathrm{thres}/2<M_\mathrm{onset}$ holds for most models, the hybrid models appear to lie on a virtual curve in Fig.~\ref{fig:pt}, which simply resembles the $\Lambda(M)$ relation of the hadronic model. Using other hadronic EoS for the density regime below the phase transition, we expect the corresponding models to be shifted along the black dashed line following the $\Lambda(M)$ curve of the hadronic EoS.

We add one more aspect corroborating and refining our line of arguments to exclude purely hadronic EoS models above $\tilde{\Lambda}_\mathrm{thres}^\mathrm{hybrid}(M_\mathrm{thres})$. We discuss in Sect.~\ref{sec:mthr} that there is a tight relation between $M_\mathrm{max}$, $\tilde{\Lambda}_\mathrm{thres}$ and $M_\mathrm{thres}$ in particular for hadronic EoSs. Using the relation for the base sample with $q=1$ (fit 43 in Tab.~\ref{tab:mthr2}), we draw curves of constant $M_\mathrm{max}$ in Fig.~\ref{fig:pt} (blue lines), see also Fig.~1 in~\cite{Bauswein2020a} showing that $M_\mathrm{max}$ increases with $M_\mathrm{thres}$ but decreases with $\tilde{\Lambda}_\mathrm{thres}$.

The blue solid curve for $M_\mathrm{max}=1.97~M_\odot$ closely follows the dashed black line separating the ``only hybrid'' area form the ``mixed'' regime. Current pulsar observations imply that $M_\mathrm{max}>1.97~M_\odot$ (corresponding to the one-sigma error bar in~\cite{Antoniadis2013}; Ref.~\cite{Cromartie2019} yields $M_\mathrm{max}>1.96~M_\odot$ within two sigma). The tight relation between $M_\mathrm{max}$, $\tilde{\Lambda}_\mathrm{thres}$ and $M_\mathrm{thres}$ thus explains why there are no purely hadronic models beyond the blue solid line or the dashed black line, respectively. We remark that the EoSs comprising the base sample have been selected by the condition that their maximum mass exceeds 1.97~$M_\odot$. One hadronic model of the base sample occurs slightly above the blue solid curve because the relation $M_\mathrm{max}(M_\mathrm{thres},\tilde{\Lambda}_\mathrm{thres})$ features small deviations (see Sect.~\ref{sec:mthr}). Otherwise, only hybrid models can lead to $(\tilde{\Lambda}_\mathrm{thres},M_\mathrm{thres})$ significantly beyond the blue curve for the reasons explained above. Generally, these considerations provide strong support for the assumption that no viable hadronic models occur in the upper left corner of the diagram and that indeed the phase transition to deconfined quark matter can be identified by $\tilde{\Lambda}_\mathrm{thres}>\tilde{\Lambda}_\mathrm{thres}^\mathrm{hybrid}(M_\mathrm{thres})$ with either the black dashed or the solid blue line as quantitative criterion.

Note that the hybrid models above the blue curve are compatible with a maximum mass of 1.97~$M_\odot$ (except for the VBAG EoS with $M_\mathrm{max}$ slightly below; see Tab.~\ref{tab:data}). It is precisely the presence of a phase transition which causes models to yield a combination of $(\tilde{\Lambda}_\mathrm{thres},M_\mathrm{thres})$ in a regime which is inaccessible by hadronic EoSs with $M_\mathrm{max}>1.97~M_\odot$.

We emphasize that the blue lines in Fig.~\ref{fig:pt} also show that the criterion for a phase transition can be updated by new or ongoing pulsar measurements which shift the lower limit on $M_\mathrm{max}$ towards higher masses~\cite{Cromartie2019}. For instance, $M_\mathrm{max}>2.1~M_\odot$ would imply a much larger region in the $\tilde{\Lambda}_\mathrm{thres}-M_\mathrm{thres}$ plane where only hybrid EoSs occur (beyond the dashed blue line labeled with $2.1~M_\odot$).

Finally, we include an additional curve in Fig.~\ref{fig:pt} to understand the location of different EoS models. For equal-mass mergers $\tilde{\Lambda}_\mathrm{thres}$ is given by $\Lambda(M_\mathrm{thres}/2)$. Moreover, it is known that the tidal deformability correlates well with the compactness $G M/(c^2R)$ of nonrotating NSs. As can be seen in Fig.~1 of the Supplemental Material of~\cite{Bauswein2020a}, for most EoS models $M_\mathrm{thres}/2$ falls roughly in the regime of constant radii in the mass-radius relations. We use the relation $\Lambda(M)=0.0093\left(\frac{GM}{c^2R}\right)^{-6}$ from~\cite{De2018} and assume a constant radius of 13.5~km, which is approximately the limit given by the inspiral GW signal from GW170817. This results in the solid red curve, which is given by $\tilde{\Lambda}_\mathrm{thres}(M_\mathrm{thres})=0.0093\left(\frac{G M_\mathrm{thres}}{2c^2 13.5~\mathrm{km}}\right)^{-6}$. In fact we find that all EoSs of the base sample are compatible with this limit in Fig.~\ref{fig:pt}. Only excluded models occur above the red curve.

\subsection{Asymmetric mergers}

\begin{figure*}% lamthres-mthres-q07-add-computed.py
\centering
\includegraphics[width=\columnwidth]{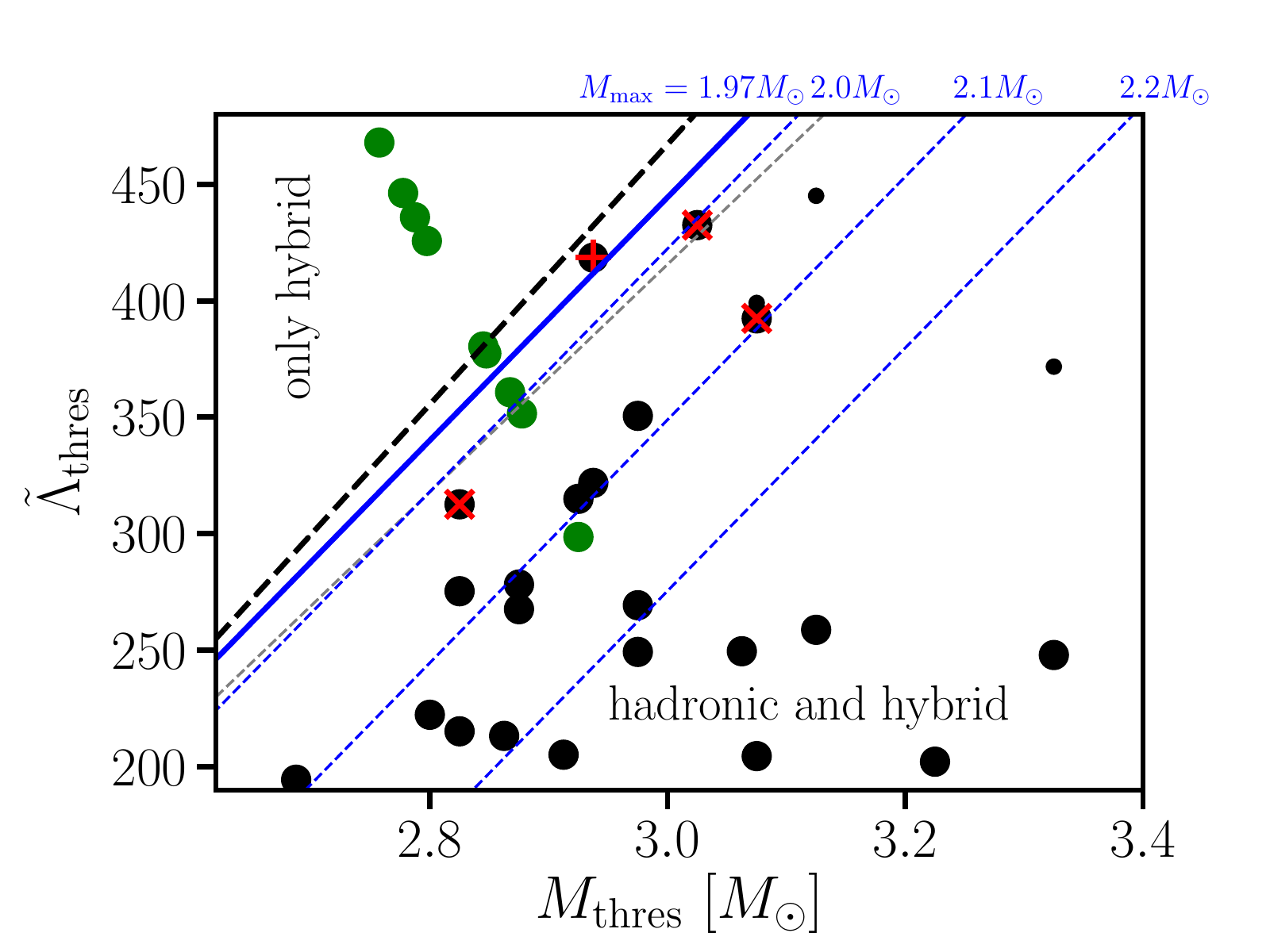} %new: execfile("lamthres-mthres-add-q085-computed.py")
\includegraphics[width=\columnwidth]{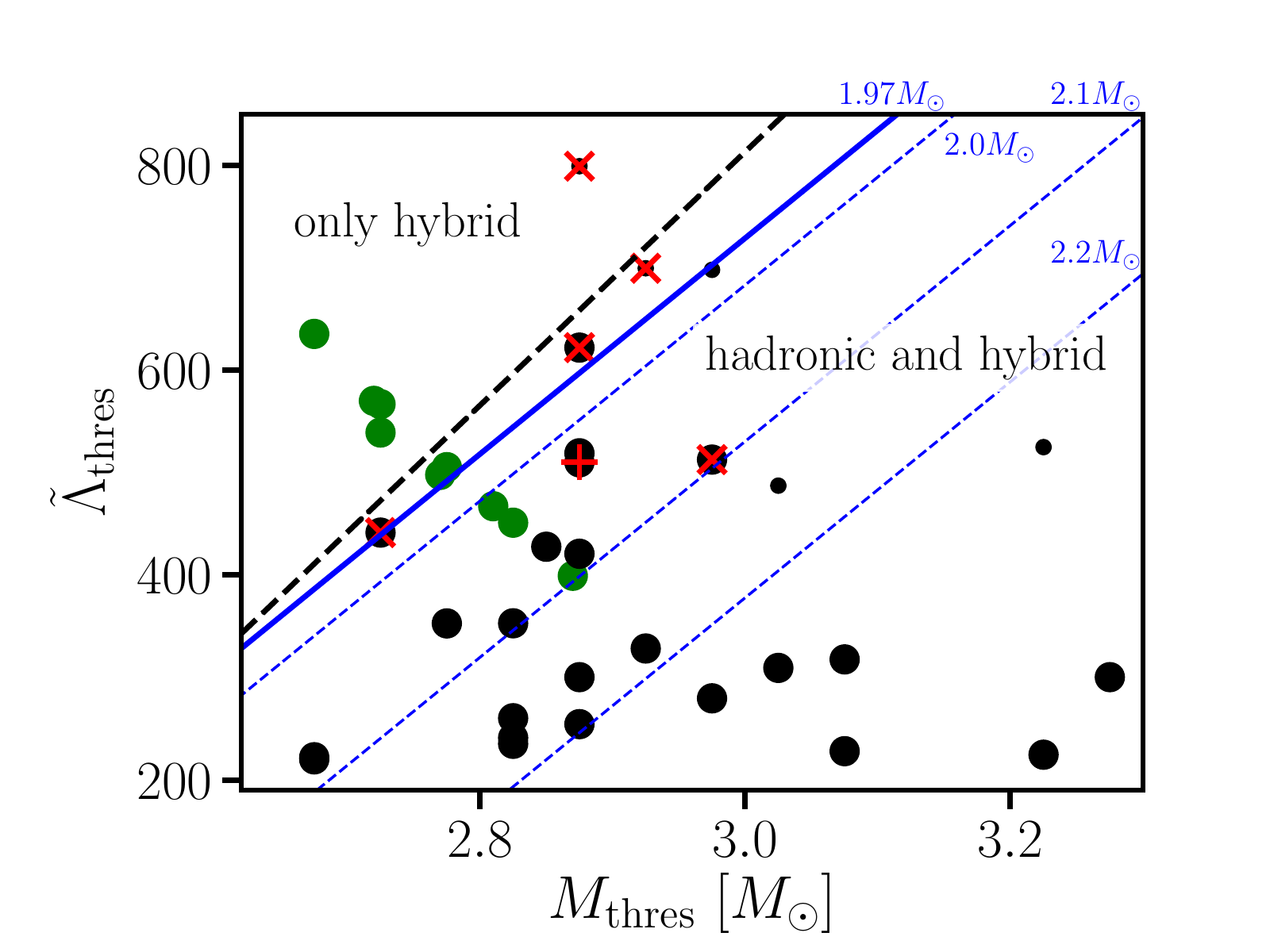}
\caption{Same as Fig.~\ref{fig:pt} but for a binary mass ratio $q=0.85$ (left panel) and $q=0.7$ (right panel). For orientation the gray dahsed line in the left panel is identical to the black dashed line in Fig.~\ref{fig:pt}.} 
\label{fig:2q07}
\end{figure*}

So far we discussed the signature of a phase transition focusing on equal-mass mergers assuming that the binary mass ratio will be measured with sufficient precision. Reasonably one can expect that small deviations from $q=1$ lead to a very similar diagram. Figures~\ref{fig:dd2f} and~\ref{fig:Dmthr08507} show that $M_\mathrm{thres}$ does not strongly vary for $q$ larger than $\sim 0.85$. As a consequence also the tidal deformability of a slightly asymmetric binary at the threshold does not change significantly. We explicitly present the results for $q=0.85$ and $q=0.7$. Figure~\ref{fig:2q07} shows $\tilde{\Lambda}_\mathrm{thres}$ as function of $M_\mathrm{thres}$ for mass ratios $q=0.85$ and $q=0.7$ to be compared to Fig~\ref{fig:pt}. Here we consider the full set of EoSs with the same meaning of the symbols as in Fig~\ref{fig:pt}.

The figure demonstrates that also for $q=0.85$ and $q=0.7$ the presence of a strong phase transition can be identified by the location of $(M_\mathrm{thres},\tilde{\Lambda}_\mathrm{thres})$ in comparison to the area spanned by purely hadronic EoSs. We infer the criteria 
\begin{equation}\label{eq:hybrid085}%lamthres-mthres-q07-computed.py
  \tilde{\Lambda}_\mathrm{thres}^\mathrm{hybrid, q=0.85}= 558 (M_\mathrm{thres}/M_\odot)-1207
\end{equation}
for $q=0.85$ and
\begin{equation}\label{eq:hybrid07}%lamthres-mthres-q07-computed.py
  \tilde{\Lambda}_\mathrm{thres}^\mathrm{hybrid, q=0.7}= 1238 (M_\mathrm{thres}/M_\odot)-2901
\end{equation}
for $q=0.7$ as indicator for a phase transition in asymmetric mergers. As for the equal-mass case some hybrid EoSs occur below the black dashed line given by~\eqref{eq:hybrid085} or~\eqref{eq:hybrid07} meaning that a measurement in this region does not inform about the presence of a phase transition.

Again, we include lines of constant $M_\mathrm{max}$ in Fig.~\ref{fig:2q07} using fit 44 and fit 45 from Tab.~\ref{tab:mthr2} for purely hadronic EoSs (blue curves). As for the equal-mass case the solid blue curve marking $M_\mathrm{max}=1.97~M_\odot$ configurations is roughly identical to the black dashed line which is drawn by hand to separate the ``only hybrid'' regime and the mixed regime. We also note that the similarity between Fig.~\ref{fig:pt} and the left panel in Fig.~\ref{fig:2q07} means that one does not require a very accurate determination of the mass ratio since one can readily formulate a common criterion being valid for the range $0.85\leq q \leq 1$ (see fit~46). For comparison, the gray dashed line in the left panel in Fig.~\ref{fig:2q07} displays the black dashed line from Fig.~\ref{fig:pt}.

In Fig.~\ref{fig:2q07} we note that there occurs one excluded hadronic EoS in the ``only hybrid'' regime at about $\tilde{\Lambda}_\mathrm{thres}\approx 800$ (only visible in the right panel and outside the plotted range in the left panel). This model is the GNH3 EoS, which has a low maximum mass of $M_\mathrm{max}=1.96~M_\odot$. The EoS is very stiff at low densities and softens strongly at higher densities due to the appearance of hyperons. For asymmetric mergers the combined tidal deformability is dominated by the tidal deformability of the lighter binary component, which in the case of GNH3 becomes very large because of the large radii of low-mass NSs described by this EoS (see mass-radius relation of this EoS e.g. in the Supplemental Material of~\cite{Bauswein2020a}). Therefore the model is strongly shifted towards higher $\tilde{\Lambda}_\mathrm{thres}$. Since the EoS is too stiff at lower densities to be compatible with current astrophysical constraints, it does not spoil the unambiguousness of the signature of hybrid models in the ``hybrid'' regime.

Finally, we remark that it may be possible to obtain an independent estimate of $\tilde{\Lambda}_\mathrm{thres}$ for instance through radius or $\Lambda$ measurements at a different mass which can be interpolated or extrapolated to $M_\mathrm{thres}$. Alternatively, one may reformulate the criteria to employ these quantities, which we leave to future work.

For asymmetric binaries with $q=0.7$ the tidal deformability at the threshold mass $\tilde{\Lambda}_\mathrm{thres}$ varies in a relatively wide range between 200 and 650 or even 800 if one includes the models of the ``excluded sample''. The latter limit may be the relevant one for fully agnostic searches. This compares to range of about $200 \leq \tilde{\Lambda}_\mathrm{thres} \leq 450 $ for equal-mass binaries and mildly asymmetric systems with $q=0.85$ (cf.~\cite{Zappa2018,Agathos2020}). These ranges are important to classify the merger product based entirely on a measurement of the tidal deformability. For $\tilde{\Lambda}<200$ one can safely assume that a prompt collapse takes place, while only for $\tilde{\Lambda}>800$ or $\tilde{\Lambda}>650$, respectively, one can presume no direct BH formation.

\subsection{General EoS properties}

\begin{figure*}% lamthres-mthres-arrows-computed.py  RM-arrows-computed.py
\centering
\includegraphics[width=\columnwidth]{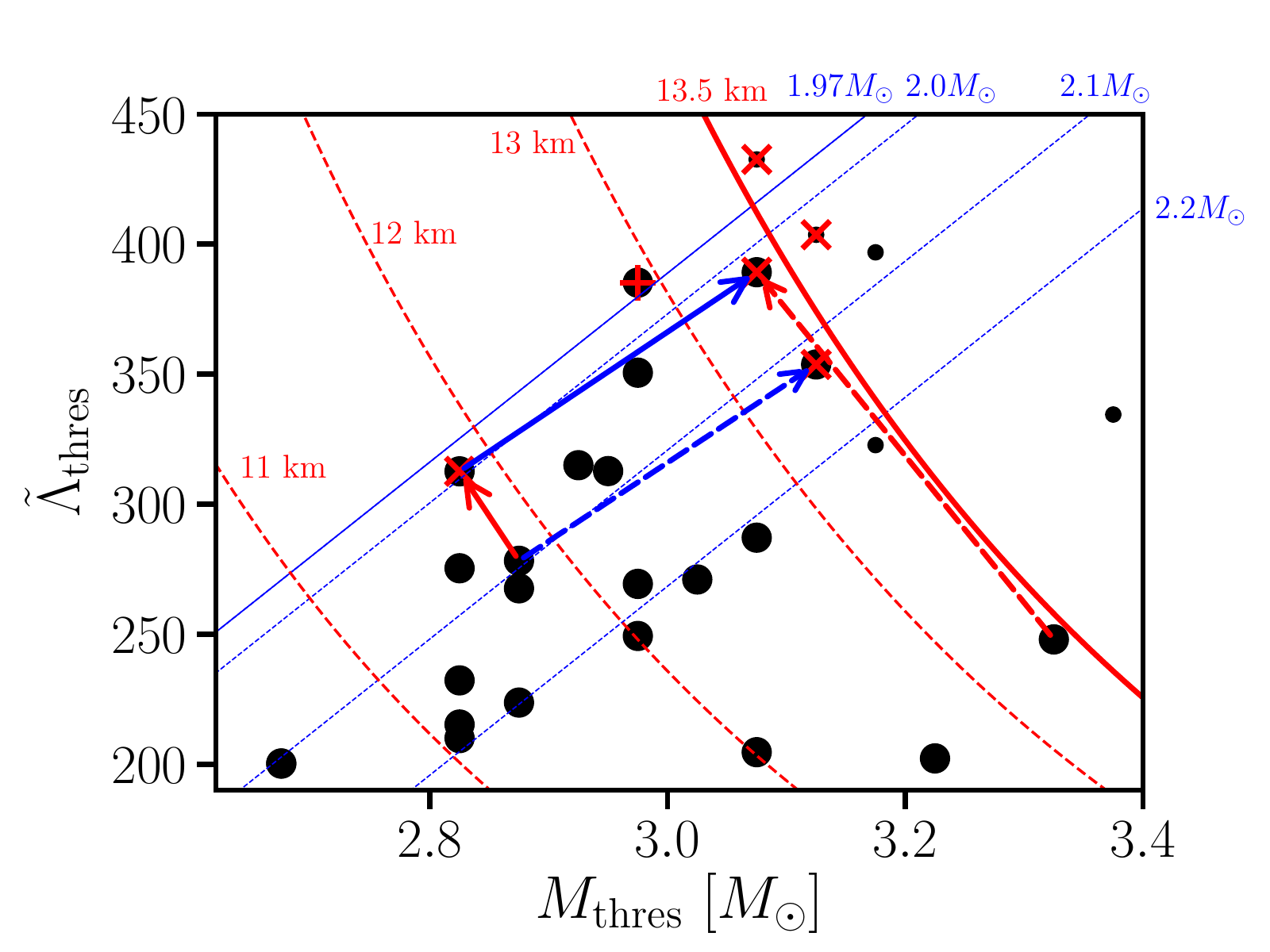}
\includegraphics[width=\columnwidth]{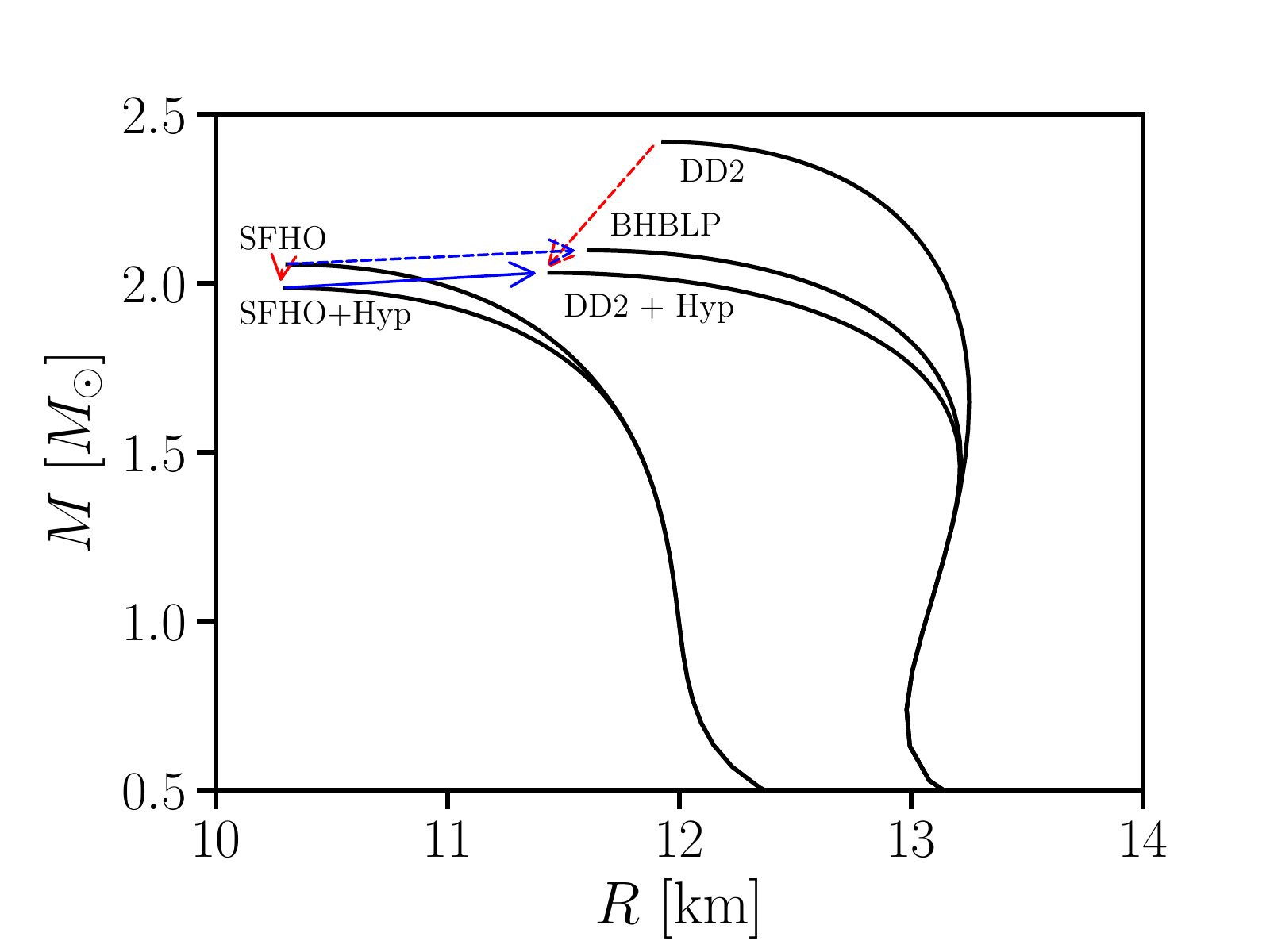}
\caption{Left panel: same as Fig.~\ref{fig:pt} with only purely hadronic models. Red arrows link EoS models with comparable
radii at $M_\mathrm{thres}/2$, but which
differ with respect to  the presence or absence of hyperons  above a certain central density and yield different $M_\mathrm{max}$. Blue arrows  link EoS models with comparable $M_\mathrm{max}$. See main text and caption of Fig.~\ref{fig:pt} for more explanations; see Tab.~I of Supplemental Material to~\cite{Bauswein2020a} for data of the individual EoS models. Right panel: mass-radius relations of nonrotating NSs for the EoS models connected with arrows in the left panel. } 
\label{fig:comp}
\end{figure*}

To understand the EoS effects in the $\tilde{\Lambda}_\mathrm{thres}-M_\mathrm{thres}$ plane, we highlight selected EoS models by connecting them with arrows in Fig.~\ref{fig:comp} (the figure in the left panel is essentially identical to Fig.~\ref{fig:pt}, but hybrid models are not shown for clarity). For comparison, the right panel displays the mass-radius relations of the same EoSs, where we connect the maximum-mass configurations with arrows following the same color scheme. 

Again, as in Fig.~\ref{fig:pt} we include lines of constant $M_\mathrm{max}$ in blue and we also draw sequences which indicate configurations of constant radius. As above we use the results of~\cite{De2018} and obtain $\tilde{\Lambda}_\mathrm{thres}(M_\mathrm{thres})=0.0093\left(\frac{G M_\mathrm{thres}}{2c^2
\bar{R}}\right)^{-6}$ with $\bar{R}$ being a chosen constant radius. Interestingly, the blue and red curves are roughly perpendicular to each other.

We observe two effects. (1) The red arrows connect a purely nucleonic EoS with the same model which includes hyperons. The hyperons occur beyond some transition density, which is why the M-R relations in the right panel are identical below some mass. The effect of the hyperons is a softening at higher densities, which leads to a reduction of the maximum mass in comparison to the nucleonic reference model, e.g.~\cite{Glendenning1982,Oertel2015,Chatterjee:2015pua}. In the left panel smaller $M_\mathrm{max}$ result in a shift towards the upper left, i.e. towards smaller $M_\mathrm{thres}$ and higher $\tilde{\Lambda}_\mathrm{thres}$. The red arrows are roughly perpendicular to the thin blue lines showing contours of constant $M_\mathrm{max}$ (fit 43 form Tab.~\ref{tab:mthr2}). This effect is somewhat comparable to the behavior of the hybrid models which also resulted in a shift towards the upper left compared to their hadronic reference model. The red arrows are roughly parallel to the curves of constant radius because both models, i.e. with and without hyperons, have the same or nearly the same radius in the mass range around $M_\mathrm{thres}/2$.

(2) The blue arrows visualize a comparison between EoSs which have roughly the same $M_\mathrm{max}$ but different radii. It is interesting to note that models towards the upper right in the $\tilde{\Lambda}_\mathrm{thres}-M_\mathrm{thres}$ diagram are those which are relatively stiff at lower densities, i.e. yield large NS radii, but strongly soften at higher densities which leads to a relatively low $M_\mathrm{max}$. 

These models accumulate towards the tip of the triangle spanned by the viable EoSs, i.e. close to the intersection between the solid red and solid blue curve. We conclude that models in this area are indicative of a strong softening at higher densities as it is typical for instance for the occurrence of hyperons. Hyperonic EoSs are marked with a red cross in the left panel of Fig.~\ref{fig:comp}. Hyperonic models like the SFHO+Hyp EoS do not exactly follow this behavior because already the purely nucleonic EoS is rather soft at lower densities. But one can clearly identify the impact of the additional EoS softening at higher densities resulting in a shift towards the upper left. The effect is somewhat less pronounced as it is apparent from the mass-radius relations in the right panel that SFHO and SFHO+Hyp are not too different. We generally conclude that the position in the $\tilde{\Lambda}_\mathrm{thres}-M_\mathrm{thres}$ plane is informative about details of the EoS.

Also, Fig.~\ref{fig:comp} illustrates that upper or lower limits on $\tilde{\Lambda}_\mathrm{thres}$, $M_\mathrm{thres}$ or $M_\mathrm{max}$ from a single measurement can already provide interesting EoS constraints. These can be quantified through the fit formulae provide in Tab.~\ref{tab:mthr2}.

\section{Univariate relations for $M_\mathrm{thres}$}\label{sec:add}

\begin{figure}% r18-mthres-computed.py
\centering
\includegraphics[width=\columnwidth]{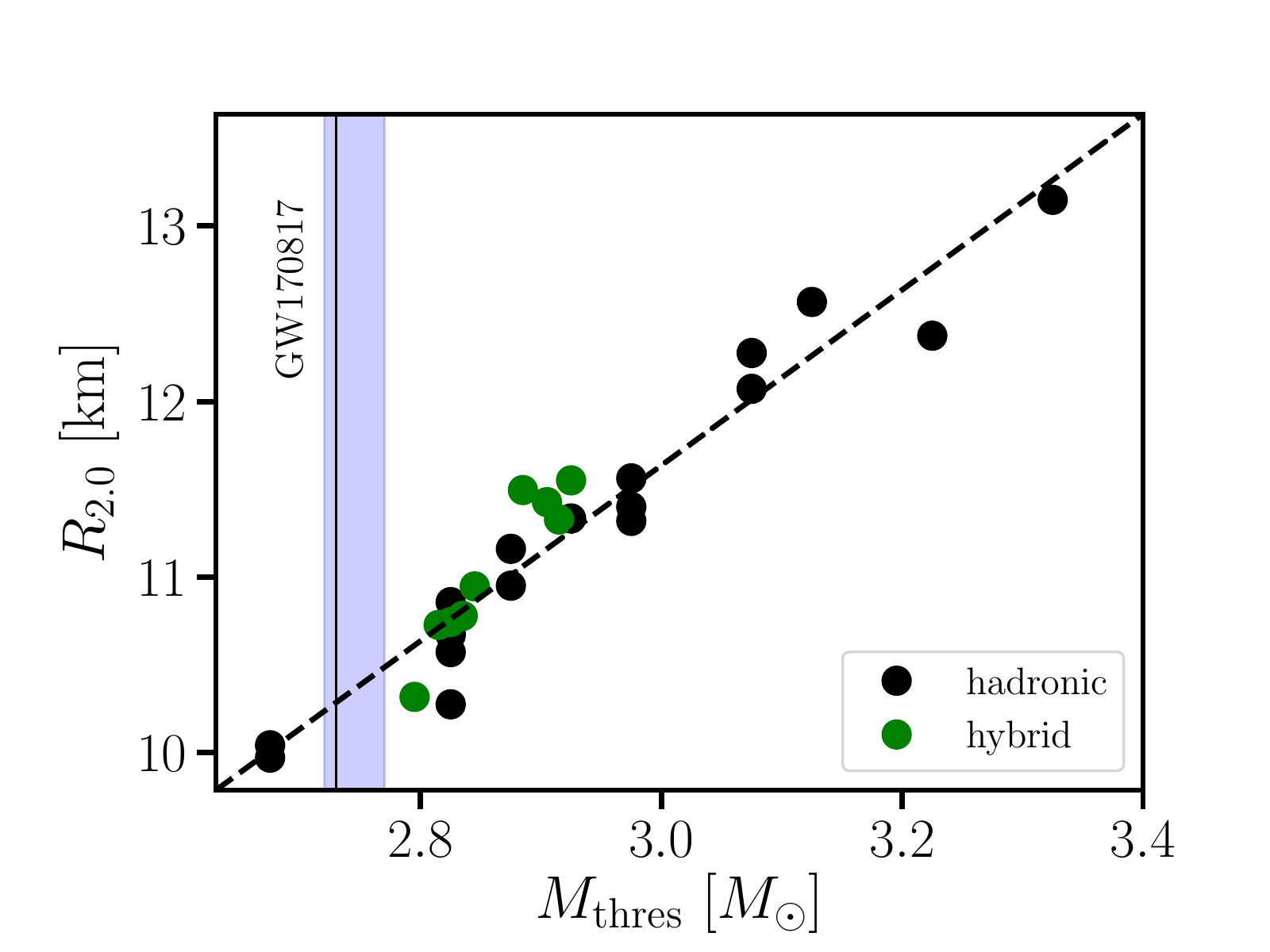}
\caption{Radius of nonrotating NSs with a mass of 2.0~$M_\odot$ as function of $M_\mathrm{thres}$ for equal-mass mergers. Black symbols refer to models of the hadronic base sample, green dots show hybrid EoSs. Dashed line is a linear fit to the data (see Tab.~\ref{tab:unirel}). The bluish band displays the total binary mass of GW170817~\cite{Abbott2019}.} 
\label{fig:mthrr18}
\end{figure}

\begin{figure}% lam18-mthres-computed.py, execfile("lam20-mthres-computed.py")
\centering
\includegraphics[width=\columnwidth]{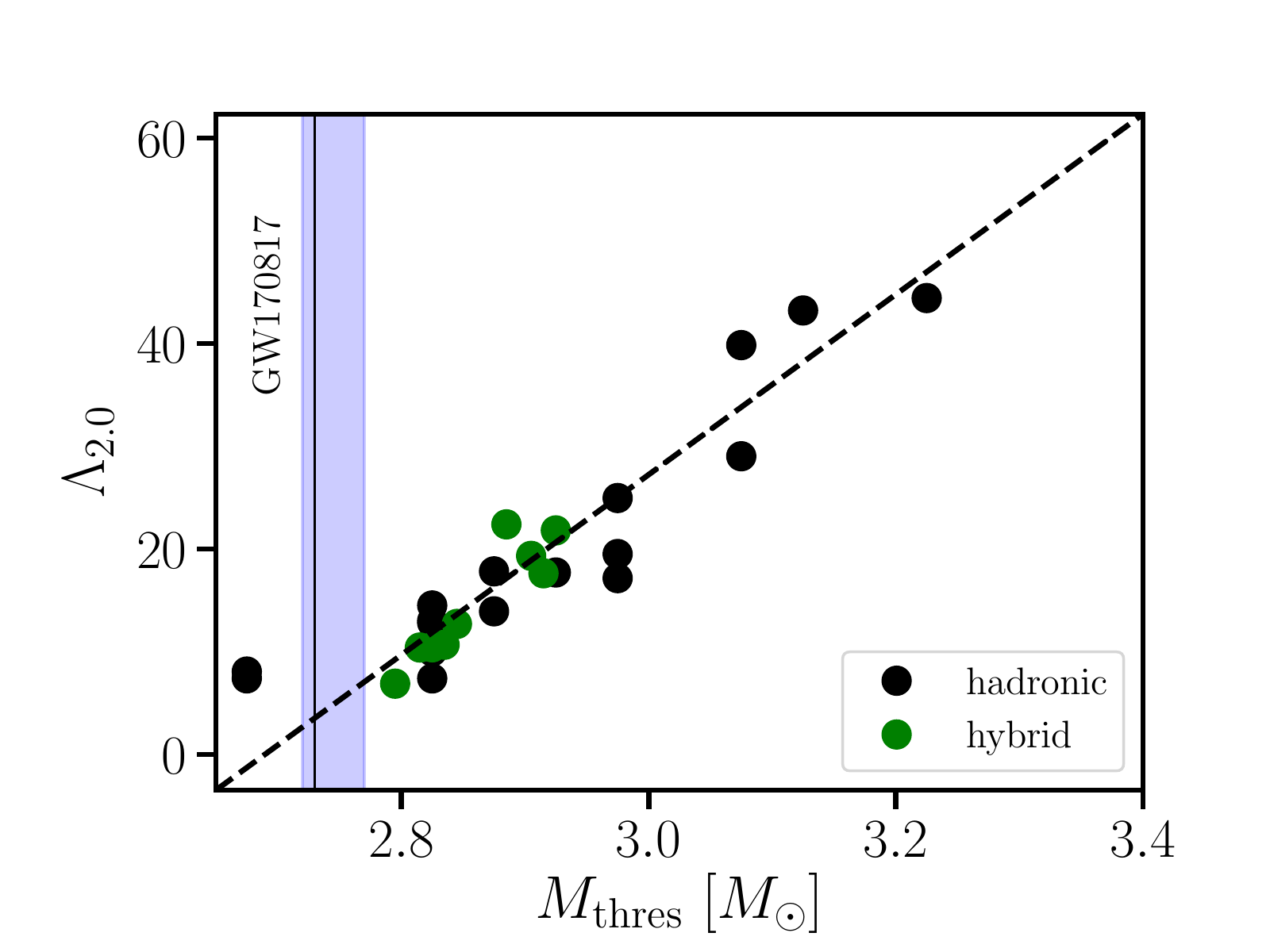}
\caption{Tidal deformability for NSs with a mass of 2.0~$M_\odot$ as function of $M_\mathrm{thres}$ for equal-mass mergers. Black symbols refer to models of the hadronic base sample, green dots show hybrid EoSs. Dashed line is a linear fit to the data (see Tab.~\ref{tab:unirel}). The bluish band displays the total binary mass of GW170817~\cite{Abbott2019}.} 
\label{fig:mthrlam18}
\end{figure}

Finally, we point out some additional relations. For instance, we find that univariate relations between the threshold mass and the radius or the tidal deformability of high-mass NSs are relatively tight, although not as tight as the bilinear relations described in Sect.~\ref{sec:mthr}. Figures~\ref{fig:mthrr18} and~\ref{fig:mthrlam18} show examples of such relations for equal-mass mergers, linking the threshold mass and stellar properties of NSs with 2.0~$M_\odot$. Dashed lines display the fits, which are given in Tab.~\ref{tab:unirel}. For these figures and the fits we employ the purely hadronic base sample and the hybrid EoSs. Apparently, the latter follow the same relation and do not increase the overall scatter. Note that in Fig.~\ref{fig:mthrlam18} the deviations from the fit become larger at higher $M_\mathrm{thres}$ implying that a $\Lambda$ determination at smaller $M_\mathrm{thres}$ will be in fact more accurate than indicated by the maximum residual of this fit.

Table~\ref{tab:unirel} provides also fits for other high-mass NS properties like $R_{1.8}$, $R_{1.9}$ or the radius $R_\mathrm{max}$ of the maximum-mass configuration\footnote{The relation with the radius of the maximum mass configuration $R_\mathrm{max}=R(M_\mathrm{max})$ is not as tight as those with $R_{1.8/1.9/2.0}$, and in particular some hybrid EoSs occur at relatively large $R_\mathrm{max}$. See also fit 27 in Tab.~\ref{tab:mthr1} for $R_\mathrm{max}$, which is significantly tighter because of the additional $M_\mathrm{max}$ dependence. Generally, univariate relations may not be completely unexpected considering the findings in~\cite{Bauswein2013,Bauswein2014a}. Combining the relations in Fig.~2 and Fig.~8 in~\cite{Bauswein2014a} suggests relations between $M_\mathrm{thres}$ and the TOV properties of high-mass NSs.}. The tightest relation is found for $R_{2.0}$ with an average deviation of only 149~m. In this relation we employ $R_\mathrm{max}$ instead of $R_{2.0}$ for EoS models with $M_\mathrm{max}<2.0~M_\odot$. Including the excluded hadronic sample would not strongly increase the scatter in the relations given in Tab.~\ref{tab:unirel}. Finally, we stress that relations for binary mass ratio $q=0.85$ and their respective tightness are almost identical to the ones given in Tab.~\ref{tab:unirel} for $q=1$. Similarly, we find that combining the results for $q=1$ and $q=0.85$ in a single relation leads to very similar fits, which describe the data nearly as well as those in Tab.~\ref{tab:unirel}. We thus do not list these fits explicitly but instead suggest to employ the given relations of equal-mass systems for the whole range $0.85 \leq q \leq 1$. However, for a binary mass ratio $q=0.7$ such relations are less tight than for equal-mass binaries.


These fits can be employed to directly constrain the radius or tidal deformability of a high-mass NS from a measurement of only $M_\mathrm{thres}$. Already, a lower or upper limit on $M_\mathrm{thres}$ may provide an interesting bound. Since radius or $\Lambda$ constraints through this type of relations rely only on a measurement of the binary masses, they may provide a useful sanity check independent of the extraction of finite-size effects from the inspiral or any other inference of NS properties\footnote{It is beyond the scope of this paper to discuss in detail the accuracy of future GW detections of high-mass binaries. But we note that adopting a factor 10 improvement of the parameter estimation in GW170817 would determine the tidal deformability to within about $\pm 30$, i.e. approximately the range in Fig.~\ref{fig:mthrlam18}, whereas $M_\mathrm{tot}$ and $q$ would be known with very good precision~\cite{Abbott2019} and would thus not contribute significantly to the error. Although this comparison may not be fully applicable to high-mass systems, we simply note that even assuming significant systematic uncertainties and a coarse determination of $M_\mathrm{thres}$ could already constrain high-mass NS properties. Clearly, extrapolations of the EoS inference at lower densities could be informative about high-mass properties as well, although this approach would rely on additional assumptions.}. A conversion of these relations to $\mathcal{M}_\mathrm{thres}$ is trivial for fixed mass ratios.

\begin{table*} 
\begin{tabular}{|l|c|c|c|c|c|}  \hline 
fit & $a$ & $b$ & max. dev. & av. dev. & $R >$ \\ \hline

%%%% base and hyb eos samples %%%%
%figure(); execfile("r19-mthres-computed.py")
%$R_{1.8}= a M_\mathrm{thres}+b$ & 4.395 & -1.108 & 0.624 & 0.235 \\ \hline
$R_{1.8}= a M_\mathrm{thres}+b$ & 4.395 $\pm$ 0.384 & -1.108 $\pm$ 1.125 & 0.624 & 0.235 & 10.22 \\ \hline

%$R_{1.9}= a M_\mathrm{thres}+b$ & 4.656 & -2.058 & 0.542 & 0.170 \\ \hline
$R_{1.9}= a M_\mathrm{thres}+b$ & 4.656 $\pm$ 0.291 & -2.058 $\pm$ 0.852 & 0.542 & 0.170 & 10.06 \\ \hline

%$R_{2.0}= a M_\mathrm{thres}+b$ & 4.998 & -3.359 & 0.486 & 0.149 \\ \hline
$R_{2.0}= a M_\mathrm{thres}+b$ & 4.998 $\pm$ 0.268 & -3.359 $\pm$ 0.783 & 0.486 & 0.149 & 9.75 \\ \hline
%$R_{2.1}= a M_\mathrm{thres}+b$ & 5.130 & -4.087 & 0.782 & 0.217 \\ \hline %21

%$R_\mathrm{max}= a M_\mathrm{thres}+b$ & 3.818 & -0.623 & 1.104 & 0.280 \\ \hline %max
$R_\mathrm{max}= a M_\mathrm{thres}+b$ & 3.818 $\pm$ 0.488 & -0.623 $\pm$ 1.427 & 1.104 & 0.280 & 8.66 \\ \hline

%execfile("lam20-mthres-computed.py") etc

$\Lambda_{1.8}= a M_\mathrm{thres}+b$ & 192.278 $\pm$ 13.867 & -500.176 $\pm$ 40.583 & 26.057 & 7.168 & -- \\ \hline
$\Lambda_{1.9}= a M_\mathrm{thres}+b$ & 132.102 $\pm$ 7.660 & -349.164 $\pm$ 22.418 & 13.957 & 4.213 & -- \\ \hline
%$\Lambda_{2.0}= a M_\mathrm{thres}+b$ & 87.720 & -235.931 & 11.963 & 3.387 \\ \hline%20
$\Lambda_{2.0}= a M_\mathrm{thres}+b$ & 87.720 $\pm$ 6.065 & -235.931 $\pm$ 17.749 & 11.963 & 3.387 & -- \\ \hline

\end{tabular} 
\caption{Univariate relations between the threshold mass $M_\mathrm{thres}$ and high-mass NS properties described by different fit formulae (first column) for binary mass ratio $q=1$. $a$ and $b$ are fit parameters with their respective variances in units such that masses are in solar mass and radii in km. The next two columns list the maximum and average deviation between fit and the underlying data (in km for functions involving radii; dimensionless for relations with the tidal deformability). These fits are obtained for equal-mass mergers employing the base sample and the hybrid sample. The last column provides the limit on the high-mass NS property if GW170817 was not a prompt collapse (see text). We do not list limits on the tidal deformability because taking into account the maximum scatter of the relations, the limit was negative. However, the figure also shows that the deviations from the fit are smaller at smaller threshold masses, which is why the relation actually does imply a certain bound. Very similar relations also with regard to their tightness are obtained for $q=0.85$ or for a data set including results for $q=1$ and $q=0.85$.}
\label{tab:unirel} 
\end{table*} 
 %label tab:unirel

We directly apply these findings to derive new constraints from the observation of GW170817~\cite{Abbott2019} assuming that the merger remnant did not directly collapse, i.e. that the measured total binary mass $M_\mathrm{tot}^\mathrm{GW170817}=2.73_{-0.01}^{+0.04}~M_\odot$ is smaller than $M_\mathrm{thres}$. Although the mass ratio was not accurately determined for GW170817 with $0.73\leq q \leq 1$, we can assume that $M_\mathrm{thres}(q=1)>M_\mathrm{tot}^\mathrm{GW170817}$ because for $q<1$ the threshold mass is either nearly equal or smaller than that of equal-mass mergers for basically all EoS models\footnote{Note that this argument might be incorrect for very small radii and large $M_\mathrm{max}$, see Fig.~\ref{fig:DMthr}. In any case we find at most a very minor deviation for other mass ratios.} (see Sect.~\ref{sec:q}). The vertical line in Fig.~\ref{fig:mthrr18} indicates the total mass of GW170817 and shows that the radius $R_{2.0}$ of a NS with 2.0~$M_\odot$ should be larger than 9.75~km. The corresponding limits derived from the other relations are given in Tab.~\ref{tab:unirel}. We adopt the lower bound of the error estimate of $M_\mathrm{tot}^\mathrm{GW170817}$ (95\% lower bound from~\cite{Abbott2019} with low-spin prior) and also subtract the maximum residual of the fit. These new EoS constraints are not very strong\footnote{Two EoS models on the left of the bluish band with $R_{2.0}\approx 10$~km are excluded. Other models with the same radius could still be marginally compatible if they deviated stronger from the dashed curve.}, but obviously the limits can be updated by any new merger observation with information on the merger outcome. The application of these relations is trivial. In particular, a prompt collapse event would establish upper bounds on the stellar parameters of high-mass NSs.

Note that these constraints differ from the one described in Sect.~\ref{sec:mthr} and in~\cite{Bauswein2017}, where we employ an additional relation between $R_{1.6}$ and $M_\mathrm{max}$, which had been constructed in a very conservative way (see~\cite{Bauswein2019c}). Here, instead we do not rely on any information about $M_\mathrm{max}$, which is the reason why the limits here are less constraining.

The univariate relations are not as tight as the bilinear fits described in Sect.~\ref{sec:mthr}, but apparently the EoS dependence of the collapse behavior is generally well captured. Stellar properties of high-mass NSs may represent natural choices to characterize the EoS in the density regime which determines the collapse. Measuring or constraining the threshold mass is thus important to assess the high-density regime noting that tidal effects during the inspiral become weaker for high-mass binaries and the inference of the EoS at higher densities from observations of binaries with lower masses relies on some kind of extrapolation or interpolation procedure using information for instance from pulsar measurements. The threshold mass instead is a very direct and thus robust messenger of the high-density EoS albeit it clearly requires more work to understand the detailed signatures indicating the outcome of a merger and affecting finally the precision of future $M_\mathrm{thres}$ constraints.  We suspect that the next observing runs will improve the statistics of the binary population properties, which will allow a better assessment of the mid-term and long-term prospects to estimate $M_\mathrm{thres}$.

\section{Summary and conclusions}\label{sec:sum}

In this study we describe the dependencies of the threshold binary mass $M_\mathrm{thres}$ for prompt BH formation in NS mergers. We consider a total of 40 different high-density EoSs including models with a phase transition to deconfined quark matter. We describe the EoS dependence of the threshold to prompt BH formation by considering relations between $M_\mathrm{thres}$ and stellar properties of nonrotating NSs. We determine $M_\mathrm{thres}$ for different binary mass ratios focusing on configurations with $q=1$, $q=0.85$ and $q=0.7$. By this we can assess the influence of $q$ in a range that may be typical for many current and future NS merger observations~\cite{Abbott2017,Abbott2020}. Our main findings are the following:

\begin{itemize}

\item We present a number of fit formulae describing $M_\mathrm{thres}$ as function of $M_\mathrm{max}$ and another stellar parameter, which can be either a NS radius or the tidal deformability. These relations are particularly tight (with deviations of only a few $0.01~M_\odot$) if we consider  $M_\mathrm{thres}$ for a fixed binary mass ratio. The tightness of the relations also depends on the set of EoSs, where one may make different assumptions on which models are considered viable. For instance, one may include models with a phase transitions to quark matter or EoSs which are incompatible with a certain confidence level of the tidal deformability inference from GW170817. Not unexpectedly, enlarged EoS samples lead to relations with a somewhat larger scatter. We derive similar relations for the chirp mass at the threshold to prompt collapse, which may be important in particular when the binary mass ratio is not well measured. We note that all these relations are bilinear and hence they are simple to invert. This is useful for instance if one employs an observation to obtain a bound on $M_\mathrm{thres}$ from which one can then derive EoS constraints, e.g. for NS radii or for $M_\mathrm{max}$. 

\item We explicitly show which limits on NS radii are implied by a merger event of a given total binary mass following the idea in~\cite{Bauswein2017}. This constraint can be immediately employed and establishes a lower limit on the radius or the tidal deformability if the merger remnant did not directly collapse. An upper limit is obtained from the measured total binary mass if a prompt collapse took place.

\item This study quantifies for the first time systematically the influence of the binary mass ratio on $M_\mathrm{thres}$. For most EoS models we find that $M_\mathrm{thres}$ decreases with the binary mass asymmetry, i.e. with decreasing $q$. We observe that the decrease of $M_\mathrm{thres}$ becomes stronger for more asymmetric binaries, whereas $M_\mathrm{thres}$ remains roughly constant in the range $0.85 \leq q \leq 1$. This means that $M_\mathrm{thres}(q)$ can be well described by higher-order polynomials. In the range $0.85 \leq q \leq 1$ the corresponding threshold mass of equal-mass binaries is a good approximation to $M_\mathrm{thres}(q)$ {with $M_\mathrm{thres}(q=0.85)$ being at most 0.1~$M_\odot$ smaller or 0.05~$M_\odot$ higher compared to $M_\mathrm{thres}(q=1)$ for a few particularly soft EoSs}. Importantly, the magnitude of the reduction of $M_\mathrm{thres}$ with decreasing $q$ depends in a systematic way on the EoS. Specifically, the difference $M_\mathrm{thres}(q=1)-M_\mathrm{thres}(q=0.7)$ can be well approximated as function of stellar parameters of nonrotating NSs. For EoSs which are stiff at lower and intermediate densities  and result in large NS radii, the impact of $q$ on the collapse behavior is more pronounced (strong reduction of $M_\mathrm{thres}(q=0.7)$ w.r.t. the equal-mass case $M_\mathrm{thres}(q=1)$). Soft EoSs yield a threshold mass which is approximately constant, or shows only a weak influence of $q$. 

\item Combining these findings we derive fit formulae for $M_\mathrm{thres}(q)$ and $\mathcal{M}_\mathrm{thres}(q)$ quantifying their EoS dependence in a range of $q$. These relations include an explicit $q$ dependence and are approximately as tight as fits for fixed mass ratio. The generalized formulae are valid for $0.7 \leq q \leq 1$, but likely yield a fair description even for mass ratios somewhat below $q<0.7$. Future work should specify the exact dependence on $q$, which follows a higher-order polynomial. The exponent may be EoS dependent on its own, but we find that a power of three provides a relatively accurate description.

\item As the $M_\mathrm{thres}$ data follows some systematic behavior, we sketch a toy model within Newtonian physics to explain at least qualitatively the impact of the mass ratio on the collapse behavior. Generally, one may understand a reduction of the threshold mass with the binary asymmetry as a consequence of asymmetric binaries of the same total mass having less angular momentum than the equal-mass binary at the same orbital distance.

\item Considering the $M_\mathrm{thres}$ data we also explore the impact of phase transitions to quark matter on the collapse behavior following our previous investigation in~\cite{Bauswein2020a}. We focus on the parameters $M_\mathrm{thres}$ and $\tilde{\Lambda}_\mathrm{thres}$. The latter is the combined tidal deformability of the binary system at the threshold mass. Both quantities can be in principle measured to some precision with a number of detections which reveal the outcome of the merger. If $\tilde{\Lambda}_\mathrm{thres}$ is relatively high compared to $M_\mathrm{thres}$, such measurements can only be explained by the presence of a strong phase transition because no purely hadronic EoS can yield such a combination of $M_\mathrm{thres}$ and $\tilde{\Lambda}_\mathrm{thres}$. This also implies that in principle already a single measurement with a constraint on $M_\mathrm{thres}$ and $\tilde{\Lambda}_\mathrm{thres}$ can provide evidence for the occurrence of the hadron-quark phase transition in NS mergers.

The impact of a phase transition on these quantities is understandable. The tidal deformability $\tilde{\Lambda}_\mathrm{thres}$ is determined by the EoS at moderate densities, i.e. by the purely hadronic regime of the EoS, and is not informative about the presence of a phase transition at higher densities. If a phase transition takes place at higher densities, the EoS effectively softens and thus yields a relatively low $M_\mathrm{thres}$. No purely hadronic EoS can introduce such a strong softening at higher densities to lead to a comparable reduction of $M_\mathrm{thres}$.

Two remarks are important. Certain combinations of $M_\mathrm{thres}$ and $\tilde{\Lambda}_\mathrm{thres}$ can point to a phase transition. However, the presence of quark matter does not necessarily lead to a combination of $M_\mathrm{thres}$ and $\tilde{\Lambda}_\mathrm{thres}$ which would be obviously indicative of a phase transition but in fact would be compatible with a purely hadronic EoS. These are hybrid models where the phase transition is relatively weak in the sense that the resulting stellar parameters (e.g. the mass-radius relation) are roughly comparable to those of purely hadronic EoSs. In an extreme case, a phase transition could even have a large latent heat, but quark matter could become very repulsive at higher densities and lead to a stabilization of the merger remnant and thus relatively large $M_\mathrm{thres}$. Note that we consider combinations of $M_\mathrm{thres}$ and $\tilde{\Lambda}_\mathrm{thres}$ for fixed mass ratio. We show that also for $q=0.85$ and $q=0.7$ the hadron-quark phase transition can lead to a characteristic imprint on these quantities as described. For a practical application one will either need to consider systems of approximately fixed $q$ or events within a small range of $q$, e.g. $0.85 \leq q \leq 1$ as in this study. Alternatively, one may find an appropriate way to combine measurements with different mass ratio, which we leave for future work. Also the specific requirements for GW data analysis should be investigated in more details. We emphasize that already a relatively coarse determination of $\tilde{\Lambda}_\mathrm{thres}$ may be sufficient to deduce the presence of a phase transition, which we consider one of the advantages of this signature.

Following these considerations for phase transitions as the most extreme example of a sudden change of EoS properties at higher densities, we generalize these findings to hadronic EoSs. We find that the combination of $M_\mathrm{thres}$ and $\tilde{\Lambda}_\mathrm{thres}$ informs about the properties of different density regimes. For instance, a relatively high $\tilde{\Lambda}_\mathrm{thres}$ and a low threshold mass generally indicate the softening of the EoS at higher densities, which may be caused by the occurrence of hyperons. 

\item Generally, our data show that only for a combined tidal deformability $\tilde{\Lambda}\gtrsim650$ no direct collapse of the merger remnant can be safely assumed if $0.7 \leq q\leq 1$. The limit may even increase to $\tilde{\Lambda}\gtrsim800$ if one takes into account EoSs of the ``excluded hadronic sample'', i.e. models with a tidal deformability incompatible with that of GW170817. For the range $0.85 \leq q \leq 1$ the bound safely excluding a prompt collapse is 450. Only for $\tilde{\Lambda}\lesssim 200$ one can safely expect that direct BH formation took place. 

\item We finally point out univariate relations between the threshold mass and stellar parameters of high-mass NSs like the radius or tidal deformability of stars with about 1.8 to 2.0~$M_\odot$. The relations are relatively tight with deviations of only a few hundred meters (but not as tight as bilinear fits describing $M_\mathrm{thres}$). Interestingly, the EoS models with phase transitions follow the same behavior as purely hadronic EoSs. Considering the tightness of these relations and their universality, we conclude that the radii of high-mass NSs represent an EoS property which very well characterizes and determines the collapse behavior of NS mergers. This is not too surprising considering the density regime in high-mass NSs and in mergers close to the threshold to prompt BH formation. We employ these relations for new EoS constraints limiting the radii of high-mass NSs. We point out that new merger observations have the potential to significantly improve these constraints, which may be employed for sanity and consistency checks within other EoS inference methods.

\end{itemize}

Generally, our findings highlight the importance to determine the threshold mass for prompt BH formation in NS mergers. This includes observational efforts to obtain $M_\mathrm{thres}$ through the observation of electromagnetic counterparts, e.g. kilonovae or gamma-ray bursts, or the detection/exclusion of postmerger GW emission. Future work should thus further improve the theoretical understanding of these processes to enable an unambiguous interpretation of observational data. It also requires appropriate strategies and instruments for electromagnetic follow-up observations. Finally, employing postmerger GWs to determine the collapse behavior relies on specific GW data analysis methods to identify or exclude the presence of a NS remnant, e.g.~\cite{Clark2014,Chatziioannou2017,Agathos2020,Haster2020}. The prospect to measure $M_\mathrm{thres}$ and to understand by this the properties of the high-density EoS including the presence of a phase transition, stresses the importance of dedicated GW instruments with increased sensitivity in the kHz range~\cite{Punturo2010,Hild2011,Miller2015,Martynov2019,nemo2020}.

\acknowledgements{Acknowledgements: We thank N.-U. F. Bastian, D. Blaschke, M. Cierniak, C. Constantinou, T. B. Fischer, G. Martinez-Pinedo, M. Oertel, M. Prakash, A. Schneider, J. Smith for providing EoS tables and helpful discussions. A.B. and G.L. acknowledge support by the European Research Council (ERC) under the European Union's Horizon 2020 research and innovation programme under grant agreement No. 759253. A.B. and S.B. acknowlege support by Deutsche Forschungsgemeinschaft (DFG, German Research Foundation) - Project-ID 279384907 - SFB 1245. A.B. and V.V. acknowledge support by DFG - Project-ID 138713538 - SFB 881 (``The Milky Way System'', subproject A10). The work of T.S. is supported by the Klaus Tschira Foundation. T.S. is Fellow of the International Max Planck Research School for Astronomy and Cosmic Physics at the University of Heidelberg (IMPRS-HD) and acknowledges financial support from IMPRS-HD. N.S. is supported by the ARIS facility of GRNET in Athens (SIMGRAV, SIMDIFF\ and BNSMERGE allocations) and the ``Aristoteles Cluster'' at AUTh, as well as by the COST actions CA16214 ``PHAROS'', CA16104 ``GWVerse'', CA17137 ``G2Net'' and CA18108 ``QG-MM''.}

%\bibliography{references}

\appendix
\section{Simulation results and EoS properties}\label{app:data}
In Tab.~\ref{tab:data} we list the data which are employed in this study. We provide information on the different EoS models and the results from the simulations for different mass ratios $q$. The simulation data for $q=1$ and $q=0.7$ are identical to Tab.~I of the Supplemental Material in~\cite{Bauswein2020a}.

\begin{figure*}% r18-mthres-computed.py
\centering
\includegraphics[width=\columnwidth,trim=40 40 40 60,clip]{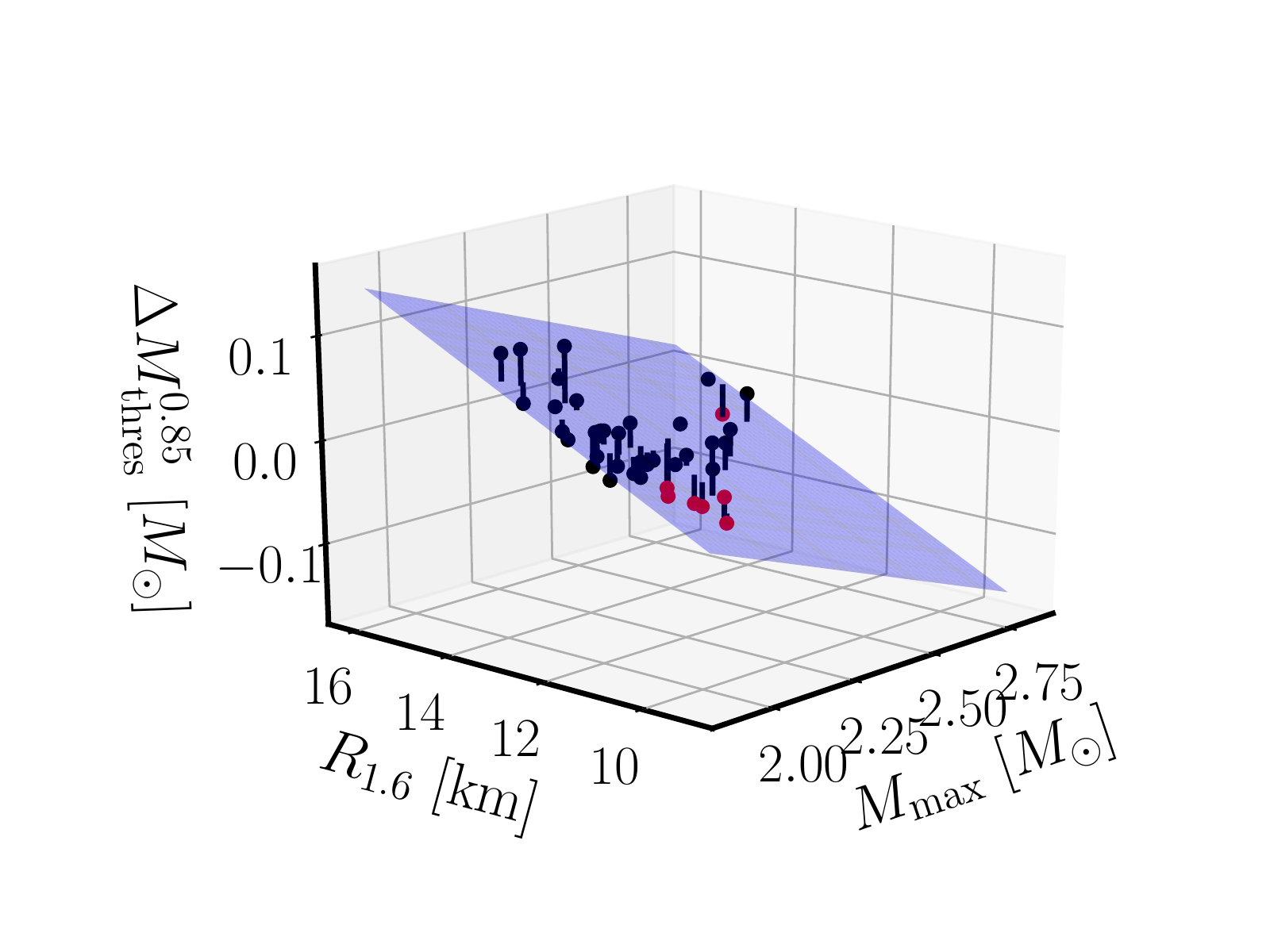}%execfile("Dmthres-r16-mmax-q08-basehybex-computed.py")
\includegraphics[width=\columnwidth,trim=40 40 40 60,clip]{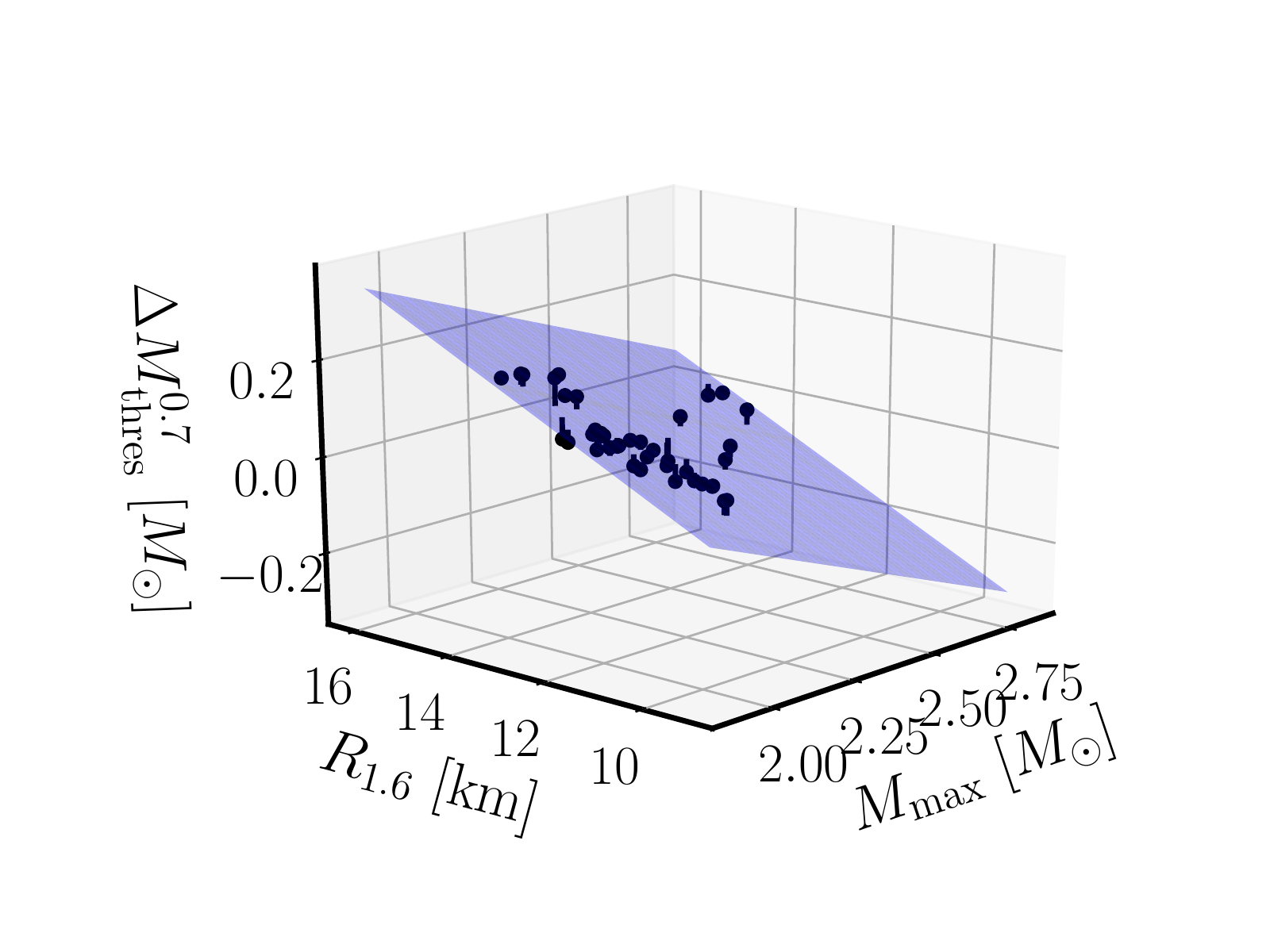}
\caption{ Difference $\Delta M_\mathrm{thres}$ between the threshold mass of equal-mass mergers and the threshold mass of asymmetric binaries with mass ratio $q=0.85$ (left panel) or $q=0.7$ (right panel) as function of $M_\mathrm{max}$ and $R_{1.6}$ (black dots). Results for the complete EoS sample are shown. The blue planes display bilinear fits as in Sect.~\ref{sec:q} with parameters $a=-0.136$, $b=2.495\times10^{-2}$ and $c=0.001$ (left panel) and $a=-0.304$, $b=5.198\times10^{-2}$ and $c=0.083$ (right panel, see Tab.~\ref{tab:Dmthr}). Deviations between the fit and the underlying data are illustrated by black lines: maximum and average deviations are 0.053~$M_\odot$ and 0.018~$M_\odot$ for $q=0.85$ in the left panel and 0.051~$M_\odot$ and 0.016~$M_\odot$ for $q=0.7$ in the right panel. The red dots show cases where $\Delta M_\mathrm{thres}$ is negative, i.e. the threshold mass of the asymmetric merger is slightly higher than that of the corresponding equal-mass case.}
\label{fig:Dmthr08507}
\end{figure*}

To assess the impact of the mass ratio $q$, we show in addition in Fig.~\ref{fig:Dmthr08507} the difference $\Delta M_\mathrm{thres}$ between the threshold mass of an equal-mass merger and an asymmetric binary (cf. Fig.~\ref{fig:DMthr}). As discussed in great detail in Sect.~\ref{sec:q}, $\Delta M_\mathrm{thres}$ clearly follows an EoS dependence. Here we show the results for the full EoS sample. The figure displays the difference between $M_\mathrm{thres}(q=1)$ and $M_\mathrm{thres}(q=0.85)$ in the left panel and the difference between $M_\mathrm{thres}(q=1)$ and $M_\mathrm{thres}(q=0.7)$ in the right panel. Both show qualitatively the same behavior.

\begin{table*} 
\begin{tabular}{|l|l|c|c|c|c|c|c|c|c|c|c|c|c|}\hline

EoS & T/B & $M_\mathrm{max}$ & $R_{1.6}$ & $\Lambda_{1.4}$ & 
$M_\mathrm{thres}$ & 
$\tilde{\Lambda}_\mathrm{thres}$ &
$M_\mathrm{thres}$ &
$\tilde{\Lambda}_\mathrm{thres}$ &
$M_\mathrm{thres}$ &
$\tilde{\Lambda}_\mathrm{thres}$ &
sample &
Ref. \\

& &  &  & & $q=1$ & $q=1$ & $q=0.85$ & $q=0.85$ & $q=0.7$ & $q=0.7$ & & \\ 

& & $(M_\odot)$ & $(\mathrm{km})$ & & $(M_\odot)$ & & $(M_\odot)$ & & $(M_\odot)$ & & & \\ \hline

% take the following from   execfile("EoS-tablew085.py")

% sp
BHBLP  &  T  & 2.100 & 13.203 & 696.1 & 3.125 & 356.4 & 3.075 & 400.9 & 2.975 & 516.8 &  b & \cite{Banik2014}  \\ \hline
DD2Y  &  T  & 2.033 & 13.182 & 695.7 & 3.075 & 392.2 & 3.025 & 440.2 & 2.875 & 626.2 &  b & \cite{Fortin2018,Marques2017}  \\ \hline
DD2  &  T  & 2.421 & 13.258 & 699.5 & 3.325 & 249.7 & 3.325 & 255.3 & 3.275 & 302.5 &  b & \cite{Hempel2010,Typel2010}  \\ \hline
DD2F  &  T  & 2.079 & 12.232 & 426.3 & 2.925 & 317.5 & 2.925 & 327.4 & 2.850 & 430.9 &  b & \cite{Typel2005,Typel2010,Alvarez-Castillo2016}  \\ \hline
APR  &  B  & 2.189 & 11.263 & 247.5 & 2.825 & 233.9 & 2.862 & 220.4 & 2.825 & 262.1 &  b & \cite{Akmal1998}  \\ \hline
BSK20  &  B  & 2.167 & 11.658 & 319.6 & 2.875 & 269.5 & 2.875 & 276.5 & 2.875 & 302.5 &  b & \cite{Goriely2010}  \\ \hline
eosUU  &  B  & 2.191 & 11.066 & 229.5 & 2.825 & 216.8 & 2.825 & 222.3 & 2.825 & 242.8 &  b & \cite{Wiringa1988}  \\ \hline
LS220  &  T  & 2.043 & 12.491 & 541.1 & 2.975 & 353.5 & 2.975 & 366.7 & 2.875 & 523.2 &  b & \cite{Lattimer1991}  \\ \hline
LS375  &  T  & 2.711 & 13.776 & 957.4 & 3.575 & 225.1 & 3.575 & 230.4 & 3.575 & 250.2 &  e & \cite{Lattimer1991}  \\ \hline
GS2  &  T  & 2.091 & 13.381 & 722.3 & 3.175 & 325.3 & 3.075 & 410.8 & 3.025 & 491.0 &  e & \cite{Shen2011}  \\ \hline
NL3  &  T  & 2.789 & 14.807 & 1369.4 & 3.775 & 230.2 & 3.812 & 221.7 & 3.775 & 259.7 &  e & \cite{Hempel2010,Lalazissis1997a}  \\ \hline
Sly4  &  B  & 2.045 & 11.533 & 294.6 & 2.825 & 277.4 & 2.825 & 285.8 & 2.775 & 355.5 &  b & \cite{Douchin2001}  \\ \hline
SFHO  &  T  & 2.058 & 11.761 & 333.9 & 2.875 & 280.3 & 2.875 & 288.2 & 2.825 & 355.5 &  b & \cite{Steiner2013}  \\ \hline
SFHOY  &  T  & 1.988 & 11.758 & 333.9 & 2.825 & 315.0 & 2.825 & 323.4 & 2.725 & 444.7 &  b & \cite{Fortin2018,Marques2017}  \\ \hline
SFHX  &  T  & 2.129 & 11.972 & 395.8 & 2.975 & 271.3 & 2.975 & 277.0 & 2.925 & 330.7 &  b & \cite{Steiner2013}  \\ \hline
TM1  &  T  & 2.212 & 14.361 & 1150.8 & 3.375 & 337.1 & 3.325 & 386.9 & 3.225 & 529.1 &  e & \cite{Sugahara1994a,Hempel2012}  \\ \hline
TMA  &  T  & 2.010 & 13.673 & 935.3 & 3.175 & 400.3 & 3.125 & 461.7 & 2.975 & 703.6 &  e & \cite{Toki1995,Hempel2012}  \\ \hline
BSK21  &  B  & 2.278 & 12.552 & 515.0 & 3.075 & 289.2 & 3.125 & 266.8 & 3.075 & 319.9 &  b & \cite{Goriely2010}  \\ \hline
GS1  &  T  & 2.753 & 14.877 & 1402.3 & 3.775 & 231.2 & 3.775 & 237.7 & 3.775 & 262.3 &  e & \cite{Shen2011}  \\ \hline
eosAU  &  B  & 2.127 & 10.365 & 150.9 & 2.675 & 201.6 & 2.712 & 189.2 & 2.675 & 223.7 &  b & \cite{Wiringa1988}  \\ \hline
WFF1  &  B  & 2.120 & 10.370 & 151.0 & 2.675 & 201.6 & 2.688 & 199.9 & 2.675 & 221.5 &  b & \cite{Wiringa1988,Read2009a}  \\ \hline
WFF2  &  B  & 2.188 & 11.057 & 223.9 & 2.825 & 211.5 & 2.800 & 229.5 & 2.825 & 236.9 &  b & \cite{Wiringa1988,Read2009a}  \\ \hline
MPA1  &  B  & 2.456 & 12.458 & 479.1 & 3.225 & 203.7 & 3.225 & 208.4 & 3.225 & 226.1 &  b & \cite{Muther1987,Read2009a}  \\ \hline
ALF2  &  B  & 1.975 & 12.628 & 569.4 & 2.975 & 388.2 & 2.938 & 428.0 & 2.875 & 513.9 &  b & \cite{Alford2005,Read2009a}  \\ \hline
H4  &  B  & 2.012 & 13.731 & 853.2 & 3.125 & 407.1 & 3.025 & 524.2 & 2.925 & 704.9 &  e & \cite{Lackey2006,Read2009a}  \\ \hline
DD2F-SF-1  &  T  & 2.136 & 12.158 & 426.3 & 2.845 & 383.3 & 2.845 & 394.4 & 2.770 & 502.0 &  h & \cite{Kaltenborn2017,Bastian2018,Fischer2018,Bauswein2019,Bastian2020}  \\ \hline
DD2F-SF-2  &  T  & 2.162 & 12.071 & 426.2 & 2.925 & 300.9 & 2.925 & 318.4 & 2.870 & 402.2 &  h & \cite{Kaltenborn2017,Bastian2018,Fischer2018,Bauswein2019,Bastian2020}  \\ \hline
DD2F-SF-3  &  T  & 2.034 & 12.205 & 426.3 & 2.825 & 401.9 & 2.788 & 451.1 & 2.720 & 574.9 &  h & \cite{Kaltenborn2017,Bastian2018,Fischer2018,Bauswein2019,Bastian2020}  \\ \hline
DD2F-SF-4  &  T  & 2.031 & 12.232 & 426.3 & 2.835 & 392.5 & 2.797 & 440.6 & 2.725 & 571.2 &  h & \cite{Kaltenborn2017,Bastian2018,Fischer2018,Bauswein2019,Bastian2020}  \\ \hline
DD2F-SF-5  &  T  & 2.040 & 11.943 & 426.3 & 2.815 & 411.5 & 2.777 & 456.8 & 2.725 & 543.6 &  h & \cite{Kaltenborn2017,Bastian2018,Fischer2018,Bauswein2019,Bastian2020}  \\ \hline
DD2F-SF-6  &  T  & 2.013 & 12.231 & 426.3 & 2.795 & 431.4 & 2.757 & 483.9 & 2.675 & 640.2 &  h & \cite{Kaltenborn2017,Bastian2018,Fischer2018,Bauswein2019,Bastian2020}  \\ \hline
DD2F-SF-7  &  T  & 2.117 & 12.232 & 426.3 & 2.905 & 332.8 & 2.868 & 374.2 & 2.825 & 454.6 &  h & \cite{Kaltenborn2017,Bastian2018,Fischer2018,Bauswein2019,Bastian2020}  \\ \hline
DD2F-SF-8  &  T  & 2.026 & 12.232 & 426.3 & 2.915 & 325.0 & 2.877 & 365.6 & 2.810 & 471.4 &  h & \cite{Kaltenborn2017,Bastian2018,Fischer2018,Bauswein2019,Bastian2020}  \\ \hline
VBAG  &  T  & 1.932 & 12.214 & 422.3 & 2.885 & 345.5 & 2.848 & 388.4 & 2.775 & 505.4 &  h & \cite{Cierniak2018}  \\ \hline
ENG  &  B  & 2.238 & 11.909 & 370.0 & 2.975 & 251.2 & 2.975 & 257.6 & 2.975 & 281.6 &  b & \cite{Engvik1996,Read2009a}  \\ \hline
APR3  &  B  & 2.365 & 11.963 & 367.2 & 3.075 & 206.0 & 3.075 & 211.0 & 3.075 & 229.6 &  b & \cite{Akmal1998,Read2009a}  \\ \hline
GNH3  &  B  & 1.961 & 13.772 & 857.8 & 3.075 & 436.7 & 2.975 & 577.0 & 2.875 & 806.1 &  e & \cite{Glendenning1985,Read2009a}  \\ \hline
SAPR  &  T  & 2.196 & 11.474 & 267.9 & 2.875 & 225.5 & 2.913 & 213.2 & 2.875 & 256.9 &  b & \cite{Schneider2019}  \\ \hline
SAPRLDP  &  T  & 2.247 & 12.369 & 449.3 & 3.025 & 271.0 & 3.062 & 257.4 & 3.025 & 309.4 &  b & \cite{Schneider2019}  \\ \hline
SSkAPR  &  T  & 2.028 & 12.304 & 442.6 & 2.950 & 312.7 & 2.938 & 331.6 & 2.875 & 420.8 &  b & \cite{Schneider2019}  \\ \hline

\end{tabular}
\caption{EoS sample considered in this work. Acronyms are given in the first column, and references in the last column. In the 12th column, EoS models are grouped in three classes as described in Sect.~\ref{sec:eos}: ``b'' for the hadronic base sample, ``e'' for the excluded hadronic set of models, and ``h'' for EoSs with a phase transition to deconfined quark matter. ``T'' and ``B'' in the second column indicate whether the EoS table provides the full temperature dependence or only a barotropic relation, respectively. Third to fifth columns list some properties of static stars for the given EoS. The next six columns provide the threshold mass $M_\mathrm{thres}$ and the combined tidal deformability of a system with $M_\mathrm{thres}$ for binary mass ratios $q=1$, $q=0.85$ and $q=0.7$.}
\label{tab:data}
\end{table*}

\section{$M_\mathrm{thres}$ relations with different NS properties}\label{sec:opt}
In Sect.~\ref{sec:mthr} we employ stellar properties of nonrotating NSs to describe the EoS dependence of the threshold binary mass $M_\mathrm{thres}$. Depending on which stellar parameters are chosen, fit formulae for $M_\mathrm{thres}$ are more or less tight. Here we summarize the influence of the chosen NS radius or tidal deformability on the accuracy of the fit.

We consider functions of the form $M_\mathrm{thres}=M_\mathrm{thres}(M_\mathrm{max},R_{X})=a M_\mathrm{max} + b R_X +c$, where $R_{X}$ is the radius of a NS with mass $X\in \{1.1, 1.2, 1.3, 1.4, 1.5, 1.6, 1.7, 1.8, 1.9, 2.0\}$ in solar masses\footnote{For a few EoS models of the base sample $M_\mathrm{max}$ is smaller than 2.0~$M_\odot$, in which case we employ $R_\mathrm{max}$ instead of $R_{2.0}$.}. We employ the hadronic base sample and the simulation data for $q=1$. Table~\ref{tab:optr} provides the fit parameters and the maximum and average deviations between fit and data, which quantify the accuracy of the relations. The table clearly shows that relations with $R_{1.6}$ or $R_{1.7}$ yield the most accurate description of the data.

\begin{table*} 
\begin{tabular}{|l|c|c|c|c|c|c|c|}  \hline 
$X$  & $a$ & $b$ & $c$ & max. dev. & av. dev. & $\sum$ sq. res. & N \\ \hline \hline

%execfile("script-RX.py")

% 1.1 & 0.646 & 0.163 & -0.409 & 0.069 & 0.029 & 0.028 & 23 \\ \hline  %  2.2429889018
% 
% 1.2 & 0.635 & 0.164 & -0.39 & 0.064 & 0.025 & 0.0222 & 23 \\ \hline  %  1.77325168359
% 
% 1.3 & 0.62 & 0.164 & -0.36 & 0.058 & 0.023 & 0.0174 & 23 \\ \hline  %  1.39415196181
% 
% 1.4 & 0.602 & 0.164 & -0.318 & 0.053 & 0.02 & 0.0139 & 23 \\ \hline  %  1.11322774428
% 
% 1.5 & 0.579 & 0.164 & -0.265 & 0.047 & 0.018 & 0.0113 & 23 \\ \hline  %  0.901051089893
% 
% 1.6 & 0.547 & 0.165 & -0.198 & 0.042 & 0.016 & 0.0096 & 23 \\ \hline  %  0.767579782257
% 
% 1.7 & 0.502 & 0.166 & -0.109 & 0.039 & 0.016 & 0.0087 & 23 \\ \hline  %  0.696793340827
% 
% 1.8 & 0.437 & 0.169 & 0.014 & 0.042 & 0.016 & 0.0086 & 23 \\ \hline  %  0.689441431909
% 
% 1.9 & 0.329 & 0.174 & 0.211 & 0.046 & 0.018 & 0.0102 & 23 \\ \hline  %  0.812575140738
% 
% 2.0 & 0.059 & 0.184 & 0.732 & 0.083 & 0.028 & 0.0273 & 23 \\ \hline  %  2.18752481439
% 
% $\mathrm{max}$ & 0.45 & 0.189 & -0.011 & 0.059 & 0.018 & 0.0137 & 23 \\ \hline  %  1.09739002287

1.1 & 0.646 $\pm$ 0.061 & 0.163 $\pm$ 0.010 & -0.409 $\pm$ 0.179 & 0.069 & 0.029 & 0.0280 & 23  \\ \hline  %  2.2429889018

1.2 & 0.635 $\pm$ 0.054 & 0.164 $\pm$ 0.009 & -0.390 $\pm$ 0.158 & 0.064 & 0.025 & 0.0222 & 23  \\ \hline  %  1.77325168359

1.3 & 0.620 $\pm$ 0.048 & 0.164 $\pm$ 0.008 & -0.360 $\pm$ 0.139 & 0.058 & 0.023 & 0.0174 & 23  \\ \hline  %  1.39415196181

1.4 & 0.602 $\pm$ 0.043 & 0.164 $\pm$ 0.007 & -0.318 $\pm$ 0.123 & 0.053 & 0.020 & 0.0139 & 23  \\ \hline  %  1.11322774428

1.5 & 0.579 $\pm$ 0.039 & 0.164 $\pm$ 0.006 & -0.265 $\pm$ 0.109 & 0.047 & 0.018 & 0.0113 & 23  \\ \hline  %  0.901051089893

1.6 & 0.547 $\pm$ 0.036 & 0.165 $\pm$ 0.006 & -0.198 $\pm$ 0.099 & 0.042 & 0.016 & 0.0096 & 23  \\ \hline  %  0.767579782257

1.7 & 0.502 $\pm$ 0.034 & 0.166 $\pm$ 0.006 & -0.109 $\pm$ 0.092 & 0.039 & 0.016 & 0.0087 & 23  \\ \hline  %  0.696793340827

1.8 & 0.437 $\pm$ 0.035 & 0.169 $\pm$ 0.006 & 0.014 $\pm$ 0.089 & 0.042 & 0.016 & 0.0086 & 23  \\ \hline  %  0.689441431909

1.9 & 0.329 $\pm$ 0.039 & 0.174 $\pm$ 0.006 & 0.211 $\pm$ 0.093 & 0.046 & 0.018 & 0.0102 & 23  \\ \hline  %  0.812575140738

2.0 & 0.059 $\pm$ 0.070 & 0.184 $\pm$ 0.011 & 0.732 $\pm$ 0.139 & 0.083 & 0.028 & 0.0273 & 23  \\ \hline  %  2.18752481439

$M_\mathrm{max}$ & 0.450 $\pm$ 0.043 & 0.189 $\pm$ 0.008 & -0.011 $\pm$ 0.114 & 0.059 & 0.018 & 0.0137 & 23  \\ \hline  %  1.09739002287

\end{tabular} 
\caption{Bilinear fits $M_\mathrm{thres}=M_\mathrm{thres}(M_\mathrm{max},R_X)= a M_\mathrm{max} + b R_X + c$ employing NS radii with different masses $X$ (in solar masses), i.e. $R_X\equiv R(X)$ (see text). $a$, $b$ and $c$ are the resulting fit parameters with units such that masses are in $M_\odot$ and radii are in~km. Fifth and sixth columns specify the maximum and average deviation between fit and the underlying data (in $M_\odot$). Last two columns give the sum of the squared residuals being minimized by the fit procedure and the number of data points included in the fit. The last line provides the data for $R_X=R_\mathrm{max}$.} 
\label{tab:optr} 
\end{table*}

The radii of lighter NSs or more massive NSs lead to fits with larger deviations. This is understandable because the radii of NSs with small masses are only sensitive to the EoS at low densities and are thus not informative about EoS properties at higher densities, which are most relevant for the collapse behavior. The radii of NSs with very high mass are characteristic of the very high density regime of the EoS. Binary systems with $M_\mathrm{tot}\approx M_\mathrm{thres}$ may or may not reach such densities which are roughly comparable to those in nonrotating NSs with very high masses in the range of 2~$M_\odot$. However, also the EoS at somewhat lower densities affects the collapse behavior for instance by determining the end of the inspiral phase and thus the amount of angular momentum in the remnant (see~\cite{Bauswein2017a}). Finally, an optimally chosen $R_\mathrm{X}$ should complement the EoS information encoded in $M_\mathrm{max}$, which characterizes the EoS properties at very high densities. This may explain that the relations become worse for $X>1.7~M_\odot$ and for $R_\mathrm{max}$ in comparison to $R_{1.6}$ or $R_{1.7}$. Moreover, for stiff EoSs high-mass radii like $R_{2.0}$ are very similar to the radii of NSs with moderate masses. For soft EoSs $R_{2.0}$ is very close to $R_\mathrm{max}$, i.e. it captures the bending of the mass-radius relation towards the marginally stable configuration. The radius of a configuration on the more horizontal branch of the mass-radius relation, i.e. in the regime with $\frac{dM}{dR}\approx 0$, is very sensitive to the chosen fiducial mass, and thus even for very similar EoSs $R_{2.0}$ may differ significantly.

Following the same approach we examine relations of the form $M_\mathrm{thres}=M_\mathrm{thres}(M_\mathrm{max},\Lambda_{X})=a M_\mathrm{max} + b \Lambda_X +c$ in Tab.~\ref{tab:optlam}. Using the tidal deformability the tightest relations are found for NSs with masses in the range of about 1.3~$M_\odot$, which is significantly smaller than the ``optimal'' mass range found for NS radii. We note that the relations involving radii are generally somewhat more accurate than fits with the tidal deformability. One might interpret this in the sense that fits with radii may represent more fundamental relations.

Generally, all relations in Tabs.~\ref{tab:optr} and~\ref{tab:optlam} are relatively tight, which is not unexpected considering that the radii of NS with different masses are not completely unrelated for a fixed EoS. Also, the tidal deformability and radii of NSs are strongly correlated by definition. In the main text we discuss $\Lambda_{1.4}$, which is closer to the combined tidal deformability of GW170817. We also focus on $R_{1.6}$ instead of $R_{1.7}$, which actually results in a marginally better description, because $R_{1.6}$ was used in previous publications.

\begin{table*} 
\begin{tabular}{|l|c|c|c|c|c|c|c|}  \hline 
$X$  & $a$ & $b$ & $c$ & max. dev. & av. dev. & $\sum$ sq. res. & N \\ \hline \hline

%execfile("script-LamX.py")

% 
% 1.1 & 0.661 & 2.192e-04 & 1.17 & 0.078 & 0.025 & 0.0228 & 23 \\ \hline  %  1.82351797551
% 
% 
% 1.2 & 0.642 & 3.453e-04 & 1.222 & 0.057 & 0.024 & 0.0195 & 23 \\ \hline  %  1.56206133408
% 
% 
% 1.3 & 0.618 & 5.299e-04 & 1.284 & 0.052 & 0.024 & 0.0187 & 23 \\ \hline  %  1.49788443572
% 
% 
% 1.4 & 0.589 & 7.973e-04 & 1.359 & 0.056 & 0.025 & 0.0201 & 23 \\ \hline  %  1.60870831873
% 
% 
% 1.5 & 0.554 & 1.188e-03 & 1.45 & 0.064 & 0.026 & 0.0229 & 23 \\ \hline  %  1.82807674286
% 
% 
% 1.6 & 0.505 & 1.771e-03 & 1.567 & 0.071 & 0.028 & 0.0265 & 23 \\ \hline  %  2.11847318917
% 
% 
% 1.7 & 0.437 & 2.661e-03 & 1.723 & 0.079 & 0.03 & 0.0305 & 23 \\ \hline  %  2.44313032644
% 
% 
% 1.8 & 0.339 & 4.078e-03 & 1.945 & 0.087 & 0.033 & 0.0352 & 23 \\ \hline  %  2.81643992873
% 
% 
% 1.9 & 0.18 & 6.472e-03 & 2.297 & 0.097 & 0.034 & 0.0421 & 23 \\ \hline  %  3.36965743242

1.1 & 0.661 $\pm$ 0.055 & $(2.192 \pm 0.1222 )\times 10^{-4}$ & 1.170 $\pm$ 0.121 & 0.078 & 0.025 & 0.0228 & 23  \\ \hline  %  1.82351797551

1.2 & 0.642 $\pm$ 0.051 & $(3.453 \pm 0.1774 )\times 10^{-4}$ & 1.222 $\pm$ 0.111 & 0.057 & 0.024 & 0.0195 & 23  \\ \hline  %  1.56206133408

1.3 & 0.618 $\pm$ 0.050 & $(5.299 \pm 0.2663 )\times 10^{-4}$ & 1.284 $\pm$ 0.108 & 0.052 & 0.024 & 0.0187 & 23  \\ \hline  %  1.49788443572

1.4 & 0.589 $\pm$ 0.052 & $(7.973 \pm 0.4159 )\times 10^{-4}$ & 1.359 $\pm$ 0.112 & 0.056 & 0.025 & 0.0201 & 23  \\ \hline  %  1.60870831873

1.5 & 0.554 $\pm$ 0.055 & $(11.88 \pm 0.6631 )\times 10^{-4}$ & 1.450 $\pm$ 0.119 & 0.064 & 0.026 & 0.0229 & 23  \\ \hline  %  1.82807674286

1.6 & 0.505 $\pm$ 0.060 & $(17.71 \pm 1.069 )\times 10^{-4}$ & 1.567 $\pm$ 0.128 & 0.071 & 0.028 & 0.0265 & 23  \\ \hline  %  2.11847318917

1.7 & 0.437 $\pm$ 0.065 & $(26.61 \pm 1.735  )\times 10^{-4}$& 1.723 $\pm$ 0.138 & 0.079 & 0.030 & 0.0305 & 23  \\ \hline  %  2.44313032644

1.8 & 0.339 $\pm$ 0.072 & $(40.78 \pm 2.874 )\times 10^{-4}$ & 1.945 $\pm$ 0.150 & 0.087 & 0.033 & 0.0352 & 23  \\ \hline  %  2.81643992873

1.9 & 0.180 $\pm$ 0.083 & $(64.72 \pm 5.039 )\times 10^{-4}$ & 2.297 $\pm$ 0.172 & 0.097 & 0.034 & 0.0421 & 23  \\ \hline  %  3.36965743242

\end{tabular} 
\caption{Bilinear fits $M_\mathrm{thres}=M_\mathrm{thres}(M_\mathrm{max},\Lambda_X)= a M_\mathrm{max} + b \Lambda_X + c$ employing the tidal deformability of NSs with different masses $X$ (in solar masses), i.e. $\Lambda_X\equiv \Lambda(X)$ (see text). $a$, $b$ and $c$ are the resulting fit parameters with their respective variances with units such that masses are in $M_\odot$ and $\Lambda$ is dimensionless. Fifth and sixth columns specify the maximum and average deviation between fit and the underlying data (in $M_\odot$). Last two columns give the sum of the squared residuals being minimized by the fit procedure and the number of data points included in the fit.} 
\label{tab:optlam} 
\end{table*}

\section{General $q$ dependent relations}\label{app:q}

In Sect.~\ref{sec:q} we describe multi-dimensional fits for the threshold binary mass $M_\mathrm{thres}$ for prompt BH formation which include an explicit dependence on the binary mass ratio $q$. We supplement these relations by additional functions which use a different ansatz to capture the dependence on the $q$. We provide here one table with more terms and one table with less terms. See Sect.~\ref{sec:q} for a more detailed discussion. The essential conclusion is that the fits with six terms (Tab.~\ref{tab:generalqallterms}) are approximately as tight as the relations with five terms discussed in Sect.~\ref{sec:q} (Tab.~\ref{tab:generalq}), whereas the relations including only one term to capture the $q$ dependence (Tab.~\ref{tab:generalqfourterms}) do not lead to a very accurate description.

Furthermore, we test which power $n$ in the terms involving $\delta q^n$ in Eq.~\eqref{eq:Mqgeneral} yields the most accurate description of the data for $q=0.7$, $q=0.85$ and $q=1$ adopting the base EoS sample. By adopting different values of $n$ we find in Tab.~\ref{tab:power} that $n=3$ minimizes the deviations between fit and data. The accuracy is not very sensitive on the exact value as long as $n$ is larger than about 1.5 to account for the fact that mass ratio effects become stronger with the asymmetry. Although the table suggests that $n\approx 5$ yields accurate relations, we caution that such high powers may not be the best choice. We determine $n$ by fits to three data points per EoS at $q=1$, $q=0.85$ and $q=0.7$. $M_\mathrm{thres}$ does not change strongly between $q=0.85$ and $q=1$. Hence, any arbitrarily high power $n$ can describe this roughly constant plateau in the range $0.85\leq q \leq 1$ and in addition capture a single data point at $q=0.7$ with a stronger decrease of $M_\mathrm{thres}$ (see Fig.~\ref{fig:dd2f}).

\begin{turnpage}
\begin{table*} 
\begin{tabular}{|l|c|c|c|c|c|c|c|c|c|c|}  \hline 
%fit / EoS sample  & $c_1$ & $c_2$ & $c_3$ & $c_4$ & $c_5$ & $c_6$ & max. dev. & av. dev. & $\sum$ sq. res. & N \\ \hline 

\multicolumn{11}{|l|}{$M_\mathrm{thres}(q,M_\mathrm{max},R_{1.6})=c_1 M_\mathrm{max} + c_2 R_{1.6} +c_3 + c_4 \delta q^3 M_\mathrm{max}+ c_5 \delta q^3 R_{1.6} +c_6 \delta q^3$}  \\ \hline 

% 
% base  & 0.569 & 0.160 & -0.184 & 9.844 & -1.644 & -3.332 & 0.067 & 0.017 & 0.0333 & 69 \\ \hline  %  0.845545072286
% basehyb  & 0.656 & 0.152 & -0.298 & 9.753 & -1.653 & -3.066 & 0.116 & 0.035 & 0.1948 & 96 \\ \hline  %  3.46281332766
% baseex  & 0.627 & 0.154 & -0.243 & 9.765 & -1.646 & -3.146 & 0.108 & 0.021 & 0.0729 & 93 \\ \hline  %  1.34069533546
% basehybex  & 0.658 & 0.155 & -0.336 & 9.742 & -1.637 & -3.231 & 0.117 & 0.033 & 0.2271 & 120 \\ \hline  %  3.18803154252

EoSs  & $c_1$ & $c_2/10^{-1}$ & $c_3/10^{-1}$ & $c_4$ & $c_5$ & $c_6$ & max. dev. & av. dev. & $\sum$ sq. res. & N \\ \hline 

b  & 0.569 $\pm$ 0.028 & $1.598 \pm 0.046 $ & $-1.845 \pm 0.778 $  & 9.844 $\pm$ 1.794 & $-1.644 \pm 0.293$ & $-3.332 \pm 4.950$ & 0.067 & 0.017 & 0.0333 & 69 \\ \hline    %  0.845545072286
b + h  & 0.656 $\pm$ 0.050 & $1.522 \pm 0.092 $ & $-2.984 \pm 1.511 $  & 9.753 $\pm$ 3.184 & $-1.653 \pm 0.583$ & $-3.066 \pm 9.619$ & 0.116 & 0.035 & 0.1948 & 96 \\ \hline    %  3.46281332766
b + e  & 0.627 $\pm$ 0.020 & $1.541 \pm 0.037 $ & $-2.426 \pm 0.485 $  & 9.765 $\pm$ 1.252 & $-1.646 \pm 0.233$ & $-3.146 \pm 3.085$ & 0.108 & 0.021 & 0.0729 & 93 \\ \hline    %  1.34069533546
b + h + e  & 0.658 $\pm$ 0.029 & $1.550 \pm 0.056 $ & $-3.363\pm 0.716 $  & 9.742 $\pm$ 1.829 & $-1.637 \pm 0.358$ & $-3.231 \pm 4.557$ & 0.117 & 0.033 & 0.2271 & 120 \\ \hline    %  3.18803154252

\hline
\hline

\multicolumn{11}{|l|}{$M_\mathrm{thres}(q,M_\mathrm{max},R_\mathrm{max})=c_1 M_\mathrm{max} + c_2 R_\mathrm{max} +c_3 + c_4 \delta q^3 M_\mathrm{max}+ c_5 \delta q^3 R_\mathrm{max} +c_6 \delta q^3$}  \\ \hline

% b  & 0.477 & 0.182 & 0.004 & 10.779 & -1.864 & -5.381 & 0.070 & 0.021 & 0.0505 & 69 \\ \hline  %  1.28220917272
% bh  & 0.591 & 0.163 & -0.049 & 10.298 & -1.477 & -8.491 & 0.101 & 0.032 & 0.1532 & 96 \\ \hline  %  2.72288659604
% be  & 0.497 & 0.186 & -0.078 & 11.098 & -1.970 & -5.011 & 0.072 & 0.022 & 0.0709 & 93 \\ \hline  %  1.30449817033
% bhe  & 0.559 & 0.179 & -0.145 & 10.443 & -1.745 & -6.062 & 0.122 & 0.031 & 0.1934 & 120 \\ \hline  %  2.71504205091

EoSs  & $c_1$ & $c_2/10^{-1}$ & $c_3/10^{-2}$ & $c_4$ & $c_5$ & $c_6$ & max. dev. & av. dev. & $\sum$ sq. res. & N \\ \hline 

b  & 0.477 $\pm$ 0.035 & $1.825 \pm 0.065 $ & $0.392 \pm 9.205 $  & 10.779 $\pm$ 2.241 & $-1.864 \pm 0.416$ & $-5.381 \pm 5.859$ & 0.070 & 0.021 & 0.0505 & 69 \\ \hline    %  1.28220917272
b + h  & 0.591 $\pm$ 0.045 & $1.626 \pm 0.086 $ & $-4.869 \pm 12.45 $  & 10.298 $\pm$ 2.837 & $-1.477 \pm 0.548$ & $-8.491 \pm 7.922$ & 0.101 & 0.032 & 0.1532 & 96 \\ \hline    %  2.72288659604
b + e  & 0.497 $\pm$ 0.021 & $1.863 \pm 0.0437 $ & $-7.794 \pm 4.570 $  & 11.098 $\pm$ 1.327 & $-1.970 \pm 0.278$ & $-5.011 \pm 2.909$ & 0.072 & 0.022 & 0.0709 & 93 \\ \hline    %  1.30449817033
b + h + e  & 0.559 $\pm$ 0.028 & $1.788 \pm 0.060 $ & $-14.48 \pm 6.235 $  & 10.443 $\pm$ 1.787 & $-1.745 \pm 0.381$ & $-6.062 \pm 3.969$ & 0.122 & 0.031 & 0.1934 & 120 \\ \hline    %  2.71504205091

\hline
\hline

\multicolumn{11}{|l|}{$M_\mathrm{thres}(q,M_\mathrm{max},\Lambda_{1.4})=c_1 M_\mathrm{max} + c_2 \Lambda_{1.4} +c_3 + c_4 \delta q^3 M_\mathrm{max}+ c_5 \delta q^3 \Lambda_{1.4} +c_6 \delta q^3$}  \\ \hline

% 
% b  & 0.611 & 0.001 & 1.327 & 9.453 & -0.008 & -18.758 & 0.080 & 0.027 & 0.0732 & 69 \\ \hline  %  1.85961324663
% bh  & 0.661 & 0.001 & 1.212 & 9.707 & -0.009 & -19.348 & 0.111 & 0.039 & 0.2071 & 96 \\ \hline  %  3.68243472345
% be  & 0.542 & 0.001 & 1.552 & 10.750 & -0.006 & -22.431 & 0.113 & 0.035 & 0.1832 & 93 \\ \hline  %  3.36884246582
% bhe  & 0.562 & 0.001 & 1.494 & 10.791 & -0.006 & -22.613 & 0.106 & 0.043 & 0.3082 & 120 \\ \hline  %  4.32572058387

EoSs  & $c_1$ & $c_2/10^{-4}$ & $c_3$ & $c_4$ & $c_5/10^{-3}$ & $c_6$ & max. dev. & av. dev. & $\sum$ sq. res. & N \\ \hline 

b  & 0.611 $\pm$ 0.042 & $7.711 \pm 0.335 $ & $1.327 \pm 0.090$  & 9.453 $\pm$ 2.651 & $-8.440 \pm 2.133$ & $-18.76\pm 5.741$ & 0.080 & 0.027 & 0.0732 & 69 \\ \hline    %  1.85961324663
b + h  & 0.661 $\pm$ 0.052 & $7.546 \pm 0.470 $ & $1.212 \pm 0.111 $  & 9.707 $\pm$ 3.283 & $-8.531 \pm 2.993$ & $-19.35 \pm 7.076$ & 0.111 & 0.039 & 0.2071 & 96 \\ \hline    %  3.68243472345
b + e  & 0.542 $\pm$ 0.033 & $5.632 \pm 0.218 $ & $1.552 \pm 0.067 $  & 10.750 $\pm$ 2.090 & $-6.125 \pm 1.385$ & $-22.43 \pm 4.277$ & 0.113 & 0.035 & 0.1832 & 93 \\ \hline    %  3.36884246582
b + h + e  & 0.562 $\pm$ 0.036 & $5.700 \pm 0.246 $ & $1.494 \pm 0.071 $  & 10.791 $\pm$ 2.263 & $-6.065 \pm 1.563$ & $-22.61 \pm 4.544$ & 0.106 & 0.043 & 0.3082 & 120 \\ \hline    %  4.32572058387

\hline
\hline

\multicolumn{11}{|l|}{$M_\mathrm{thres}(q,M_\mathrm{max},\tilde{\Lambda}_\mathrm{thres})=c_1 M_\mathrm{max} + c_2 \tilde{\Lambda}_\mathrm{thres} +c_3 + c_4 \delta q^3 M_\mathrm{max}+ c_5 \delta q^3 \tilde{\Lambda}_\mathrm{thres} +c_6 \delta q^3$}  \\ \hline

% b  & 1.408 & 0.002 & -0.740 & 1.738 & -0.054 & 6.801 & 0.109 & 0.040 & 0.1610 & 69 \\ \hline  %  4.0892359788
% bh  & 1.067 & 0.001 & 0.365 & 6.082 & -0.020 & -10.219 & 0.186 & 0.078 & 0.7911 & 96 \\ \hline  %  14.0635232959
% be  & 1.424 & 0.002 & -0.697 & 2.315 & -0.049 & 4.087 & 0.278 & 0.056 & 0.4990 & 93 \\ \hline  %  9.17658013163
% bhe  & 1.354 & 0.001 & -0.372 & 4.387 & -0.029 & -5.295 & 0.269 & 0.086 & 1.3227 & 120 \\ \hline  %  18.5641767286
% 

EoSs  & $c_1$ & $c_2/10^{-3}$ & $c_3/10^{-1}$ & $c_4$ & $c_5/10^{-2}$ & $c_6$ & max. dev. & av. dev. & $\sum$ sq. res. & N \\ \hline 

b  & 1.408 $\pm$ 0.079 & $2.392 \pm 0.159 $ & $-7.398 \pm 2.021 $  & 1.738 $\pm$ 5.120 & $-5.370 \pm 0.760$ & $6.801 \pm 12.64$ & 0.109 & 0.040 & 0.1610 & 69 \\ \hline    %  4.0892359788
b + h  & 1.067 $\pm$ 0.135 & $0.950 \pm 0.215 $ & $3.646 \pm 3.348 $  & 6.082 $\pm$ 8.820 & $-2.037 \pm 1.064$ & $-10.22 \pm 21.44$ & 0.186 & 0.078 & 0.7911 & 96 \\ \hline    %  14.0635232959
b + e  & 1.424 $\pm$ 0.057 & $2.142 \pm 0.148 $ & $-6.965 \pm 1.543 $  & 2.315 $\pm$ 3.633 & $-4.886 \pm 0.685$ & $4.087 \pm 9.442$ & 0.278 & 0.056 & 0.4990 & 93 \\ \hline    %  9.17658013163
b + h + e  & 1.354 $\pm$ 0.079 & $1.454 \pm 0.187 $ & $-3.722 \pm 2.120 $  & 4.387 $\pm$ 5.053 & $-2.948 \pm 0.884$ & $-5.295 \pm 13.08$ & 0.269 & 0.086 & 1.3227 & 120 \\ \hline    %  18.5641767286

\hline

\end{tabular} 
\caption{Different fits describing the EoS dependence of the threshold binary mass $M_\mathrm{thres}$ for prompt BH formation including an explicit dependence on the binary mass ratio $q$ through $\delta q=1-q$ (see main text). First column specifies the set of EoSs used for the fit (``b'' $\equiv$ hadronic base sample (a), ``e'' $\equiv$ exlcuded hadronic sample (b), ``h'' $\equiv$ hybrid sample (c) as defined  in Sect.~\ref{sec:eos}). Fit parameters $c_i$ and their respective variances are given in second to sixth columns. The units of the fit parameters $c_i$ are such that masses are in $M_\odot$, radii in km and tidal deformabilities dimensionless. These relations are obtained by least-square fits to the $M_\mathrm{thres}$ data for $q=1$, $q=0.85$ and $q=0.7$.}
\label{tab:generalqallterms} 
\end{table*} 

\end{turnpage}

\begin{table*} 
\begin{tabular}{|l|c|c|c|c|c|c|c|c|}  \hline 
%fit / EoS sample  & $c_1$ & $c_2$ & $c_3$ & $c_4$ & max. dev. & av. dev. & $\sum$ sq. res. & N \\ \hline 

\multicolumn{9}{|l|}{$M_\mathrm{thres}(q,M_\mathrm{max},R_{1.6})=c_1 M_\mathrm{max} + c_2 R_{1.6} +c_3 + c_4 \delta q^3 R_{1.6}$}  \\ \hline
% 
% base  & 0.669 & 0.145 & -0.218 & -0.150 & 0.117 & 0.023 & 0.0599 & 69 \\ \hline  %  1.47325848448
% basehyb  & 0.754 & 0.137 & -0.329 & -0.183 & 0.131 & 0.040 & 0.2276 & 96 \\ \hline  %  3.95885625142
% baseex  & 0.725 & 0.139 & -0.274 & -0.185 & 0.120 & 0.031 & 0.1327 & 93 \\ \hline  %  2.38619110105
% basehybex  & 0.757 & 0.140 & -0.369 & -0.202 & 0.128 & 0.041 & 0.2907 & 120 \\ \hline  %  4.01032137537

EoSs  & $c_1$ & $c_2$ & $c_3$ & $c_4$ & max. dev. & av. dev. & $\sum$ sq. res. & N \\ \hline

b  & 0.669 $\pm$ 0.029 & $0.145 \pm 0.005 $ & $-0.218 \pm 0.080 $  & -0.150 $\pm$ 0.026 &  0.117 & 0.023 & 0.0599 & 69 \\ \hline    %  1.47325848448
b + h  & 0.754 $\pm$ 0.041 & $0.137 \pm 0.008 $ & $-0.329 \pm 0.125 $  & -0.183 $\pm$ 0.035 &  0.131 & 0.040 & 0.2276 & 96 \\ \hline    %  3.95885625142
b + e  & 0.725 $\pm$ 0.020 & $0.139 \pm 0.004 $ & $-0.274 \pm 0.050 $  & -0.185 $\pm$ 0.027 &  0.120 & 0.031 & 0.1327 & 93 \\ \hline    %  2.38619110105
b + h + e  & 0.757 $\pm$ 0.025 & $0.140 \pm 0.005 $ & $-0.369 \pm 0.062 $  & -0.202 $\pm$ 0.031 &  0.128 & 0.041 & 0.2907 & 120 \\ \hline    %  4.01032137537

\hline
\hline

\multicolumn{9}{|l|}{$M_\mathrm{thres}(q,M_\mathrm{max},R_\mathrm{max})=c_1 M_\mathrm{max} + c_2 R_\mathrm{max} +c_3 + c_4 \delta q^3 R_\mathrm{max}$}  \\ \hline

% b  & 0.586 & 0.165 & -0.051 & -0.169 & 0.099 & 0.026 & 0.0768 & 69 \\ \hline  %  1.89086875931
% bh  & 0.696 & 0.150 & -0.135 & -0.207 & 0.107 & 0.037 & 0.1818 & 96 \\ \hline  %  3.16152750464
% be  & 0.609 & 0.168 & -0.129 & -0.208 & 0.104 & 0.030 & 0.1315 & 93 \\ \hline  %  2.36418278965
% bhe  & 0.665 & 0.163 & -0.206 & -0.228 & 0.131 & 0.037 & 0.2531 & 120 \\ \hline  %  3.49111228042

EoSs  & $c_1$ & $c_2$ & $c_3$ & $c_4$ & max. dev. & av. dev. & $\sum$ sq. res. & N \\ \hline

b  & 0.586 $\pm$ 0.033 & $0.165 \pm 0.006 $ & $-0.051 \pm 0.087 $  & -0.169 $\pm$ 0.033 &  0.099 & 0.026 & 0.0768 & 69 \\ \hline    %  1.89086875931
b + h  & 0.696 $\pm$ 0.037 & $0.150 \pm 0.007 $ & $-0.135 \pm 0.104 $  & -0.207 $\pm$ 0.036 &  0.107 & 0.037 & 0.1818 & 96 \\ \hline    %  3.16152750464
b + e  & 0.609 $\pm$ 0.022 & $0.168 \pm 0.005 $ & $-0.129 \pm 0.048 $  & -0.208 $\pm$ 0.030 &  0.104 & 0.030 & 0.1315 & 93 \\ \hline    %  2.36418278965
b + h + e  & 0.665 $\pm$ 0.025 & $0.163 \pm 0.005 $ & $-0.206 \pm 0.055 $  & -0.228 $\pm$ 0.033 &  0.131 & 0.037 & 0.2531 & 120 \\ \hline    %  3.49111228042

\hline
\hline

\multicolumn{9}{|l|}{$M_\mathrm{thres}(q,M_\mathrm{max},\Lambda_{1.4})=c_1 M_\mathrm{max} + c_2 \Lambda_{1.4} +c_3 + c_4 \delta q^3 \Lambda_{1.4}$}  \\ \hline

% b  & 0.707 & 0.001 & 1.137 & -0.005 & 0.115 & 0.029 & 0.0907 & 69 \\ \hline  %  2.23183352496
% bh  & 0.759 & 0.001 & 1.016 & -0.006 & 0.111 & 0.041 & 0.2292 & 96 \\ \hline  %  3.98545454714
% be  & 0.651 & 0.001 & 1.325 & -0.004 & 0.114 & 0.043 & 0.2404 & 93 \\ \hline  %  4.32244517707
% bhe  & 0.671 & 0.001 & 1.265 & -0.004 & 0.126 & 0.048 & 0.3760 & 120 \\ \hline  %  5.18609013798

EoSs  & $c_1$ & $c_2/10^{-4}$ & $c_3$ & $c_4/10^{-3}$ & max. dev. & av. dev. & $\sum$ sq. res. & N \\ \hline 

b  & 0.707 $\pm$ 0.035 & $7.346 \pm 0.299 $ & $1.137 \pm 0.077 $  & -4.834 $\pm$ 0.881 &  0.115 & 0.029 & 0.0907 & 69 \\ \hline    %  2.23183352496
b + h  & 0.759 $\pm$ 0.041 & $7.250 \pm 0.391 $ & $1.016 \pm 0.089 $  & -5.608 $\pm$ 0.999 &  0.111 & 0.041 & 0.2292 & 96 \\ \hline    %  3.98545454714
b + e  & 0.651 $\pm$ 0.029 & $5.378 \pm 0.203 $ & $1.325 \pm 0.059 $  & -3.618$\pm$ 0.696 &  0.114 & 0.043 & 0.2404 & 93 \\ \hline    %  4.32244517707
b + h + e  & 0.671 $\pm$ 0.030 & $5.497 \pm 0.220 $ & $1.265 \pm 0.060 $  & -4.060 $\pm$ 0.718 &  0.126 & 0.048 & 0.3760 & 120 \\ \hline    %  5.18609013798

\hline
\hline

\multicolumn{9}{|l|}{$M_\mathrm{thres}(q,M_\mathrm{max},\tilde{\Lambda}_\mathrm{thres})=c_1 M_\mathrm{max} + c_2 \tilde{\Lambda}_\mathrm{thres} +c_3 + c_4 \delta q^3 \tilde{\Lambda}_\mathrm{thres}$}  \\ \hline

% b  & 1.379 & 0.002 & -0.515 & -0.021 & 0.194 & 0.051 & 0.2502 & 69 \\ \hline  %  6.15786557719
% bh  & 1.110 & 0.001 & 0.316 & -0.013 & 0.190 & 0.079 & 0.8169 & 96 \\ \hline  %  14.2071533766
% be  & 1.421 & 0.002 & -0.549 & -0.024 & 0.214 & 0.066 & 0.6468 & 93 \\ \hline  %  11.6280431852
% bhe  & 1.384 & 0.001 & -0.371 & -0.019 & 0.266 & 0.089 & 1.3881 & 120 \\ \hline  %  19.1459515485

EoSs  & $c_1$ & $c_2/10^{-3}$ & $c_3$ & $c_4/10^{-2}$ & max. dev. & av. dev. & $\sum$ sq. res. & N \\ \hline 

b  & 1.379 $\pm$ 0.076 & $1.832 \pm 0.137 $ & $-0.515 \pm 0.189 $  & -2.127 $\pm$ 0.225 &  0.194 & 0.051 & 0.2502 & 69 \\ \hline    %  6.15786557719
b + h  & 1.110 $\pm$ 0.107 & $0.819 \pm 0.159 $ & $0.316 \pm 0.261 $  & -1.327 $\pm$ 0.262 &  0.190 & 0.079 & 0.8169 & 96 \\ \hline    %  14.2071533766
b + e  & 1.421 $\pm$ 0.049 & $1.700 \pm 0.117 $ & $-0.549 \pm 0.130 $  & -2.394 $\pm$ 0.237 &  0.214 & 0.066 & 0.6468 & 93 \\ \hline    %  11.6280431852
b + h + e  & 1.384 $\pm$ 0.062 & $1.264 \pm 0.136 $ & $-0.371 \pm 0.163 $  & -1.897 $\pm$ 0.256 &  0.266 & 0.089 & 1.3881 & 120 \\ \hline    %  19.1459515485

\hline

\end{tabular} 
\caption{Different fits describing the EoS dependence of the threshold binary mass $M_\mathrm{thres}$ for prompt BH formation including an explicit dependence on the binary mass ratio $q$ through $\delta q=1-q$ (see main text). First column specifies the set of EoSs used for the fit (``b'' $\equiv$ hadronic base sample (a), ``e'' $\equiv$ exlcuded hadronic sample (b), ``h'' $\equiv$ hybrid sample (c) as defined  in Sect.~\ref{sec:eos}). Fit parameters $c_i$ and their respective variances are given in second to sixth columns. The units of the fit parameters $c_i$ are such that masses are in $M_\odot$, radii in km and tidal deformabilities dimensionless. These relations are obtained by least-square fits to the $M_\mathrm{thres}$ data for $q=1$, $q=0.85$ and $q=0.7$.} 
\label{tab:generalqfourterms} 
\end{table*}

\begin{table*} 
\begin{tabular}{|l|c|c|c|c|c|c|c|c|c|c|}  \hline

\multicolumn{11}{|l|}{$M_\mathrm{thres}(q,M_\mathrm{max},R_{1.6})=c_1 M_\mathrm{max} + c_2 R_{1.6} +c_3 + c_4 \delta q^n M_\mathrm{max}+ c_5 \delta q^n R_{1.6}$}  \\ \hline %close(); execfile("script-generalq-Mthres-checkpower.py")

EoSs  & $n$ & $c_1$ & $c_2$ & $c_3$ & $c_4$ & $c_5$ & max. dev. & av. dev. & $\sum$ sq. res. & N \\ \hline 

% base  & 1.0 & 0.537 & 0.169 & -0.218 & 0.883 & -0.172 & 0.0806 & 0.0187 & 0.0415 & 69  \\ \hline  %  1.03644319797
% base  & 1.5 & 0.549 & 0.167 & -0.218 & 1.621 & -0.317 & 0.0716 & 0.0177 & 0.0361 & 69  \\ \hline  %  0.902911715938
% base  & 2.0 & 0.561 & 0.164 & -0.218 & 2.876 & -0.564 & 0.0677 & 0.0172 & 0.0341 & 69  \\ \hline  %  0.852441584388
% base  & 2.5 & 0.571 & 0.162 & -0.218 & 5.074 & -0.996 & 0.0663 & 0.0171 & 0.0335 & 69  \\ \hline  %  0.838428827683
% base  & \bf 3.0 & 0.578 & 0.161 & -0.218 & 8.987 & -1.767 & \bf 0.0660 & \bf 0.0171 & 0.0335 & 69  \\ \hline  %  0.838225448368
% base  & 3.5 & 0.583 & 0.160 & -0.218 & 16.018 & -3.153 & 0.0659 & 0.0172 & 0.0337 & 69  \\ \hline  %  0.842372111854
% base  & 4.0 & 0.587 & 0.159 & -0.218 & 28.719 & -5.658 & 0.0661 & 0.0173 & 0.0339 & 69  \\ \hline  %  0.84722913814
% base  & 4.5 & 0.589 & 0.159 & -0.218 & 51.738 & -10.197 & 0.0662 & 0.0174 & 0.0341 & 69  \\ \hline  %  0.85154575226
% base  & 5.0 & 0.591 & 0.159 & -0.218 & 93.549 & -18.444 & 0.0663 & 0.0175 & 0.0342 & 69  \\ \hline  %  0.855010647553

base  & 1.0 & 0.537 $\pm$ 0.032 & $0.169 \pm 0.005 $ & $-0.218 \pm 0.066 $ & 0.883 $\pm$ 0.138 & -0.172 $\pm$ 0.025 &  0.081 & 0.019 & 0.0415 & 69 \\ \hline    %  1.03644319797
base  & 1.5 & 0.549 $\pm$ 0.028 & $0.167 \pm 0.005 $ & $-0.218 \pm 0.062 $ & 1.621 $\pm$ 0.232 & -0.317 $\pm$ 0.042 &  0.072 & 0.018 & 0.0361 & 69 \\ \hline    %  0.902911715938
base  & 2.0 & 0.561 $\pm$ 0.027 & $0.164 \pm 0.004 $ & $-0.218 \pm 0.060 $ & 2.876 $\pm$ 0.402 & -0.564 $\pm$ 0.073 &  0.068 & 0.017 & 0.0341 & 69 \\ \hline    %  0.852441584388
base  & 2.5 & 0.571 $\pm$ 0.026 & $0.162 \pm 0.004 $ & $-0.218 \pm 0.060 $ & 5.074 $\pm$ 0.709 & -0.996 $\pm$ 0.128 &  0.066 & 0.017 & 0.0335 & 69 \\ \hline    %  0.838428827683
base  & 3.0 & 0.578 $\pm$ 0.025 & $0.161 \pm 0.004 $ & $-0.218 \pm 0.060 $ & 8.987 $\pm$ 1.268 & -1.767 $\pm$ 0.229 &  0.066 & 0.017 & 0.0335 & 69 \\ \hline    %  0.838225448368
base  & 3.5 & 0.583 $\pm$ 0.025 & $0.160 \pm 0.004 $ & $-0.218 \pm 0.060 $ & 16.018 $\pm$ 2.284 & -3.153 $\pm$ 0.412 &  0.066 & 0.017 & 0.0337 & 69 \\ \hline    %  0.842372111854
base  & 4.0 & 0.587 $\pm$ 0.025 & $0.159 \pm 0.004 $ & $-0.218 \pm 0.060 $ & 28.719 $\pm$ 4.133 & -5.658 $\pm$ 0.746 &  0.066 & 0.017 & 0.0339 & 69 \\ \hline    %  0.84722913814
base  & 4.5 & 0.589 $\pm$ 0.025 & $0.159 \pm 0.004 $ & $-0.218 \pm 0.060 $ & 51.738 $\pm$ 7.499 & -10.197 $\pm$ 1.354 &  0.066 & 0.017 & 0.0341 & 69 \\ \hline    %  0.85154575226
base  & 5.0 & 0.591 $\pm$ 0.025 & $0.159 \pm 0.004 $ & $-0.218 \pm 0.060 $ & 93.549 $\pm$ 13.635 & -18.444 $\pm$ 2.463 &  0.066 & 0.017 & 0.0342 & 69 \\ \hline    %  0.855010647553

\hline

\end{tabular} 
\caption{Different fits describing the EoS dependence of the threshold binary mass $M_\mathrm{thres}$ for prompt BH formation including an explicit dependence on the binary mass ratio $q$ through $\delta q=1-q$ to power $n$ (see main text). The units of the fit parameters $c_i$ and their respective variances are such that masses are in $M_\odot$ and radii in km. These relations are obtained by least-square fits to the $M_\mathrm{thres}$ data for $q=1$, $q=0.85$ and $q=0.7$ considering the base EoS sample and systematically varied powers $n$. Eighth and ninth columns specify the maximum and average deviation between fit and the underlying data revealing that a power of $n=3$ results in the best description of the data (highlighted by boldface entries). Last two columns give the sum of the squared residuals being minimized by the fit procedure and the number of data points included in the fit.}
\label{tab:power} 
\end{table*}

\end{document}